\documentclass[cernpreprint, atlasdraft=false, texlive=2020, UKenglish,paper=a4, texmf, orcidlogo]{atlasdoc}
 
\usepackage{atlaspackage}
\usepackage{atlasbiblatex}[showdoi=true]
 
\usepackage{atlasphysics}
\usepackage{wasysym} 
\usepackage{tikz}
\usepackage{rotating}
 
\ifthenelse{\boolean{AtlasDraft}}{
}{}
 
\graphicspath{{logos/}{figures/}}
 
\usepackage{ANA-STDM-2018-08-PAPER-defs}

\AtlasTitle{Measurement of the total cross section and $\rho$-parameter from elastic scattering in $pp$ collisions at $\sqrt{s}=13$ TeV  with the ATLAS detector}

\AtlasAbstract{
In a special run of the LHC with $\beta^\star = 2.5$~km, proton--proton elastic-scattering events were recorded at $\sqrt{s} = 13$~TeV with an integrated luminosity of $340~\upmu \textrm{b}^{-1}$ using the ALFA subdetector of ATLAS in 2016. The elastic cross section was measured differentially in the Mandelstam $t$ variable in the range from $-t = 2.5 \cdot  10^{-4}$~GeV$^{2}$ to $-t = 0.46$~GeV$^{2}$ using 6.9 million elastic-scattering candidates.
This paper presents measurements of the total cross section $\sigma_{\textrm{tot}}$, parameters of the nuclear slope, and
the $\rho$-parameter defined as the ratio of the real part to the imaginary part of the elastic-scattering amplitude in the limit $t \rightarrow 0$.
These parameters are determined from a fit to the differential elastic cross section using
the optical theorem and different parameterizations of the $t$-dependence. The results for
$\sigma_{\textrm{tot}}$ and $\rho$ are
\begin{equation*}
\sigma_{\textrm{tot}}(pp\rightarrow X) =  \mbox{104.7} \pm 1.1 \; \mbox{mb} , \; \; \;
\rho =  \mbox{0.098} \pm 0.011 .
\end{equation*}
The uncertainty in $\sigma_{\textrm{tot}}$ is dominated by the luminosity measurement, and in $\rho$ by imperfect knowledge of the detector alignment and by modelling of the nuclear amplitude.
}
 
\author{The ATLAS Collaboration}
 
\AtlasRefCode{STDM-2018-08}
 
\PreprintIdNumber{CERN-EP-2022-129}

\HepDataRecord{128017}
 
\AtlasJournal{EPJC}
\AtlasJournalRef{\EPJC 83 (2023) 441 789}
\AtlasDOI{10.1140/epjc/s10052-023-11436-8}

\hypersetup{pdftitle={ATLAS document},pdfauthor={The ATLAS Collaboration}}
 
\begin{document}
 
\maketitle
 
\tableofcontents

% The next lines are included from the .//Introduction.tex input file
 
\section{Introduction}
\label{sec:Intro}
Measurements of elastic scattering at hadron colliders give unique experimental access to non-perturbative dynamics, which cannot be calculated from first principles. Of particular importance are the total hadronic cross section, $\sigmatot$,  and the ratio of the real part to the imaginary part of the elastic-scattering amplitude ($\rho$-parameter), which probes Coulomb--nuclear interference (CNI). These  observables are related
by dispersion relations derived from foundational unitarity and analyticity arguments for scattering amplitudes.  Dispersion relations connect the $\rho$-parameter at a certain energy to the energy evolution of $\sigmatot$ both below and above this energy. The $\rho$-parameter at the LHC energy has recently received significant interest because of a measurement at $\rts=13~\TeV$ by the TOTEM experiment~\cite{TOTEM_2p5km} which measured a lower value of the  $\rho$-parameter than would be expected assuming a $\ln^{2}s$ rise of the total cross section, $s$ being the centre-of-mass energy squared.
 
The $\rho$-parameter is sensitive not only to the high-energy evolution of the total hadronic cross section but also to the
fundamental structure of the elastic-scattering amplitude. Traditionally, the elastic-scattering amplitude at
energies well above $100~\GeV$ has been thought to be dominated by an exchange of Pomerons in the $t$-channel (see e.g.\ Ref.~\cite{compete}). In QCD the Pomeron is
represented by a two-gluon colourless state with spin--parity--charge quantum numbers J$^{\textrm{PC}} = 0^{++}$. The additional possible presence of a
three-gluon colourless state with J$^{\textrm{PC}} = 1^{--}$, the so-called Odderon, can also influence the value of
the $\rho$-parameter. Thus, measurements of the $\rho$-parameter at the highest energy of
the LHC are essential.
 
The $\rho$-parameter is defined as the ratio of the real part to the imaginary part of the elastic-scattering amplitude in the limit $t\rightarrow 0$, i.e.
\begin{equation*}
\rho = \left. \frac{\textrm{Re}[f_{\textrm{el}}(t)]}{\textrm{Im}[f_{\textrm{el}}(t)]}\right|_{t\rightarrow 0} ,\label{eq:rho}
\end{equation*}
\noindent where $f_{\textrm{el}}(t)$ is the elastic-scattering amplitude and where $t$ stands for the four-momentum transfer in the reaction.
 
Normally, the $\rho$-parameter is determined by measuring the differential elastic cross section at such small
values of $|t|$ that the amplitude is sensitive both to the Coulomb amplitude and the strong amplitude, and
thus also to the interference between the two. The phase of the Coulomb amplitude is known,
and therefore the value of $\rho$ can be extracted from the measured size of the interference term.
 
The ATLAS experiment has previously measured elastic scattering at 7 and $8~\TeV$~\cite{STDM-2013-10,STDM-2015-22} using the ATLAS Roman Pot system ALFA~\cite{ALFA}. However, those measurements did not extend to the region of very small $|t|$-values where the differential cross section is sensitive to the  $\rho$-parameter. Such  small $|t|$-values require measurements of angles in the microradian range, which in turn need
even smaller divergence  of the beam at the interaction point (IP). Moreover, they require that the vertically movable Roman Pot detectors approach to within millimetres of the beam.
 
This paper presents a new measurement using $pp$ collision data at $\sqrt{s}= 13$~\TeV, corresponding to an integrated luminosity of 340~$\upmu$b$^{-1}$.
For the first time, the ATLAS measurement is extended, by an order of magnitude lower in $-t$, to such small scattering angles that the Coulomb interaction starts to play a role.
Here, the differential elastic cross section is measured down  to $-t=2.5 \cdot 10^{-4}~\GeV^2$. The acceptance for elastic events is very small
for such low values of $|t|$ and if an acceptance greater than 10\% is required, the lower limit is  $-t=4.5 \cdot 10^{-4}~\GeV^2$.
The needed small divergence of the beam at the IP is achieved by using very high-$\betastar$ optics\footnote{The $\beta$-function determines the variation of
the beam envelope around the LHC ring and depends on the focusing properties of the magnetic lattice. $\betastar$ is the value of the $\beta$-function at the IP. The beam divergence at the IP is about $0.3~\upmu$rad for $\betastar=2.5$~km.} ($\betastar=2.5$~km), producing a large beam
spot size but very small beam divergence.
 
The measurements rely upon the very accurate luminosity determination that ATLAS provides for all cross-section measurements~\cite{lumi13TeV}.
The TOTEM experiment has a different approach, using the so-called luminosity-independent method for the normalization of the differential cross section. The main uncertainty in using this method comes from the Monte Carlo estimate of the non-measurable forward diffractive cross section.
In addition, ATLAS has introduced data-driven methods to constrain the parameters of the detector alignment. The alignment uncertainty and the luminosity uncertainty together constitute the dominant uncertainty in the measurement of the differential elastic cross section.
 
The optical theorem connects the hadronic component of the total cross section $\sigmatot$ to the imaginary part of the scattering amplitude in the forward direction,
\begin{equation}
\sigmatot=4\pi \, \textrm{Im} \, [ f_{\textrm{el}}\left(t)]\right|_{t\rightarrow 0} .
\label{eq:OpticalTheorem}
\end{equation}
 
In the same manner as in the ATLAS measurements at $7~\TeV$ and $8~\TeV$, the optical theorem is also used in the present analysis to determine $\sigmatot$  by extrapolating  the differential cross section to  $t\rightarrow 0$.
The total hadronic elastic cross section, in turn,  is  obtained from  an integration of the measured differential elastic cross section.
Using the trivial relation, $\sigmainel = \sigmatot-\sigmael$, the total inelastic cross section can also be calculated.
 
Another aspect of interest is whether  the differential elastic cross section can  be described by a simple exponent in the region where the strong interaction dominates  or if higher-order corrections are needed (for an overview of the experimental situation, see~the introduction of Ref.~\cite{TOTEM_nonexp}). Such tendencies were already observed at the Intersecting Storage Rings (ISR) but the sample size was insufficient to draw a definite conclusion. With the large sample available here at 13~\TeV, ATLAS has an opportunity to establish such a non-exponential behaviour. TOTEM has reported observations of non-exponential behaviour at both $8~\TeV$~\cite{TOTEM_nonexp} and $13~\TeV$~\cite{TOTEM_13TeV}.
 
The paper is organized in the following way. Section~\ref{sec:Exp} presents the experimental set-up and includes a brief description of the ALFA subdetector. The data sets and the data-taking conditions are discussed in Section~\ref{sec:Meth}, where the measurement method and detector alignment are also explained.
The theoretical framework and simulation tools for the description of elastic scattering are
presented in Section~\ref{sec:Sim}.
The main points of the data analysis are discussed in Section~\ref{sec:Analysis}.
Section~\ref{sec:Results} is devoted to the determination of the differential elastic cross section
and the physics parameters. The interpretation of the results is discussed in
Section~\ref{sec:Interpretation}. The conclusions are given in Section~\ref{sec:Conclusion}.

% End of text imported from the .//Introduction.tex input file

% The next lines are included from the .//Exp.tex input file
\section{Experimental set-up}\label{sec:Exp}
The ALFA detector is a specific part of the ATLAS experiment~\cite{atlas1} designed to measure the elastic
scattering of protons. Because the protons are elastically scattered at very small angles the tracking
detectors need to be placed close to the beam and far from the IP.
For this purpose two stations with tracking detectors are located on both sides of the central
ATLAS detector.
 
\begin{figure}[hbt!]
\centering
\includegraphics[width=0.6\textwidth]{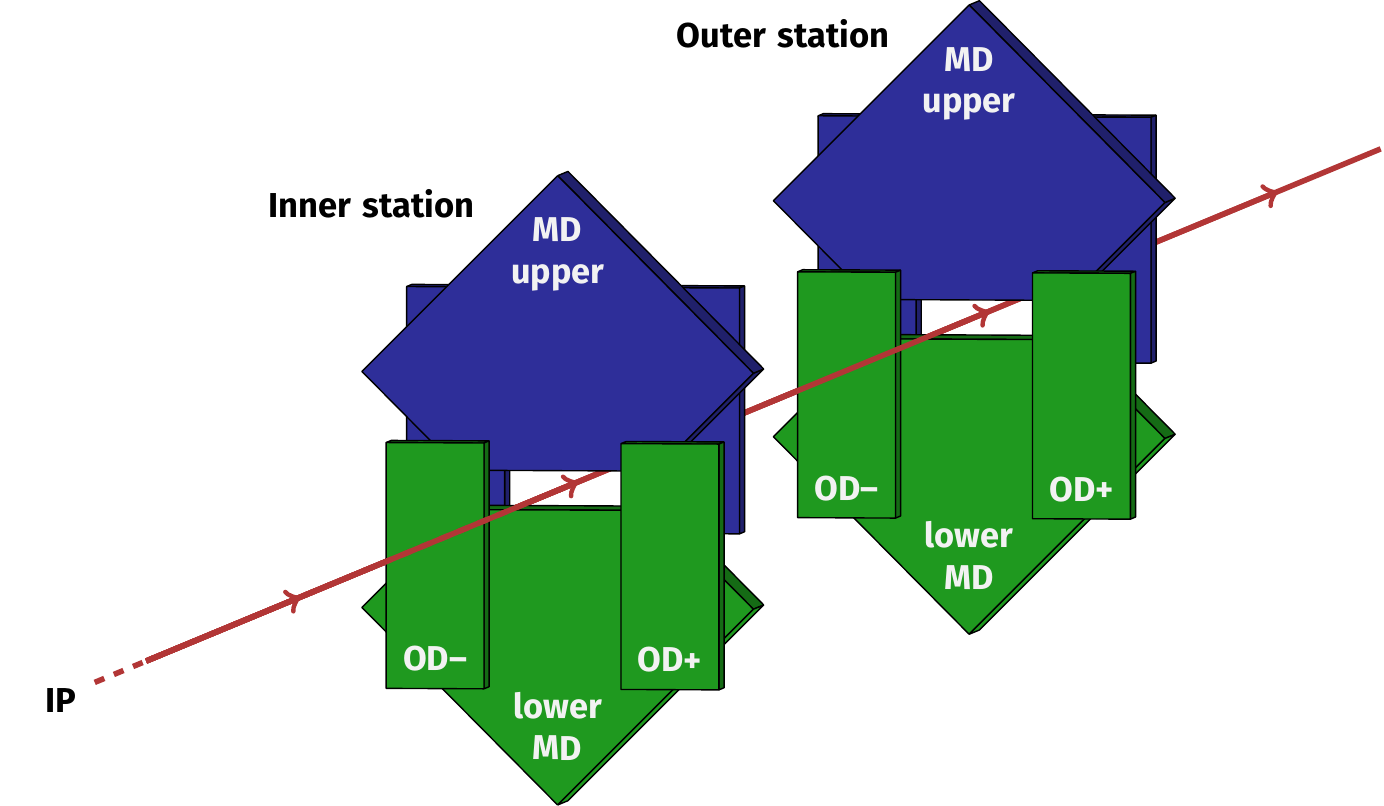}
\caption{The layout of the ALFA detectors on one side of ATLAS: the main tracking detectors MD and the
overlap detectors OD$+$ and OD$-$ on their right and left sides, respectively. The blue-coloured MDs and ODs are
housed in the upper RPs, while the green-coloured ones are in the lower RPs.}
\label{fig:ALFA_3D}
\end{figure}

The conceptual layout of the detectors in the two stations on one side of the ATLAS detector is shown in
Figure~\ref{fig:ALFA_3D}. In each station, two main
tracking detectors (MDs) measure the trajectory of scattered protons in the upper or lower
half of the station.  Overlap detectors (ODs) are placed on the right and left sides of each MD.
They are used to measure the vertical distance between the upper and lower MDs by recording halo protons passing through
the left-side or right-side ODs on the two MDs. The MDs and ODs are supplemented by trigger counters made of 3~mm scintillator tiles.
The detectors are housed in so-called Roman Pots (RPs), an upper one and a lower one, which are movable
and can approach the circulating beam in the vertical direction to within 1~mm.
Due to the layout of the LHC accelerator, with two parallel beam pipes close to each other in the horizontal plane, a choice of vertically moving RPs was natural.
 
Each MD consists of 10 modules with 64 scintillating fibres of 0.5~mm size glued on both the front and back sides of a titanium support plate. The fibres on both sides are arranged orthogonally in a $uv$-geometry at $\pm45^\circ$ to the $y$-axis. The module layers are staggered in steps of a tenth of the fibre size. The spatial resolution of the MDs is measured to be about $32~\upmu$m.
The ODs consists of only three layers of 0.5~mm fibres.
A detailed description of the ALFA detector can be found in Ref.~\cite{ALFA}.
 
\begin{figure}[hbt!]
\centering
\includegraphics[width=\textwidth]{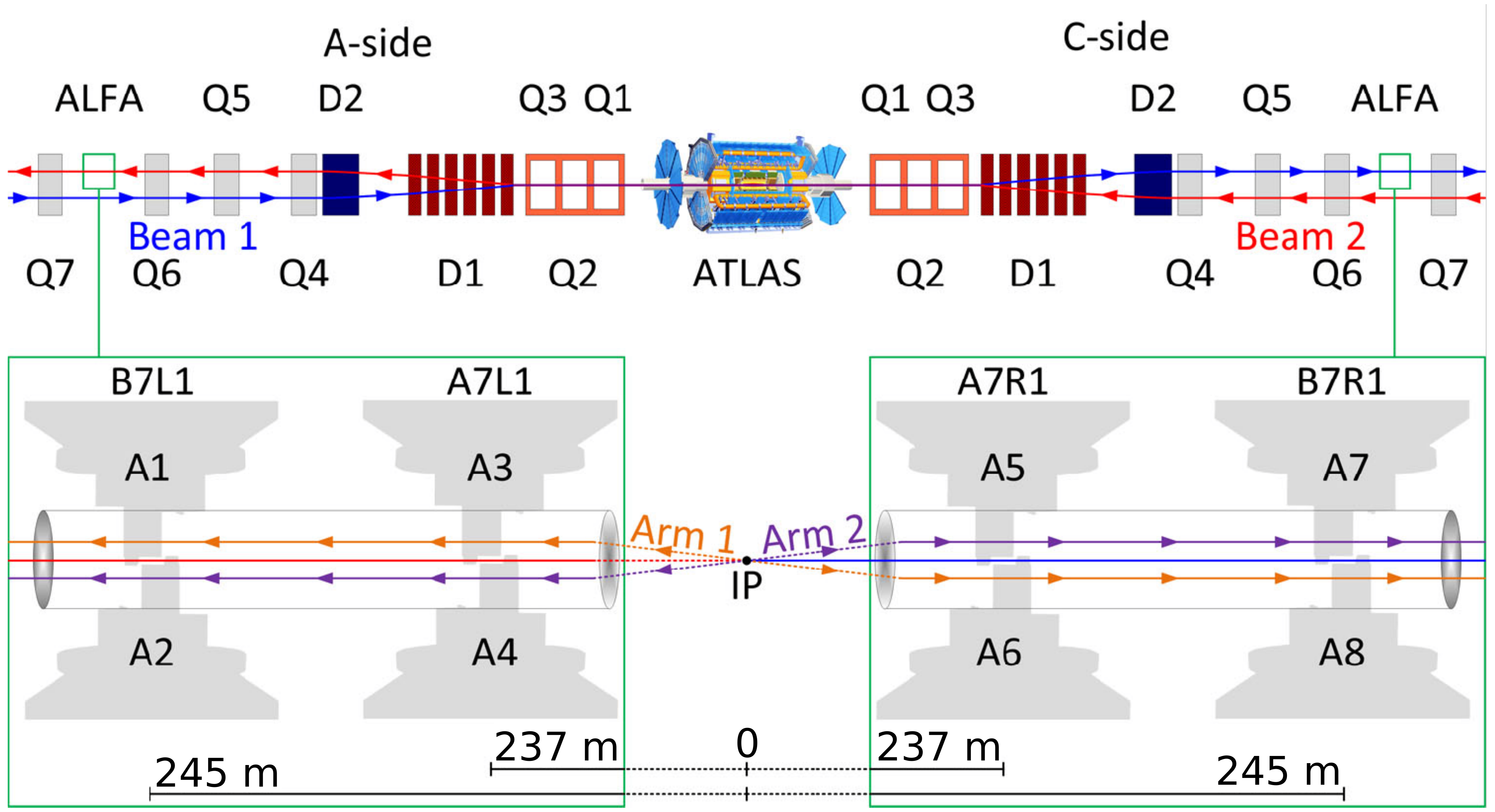}
\caption{A sketch of the experimental set-up, not to scale, showing the
positions of the ALFA Roman Pot stations in the outgoing LHC beams, and the quadrupole
(Q1--Q6) and dipole (D1--D2) magnets situated between the IP and ALFA.
The ALFA detectors are numbered A1--A8, and are combined into inner stations A7R1 and A7L1,
which are closer to the IP, and outer stations B7R1 and B7L1.
The arrows in the top panel indicate the beam directions and in the bottom panel the scattered
proton directions. The stations A7R1 and B7R1 measure the scattered protons along the outgoing
LHC beam 1 line (C-side) while stations A7L1 and B7L1 cover the outgoing LHC beam 2 line (A-side).}
\label{fig:alfa_layout}
\end{figure}

The schematic layout of the ALFA stations in the LHC is shown in Figure~\ref{fig:alfa_layout}.
Two stations at distances of 237~m and 245~m on each side of the interaction point ensure the
reconstruction of the scattered proton's trajectory.
The stations on the right side of ATLAS measure the scattered protons along the outgoing LHC beam~1~line
(C-side), while stations on the left side cover the outgoing LHC beam~2~line (A-side).
The eight MDs form two independent spectrometer arms.
Arm 1 consists of the upper detectors in the two
stations on the left side combined with the lower detectors in the two stations on the right side
of the IP. Arm 2 consists of the two lower detectors on the left side combined with the two upper detectors on the right side.
The preselection of elastic-scattering events requires triggers to have fired in at least one MD on either side of a
spectrometer arm.
 
In the LHC Long Shutdown period from 2013 to 2014 (known as LS1) some major technical investments were
made to ensure data taking with ALFA in Run~2.
Dangerous heating of the fiber detectors up to damage level was
observed at higher beam intensities in Run~1. To protect the detectors,
all stations were equipped with an active air cooling system and
additional passive cooling components.
Another modification was necessary to reach the high-$\beta^*$ value of 2.5~km:
two additional water-cooled power cables were installed to allow
more flexible quadrupole operation.
 
For physics analysis the most relevant modification was the relocation of the
outer stations B7L1 and B7R1. The new positions increased the distance
between inner and outer stations from 4~m to about 8~m. As a consequence
the track angular resolution improved significantly.

% End of text imported from the .//Exp.tex input file

% The next lines are included from the .//Meth.tex input file
\section{Experimental method}
\label{sec:Meth}
\subsection{Measurement principle}
\label{sec:measure}
The data were recorded with special beam optics characterized by
a $\betastar$ of 2.5~km \cite{Optics_2p5km} at the IP resulting in small beam divergence and
providing parallel-to-point focusing in the vertical plane.
In parallel-to-point beam optics
the betatron oscillation has a phase advance $\Psi$ of $90^\circ$ between the IP
and the RPs, such that all particles scattered
at the same angle are focused to the same position in the detector, independent of their production vertex position.
This focusing is only achieved in the vertical plane.
 
The beam optics parameters are needed for the reconstruction of the scattering angle
$\theta^\star$ at the IP.
In elastic scattering at high energies the four-momentum transfer $t$ is calculated from $\theta^\star$ by:
\begin{equation}
\label{eq:t-basic}
-t = \left(\theta^\star \times p \right)^2  ,
\end{equation}
where $p$ is the nominal LHC beam momentum of $6.5~\TeV$ \cite{LHC_Energy} and
$\theta^\star$ is reconstructed from the proton trajectories in ALFA.
The trajectories are measured in the beam coordinate system, where the transverse positions
$x$ and $y$ are determined relative to the nominal orbit.
A formalism based on transport matrices relates the
positions and angles of particles at two different points of the magnetic lattice.
 
The trajectory ($w(\xi)$, $\theta_w(\xi)$), where $w\in\{x,y\}$ is the transverse position at a
distance $\xi$ from the IP and $\theta_w$ is the angle in the $w$ direction
between the particle trajectory and the nominal orbit, is
given by the transport matrix $\mathbf{M}$ and the coordinates at
the IP ($w^\star$, $\theta_w^\star$):
 
\begin{equation}\label{eq:transport}
\left(\begin{array}{c}
w(\xi) \\
\theta_w(\xi) \\
\end{array}\right) =
\mathbf{M}
\left(\begin{array}{c}
w^\star \\
\theta_w^\star \\
\end{array}\right) =
\left(\begin{array}{cc}
M_{11} & M_{12} \\
M_{21} & M_{22}
\end{array}\right)
\left(\begin{array}{c}
w^\star \\
\theta_w^\star \\
\end{array}\right)  .
\end{equation}
Here the elements of the transport matrix can be calculated from the
optical function $\beta$ and its derivative with respect to $\xi$ and $\Psi$.
The transport matrix $\mathbf{M}$ must be calculated separately in $x$ and $y$
and depends on the longitudinal position $\xi$; the corresponding indices have been
dropped for clarity.
The focusing properties of the beam optics in the vertical plane enable
a reconstruction of the scattering angle with good precision using only $M_{12}$,
but in the horizontal plane the phase advance is
close to $180^{\circ}$ and different reconstruction methods are investigated.
 
The ALFA detector was designed to use the `subtraction' method, exploiting the
fact that for elastic scattering the particles are back-to-back, so that the scattering angles on the A-side and C-side
are the same in magnitude and opposite in sign, and that the protons originate from the same vertex.
The beam optics was optimized to maximize the lever arm  $M_{12}$ in the vertical plane to access the
smallest possible scattering angle. The positions measured with ALFA on the A-side and C-side
of ATLAS have the same magnitude to within 50--100~$\upmu$m but opposite signs, and in the subtraction method the
scattering angle is calculated according to:
\begin{equation}
\label{eq:subtraction}
\theta^\star_w = \frac{w_\mathrm{A} - w_\mathrm{C}}{M_{12,\mathrm{A}} + M_{12,\mathrm{C}}} .
\end{equation}
The measurements from inner and outer stations are averaged in $y$ to obtain the final value of $\theta^\star$.
The subtraction method is the nominal method in both planes and yields the best $t$-resolution.
In contrast to previous analyses using $90$~m optics \cite{STDM-2013-10,STDM-2015-22}, the lever arm $M_{12}$
in the horizontal plane at the inner stations is so small that its contribution to the scattering angle
determination introduces an unacceptable degradation of the resolution. Therefore, a method named `subtraction light'
is used instead, which uses $y$ everywhere in the $\theta^\star$-reconstruction but
ignores the $x$-measurement at the inner stations and only uses $x$ at the outer stations.
However, the $x$ measurement is used for event selection purposes
and for the local track angle used in the method described below.
 
An alternative method for the reconstruction of the horizontal scattering angle uses the
`local angle' $\theta_w$ of the tracks measured between the inner and outer stations on the same side:
\begin{equation}
\label{eq:localangle}
\theta^\star_w = \frac{\theta_{w,\mathrm{A}} - \theta_{w,\mathrm{C}}}{M_{22,\mathrm{A}} + M_{22,\mathrm{C}}} .
\end{equation}
Another method performs a `local subtraction' of measurements at the inner station
at 237~m and the outer station at 245~m, separately on the A-side and C-side, before combining the two sides:
\begin{equation*}
\theta^\star_{w,S} = \frac{M_{11,S}^{245} \times w_{237,S} - M_{11,S}^{237} \times w_{245,S}}{M_{11,S}^{245} \times M_{12,S}^{237} - M_{11,S}^{237} \times M_{12,S}^{245}} \; \; , \; S=\mbox{A, C} .
\end{equation*}
Finally, the `lattice' method uses both the measured positions and the local angle to reconstruct the scattering angle
by the inversion of the transport matrix
\begin{equation*}\label{eq:lattice_gen}
\left(\begin{array}{c}
w^\star \\
\theta_w^\star
\end{array}\right) =
\mathbf{M^{-1}}
\left(\begin{array}{c}
w \\
\theta_w \\
\end{array}\right)  ,
\end{equation*}
and from the second row of the inverted matrix the scattering angle is determined to be
\begin{equation*}
\theta^\star_{w} = M_{21}^{-1} \times w + M_{22}^{-1} \times \theta_w  .
\end{equation*}
The lattice method yields a scattering angle measurement at each station, and the average value
is taken for the $t$-reconstruction.
All methods using the local angle suffer from limited resolution due to a moderate
angular resolution of about 5~$\upmu$rad. These alternative methods are nevertheless used to
cross-check the subtraction method and determine beam optics parameters.
 
For all methods, $t$ is calculated from the scattering angles as follows:
\begin{eqnarray}\label{eq:txy}
- t & = &\left((\theta^\star_{x})^2 + (\theta^\star_{y})^2  \right)p^2  ,
\end{eqnarray}
where $\theta^\star_{y}$ is always reconstructed with the subtraction method, because of the parallel-to-point focusing
in the vertical plane, while the four methods are used for $\theta^\star_{x}$.
 
\subsection{Data taking}
\label{sec:Data}
Since the high-$\betastar$ runs are very different from the standard LHC runs, a few test fills were
performed to find acceptable beam parameter settings.
The minimum accessible value $t_{\textrm{min}}$ is influenced by three parameters:
the $\beta$-function, the distance of the detectors from the beam trajectory,
and the beam emittance. It is expressed in the following formula:
\begin{equation*}
|t_{\textrm{min}}| = \frac{p^2 \cdot n^2 \cdot \epsilon_N}{\betastar} ,
\end{equation*}
with the beam momentum $p$, the detector transverse position as a multiple $n$ of the beam width
$\sigma = \sqrt{\epsilon \cdot \beta}$, the (normalized) emittance $\epsilon$ ($\epsilon_N$ and
the value of the $\beta$-function at the IP, $\betastar$.
 
To achieve a $|t_{\textrm{min}}|$ value of a few times 10$^{-4}~\GeV^2$ for the given beam momentum of $6.5~\TeV$ and
$\betastar = 2.5$~km, the detectors need to be placed at a distance of $3\sigma$ from the
beam trajectory. The target value for the emittance was about 1~$\upmu$m.
 
To achieve acceptable conditions for recording physics data, several collimator settings were tested. The tight collimator positions induced shower particles by interactions with halo particles. The main challenge in the
data taking was the handling of the rapid increase of background for elastic triggers.
 
The final procedure was to scrape the beam with the primary vertical collimators to $2\sigma$ of the
beam width and position the RPs at $3\sigma$. The collimators were then retracted to $2.5\sigma$
to reduce the impact of shower particles from their edges. In addition the secondary vertical and horizontal
collimators were used to optimize the background conditions.
 
Due to LHC machine protection requirements, the total beam intensity in the high-$\betastar$ runs is limited to
$3 \cdot 10^{11}$ protons per beam. A filling scheme of five bunches with $6 \cdot 10^{10}$ protons each was
used in all fills. For background studies in the first fill, one pair of non-colliding bunches was used.
 
After the LHC was filled, the energy was ramped up to $6.5~\TeV$ and $\betastar$ was de-squeezed to 2.5~km.
The next step was the beam-based alignment (BBA). In this procedure, the RP windows scrape
the beam edge and the resulting signals in the beam loss monitors~\cite{BLM} are used to determine the beam
trajectory. All RP and collimator positions are given in terms of the beam width $\sigma$ relative to
the measured beam trajectory.
 
When the BBA procedure was finished, the collimators were moved to their final positions and the RPs were
set at the $3\sigma$ positions, corresponding to a distance between the RP window facing the beam and
the beam orbit of about 0.5~mm.
The detectors themselves were about 0.3 to 0.4~mm further away from the beam orbit due to the RP window
thickness and the gap towards the detector edge.
 
The evolution of the background fraction was
monitored by the ratio of background to elastic trigger rates. If the background rate reached the
level of the elastic rate the RPs were retracted slightly and a re-scraping to a beam width of
$2\sigma$ was performed. Depending on the beam intensity, this procedure was repeated after 30
to 90~minutes. The fill was dumped when the elastic rate fell below 1~Hz.
 
In four fills (5313, 5315, 5317, 5321), which each lasted between 10 and 24~hours, a total integrated
luminosity of about 340~$\upmu$b$^{-1}$ was collected. The luminosity and the recorded number of elastic-scattering
candidates for
the individual fills and runs are given in Section~\ref{sec:Event_selection}.
 
The emittance was regularly measured by wire scans and a dedicated synchrotron light monitor. After
injection, typical values of the horizontal and vertical emittances were below 1~$\upmu$m. After ramp-up
and de-squeezing, these values grew to about 1.1~$\upmu$m for the vertical emittance and about 2.5~$\upmu$m for the
horizontal emittance. The measured emittance and its time evolution are taken into account in the
event simulation model described in Section~\ref{sec:ElsA}.
 
A dedicated trigger menu was used to record the sample of
elastic interactions, and also the various diffractive and background samples
needed for systematic studies.
The collection of elastic
events is based on conditions which require in each spectrometer arm a trigger signal in at least one
detector in the stations at opposite sides of the central ATLAS detector. For background
studies, a similar trigger pattern was used but it combined all upper or lower detectors of all stations,
which excludes elastic events by construction. The trigger efficiency was derived from a minimum-bias sample,
which is based on a single trigger signal from any ALFA detector. For the selected data
sample, the data acquisition dead time was below $0.3\%$ and the trigger efficiency was above $99.97\%$.
An extensive software suite~\cite{ATL-SOFT-PUB-2021-001} is used in the reconstruction of data, in detector operations, and in the trigger and data acquisition systems of the experiment.
 
\subsection{Track reconstruction and alignment}
\label{sec:align}
The tracks of the elastically scattered protons are reconstructed in the ALFA main detectors.
The tracks are reconstructed individually in each ALFA detector.
The reconstruction method takes advantage of the fact that the tracks left by the
elastically scattered protons are almost parallel to the beam.
Then, one can neglect the slope of the trajectory and focus on the determination of its position only.
Because the layers are staggered, the position is constrained by the geometrical overlap of the fired fibres projected separately on the $u$ and $v$ directions.
A track is required to consist of hits in at least three overlapping fibres in each direction
(an elastically scattered proton typically fires 18--19 fibres in a detector).
 
Reconstruction of the event kinematics requires knowledge of the transverse position of the tracks relative to the beam.
A transformation from the detector-related $u$ and $v$ coordinates to the $x$ and $y$ coordinates relative to the beam requires a good understanding of the detector position:
the rotation of the detector around the beam axis,\footnote{Other possible rotations only affect track positions at the level of rotation angle squared and are neglected.}
the horizontal offset between the centre of the detector and the centre of the beam,
and the vertical distance between the detector and the beam.
 
Information about the horizontal offset and the rotation angle is obtained from the analysis of distributions of elastically scattered protons
(see Section \ref{sec:Event_selection} for details of the event selection).
The $(x,y)$ distribution of the positions of these protons at a given RP station has an elliptical shape centred around the beam and is elongated in the $y$ direction.
The two detectors of the station, upper and lower, measure fragments of this ellipse.
For a correctly calculated horizontal offset, the average value of the $x$ position, $\left< x \right>$,  should be zero.
For a correctly determined rotation of the detector, no correlation should be observed between $x$ and $y$.
The alignment procedures are based on measurements of $\left< x \right>$ and how $\left< x \right>$ depends on $y$.
The resulting alignment corrections are applied iteratively until the detectors are fully aligned.
 
The information about the vertical distance between each detector and the centre of the beam is obtained from several sources.
First, the distance between the upper and lower detectors is measured using data collected by the ALFA overlap
detectors, which are placed on the sides of the main detectors and extend below the beam for the upper RP and
above the beam for the lower RP \cite{ALFA}.
A simultaneous measurement of traversing particles by the upper and lower detectors allows the determination of their relative position in the vertical direction.
 
The measurement using the overlap detectors provides the vertical distance between the upper and lower detectors, i.e.\ the sum of their distances to the beam centre.
The determination of the position of the beam between the detectors is again based on a basic property of the $(x,y)$ distribution of elastically scattered protons, namely that it has an up--down mirror symmetry.
Part of the alignment procedure is to search for a vertical beam offset that equalizes the $y$ distributions (corrected for the reconstruction efficiency, see Section \ref{sec:RecoEff})  in the upper and lower detectors.
 
\begin{table}[t]
\begin{center}
\caption{
The results of the first step of the alignment analysis.
The ranges of values and uncertainties obtained for different detectors and data-taking periods are presented.
}
\begin{tabular}{lcccc}
\hline
Alignment parameter & Rotation angle & Horizontal offset & Vertical distance & Vertical offset \\
\hline
Range of values & $-2$ to $+4$~mrad & $-300$ to $+600$~$\upmu$m & 1.64 to 1.93~mm & $-90$ to $+70$~$\upmu$m \\
Range of uncertainty & 0.2 to 0.6~mrad & 2 to 13~$\upmu$m & $\sim$30~$\upmu$m & 10 to 60~$\upmu$m \\
\hline
\end{tabular}
\label{tab:alignment}
\end{center}
\end{table}
 
The alignment analysis presented above was performed separately for different detectors and data-taking periods, resulting in a range of values and corresponding uncertainties for each parameter.
Table~\ref{tab:alignment} summarizes the obtained results.
 
The dominant systematic uncertainty for the rotation angle originates from ignoring the underlying fibre structure
of the detectors.
This structure can induce a bias in the reconstructed hit pattern at a scale comparable to the spatial resolution.
The magnitude of this effect is estimated by varying the fiducial volume in which the alignment analysis is performed.
The statistical uncertainty is at a similar level.
The total uncertainty of the horizontal offset is dominated by the systematic component, evaluated by changing the method
of extracting the centre of the distribution.
 
The uncertainty for the vertical distance is dominated by the systematic component originating from our imperfect knowledge of how the overlap detectors are positioned relative to the main detectors,
from the choice of statistical method used to extract the distance value, and from the event selection.
The dominant source of systematic uncertainty for the vertical offset is related to the choice of method for testing the compatibility of the distributions, and this uncertainty is similar in size to the statistical uncertainty.

The next step of the alignment analysis exploits the fact that the kinematics of an elastic-scattering event are fully described by only two parameters, meaning that a measurement of the proton position in one detector fully constrains the kinematics.
This measurement can then be extrapolated to all other detectors of the same elastic arm.
The properties of the LHC optics used for these measurements, especially its parallel-to-point focusing in the vertical plane, make this extrapolation very precise in the $y$ coordinate.
Comparing the extrapolated position with the measured one provides further constraints on the vertical alignment.
The analysis showed a need for distance corrections at different stations of up to 55~$\upmu$m, which is
larger than the estimated distance uncertainties, suggesting that they are underestimated.
One possible explanation is a non-zero average angle of the particles traversing the overlap detectors,
which would lead to a bias in the distance measurement.
 
The above steps of the alignment analysis are based on the same principles and techniques as used in the previous ALFA measurements \cite{STDM-2013-10,STDM-2015-22}.
However, the final analysis presented in this paper is very sensitive to the distance uncertainties.
Therefore, additional steps for vertical alignment were introduced to ensure sufficient precision.
Once the relative positions of all detectors within each elastic arm are fixed using the precise extrapolations discussed above,
the two remaining parameters are the global vertical distance between the two arms and the global vertical position of the beam between them.
 
The global vertical distance is found by performing the complete analysis presented in this paper (see
Section~\ref{sec:Analysis}) assuming different values of the global vertical distance.
A fit to the differential elastic cross section is performed using only statistical uncertainties.
The $\chi^2$ of the fit, considered as a function of the global vertical distance, has a clear minimum whose position determines the final vertical distance used in the analysis.
This correction is found to be 86~$\upmu$m. It is added to the distance determined by averaging the distance
measurements from individual stations.
The uncertainty is evaluated to be 22~$\upmu$m, with the largest contribution originating from the variation
of the luminosity  within its uncertainty (see Section \ref{sec:Luminosity}).
 
It is worth noting that the sensitivity of the $\chi^2$ value to distance originates from the very good acceptance of the ALFA detectors down to the region dominated by the Coulomb interaction, which is well understood.
Performing the analysis with misaligned detectors affects the measured distribution,
making it incompatible with our knowledge of the physics that governs the behaviour of
the $t$-spectrum in this range, thus increasing the $\chi^2$.
 
As the final step of the alignment, the vertical position of the beam between the elastic arms is fine-tuned by equalizing the $t$-spectra measured in the two arms.
The principle is similar to that for the measurement per station discussed above, but when performed at the arms level the resulting uncertainty is improved to 4--15~$\upmu$m, with corrections of the order of 12--27~$\upmu$m
(depending on the data-taking period).
The uncertainty is dominated by two systematic components of similar size related to the details of testing the compatibility of the distributions and to the uncertainty of the reconstruction efficiency in the two arms.

% End of text imported from the .//Meth.tex input file

% The next lines are included from the .//Sim.tex input file
\section{Simulation model for elastic scattering}
\label{sec:Sim}
\subsection{Theoretical predictions}
\label{sec:Theory}
Elastic scattering is related to the total cross section through the optical theorem (Eq.~\eqref{eq:OpticalTheorem})
and the differential elastic cross section is obtained from the scattering amplitudes of the
contributing diagrams:
\begin{equation}
\label{eq:elamplitudes}
\frac{\mathrm{d}\sigma}{\mathrm{d}t} = \frac{1}{16\pi}\left|f_{\mathrm{N}}(t) + f_{\mathrm{C}}(t)\mathrm{e}^{\mathrm{i}\alpha\phi(t)}\right|^2  .
\end{equation}
Here, $f_{\mathrm{N}}$ is the purely strong-interaction amplitude, $f_{\mathrm{C}}$ is the Coulomb amplitude,
$\alpha$ is the fine-structure constant, and
a phase $\phi$ is induced by long-range Coulomb interactions~\cite{Bethe,WestAndYennie}.
In the simplest model elaborated in Ref.~\cite{WestAndYennie} the individual amplitudes are given by
\begin{eqnarray}
f_{\mathrm{C}}(t) & = & -8\pi\alpha\hbar c\frac{G^2(t)}{|t|}  , \label{eq:twoamplitudes}\\
f_{\mathrm{N}}(t) & = & \left(\rho + \mathrm{i}\right)\frac{\sigmatot}{\hbar c}\mathrm{e}^\frac{-B|t|}{2}  , \nonumber
\end{eqnarray}
where $G$ is the electric form factor of the proton and $B$ is the nuclear slope.
The value of $f_{\mathrm{N}}(0)$ follows from the optical theorem, while the $t$-dependence is
the simplest parameterization, which is valid only at small $|t|$.
A possible generalization of Eq.~\eqref{eq:twoamplitudes} that incorporates a $t$-dependent slope consists of
the introduction of additional terms $C$ proportional to $t^{2}$ and $D$ proportional to $|t|^{3}$
\begin{equation}
\label{eq:Cterm}
f_{\mathrm{N}}(t)  =  \left(\rho + \mathrm{i}\right)\frac{\sigmatot}{\hbar c}\mathrm{e}^\frac{-B|t| - Ct^{2} - D|t|^{3}}{2}  ,
\end{equation}
parameterizing the curvature of the $t$-spectrum at large $|t|$. Their effect is most visible in between $-t = 0.05~\GeV^2$ and
$-t = 0.2~\GeV^2$. At smaller $|t|$, the slope of the $t$-spectrum is essentially constant and well parameterized by $\exp{(-B|t|/2)}$, and
at very large $|t|$ beyond $0.2~\GeV^{2}$ the form of the spectrum changes when approaching the diffractive \textit{dip}, which is a local minimum in the $t$-spectrum generated
by the interference of diffractive amplitudes. It is located around
$0.5~\GeV^2$ in $13~\TeV$ $pp$ collisions and followed by a local maximum, called the \textit{bump}. At yet larger $|t|$-values the spectrum continues to fall steeply.
 
Depending on the number of terms included
in the exponential function, three different models are considered for parameterization of the nuclear amplitude; they are referred to as the $B$-model with validity up to $-t=0.04~\GeV^2$, the $BC$-model with validity up to $-t=0.1~\GeV^2$ and the full $BCD$-model with
validity up to $-t=0.2~\GeV^2$. The validity ranges are approximate and were empirically determined by studying
the quality of a fit to the data.
 
The theoretical form of the $t$-dependence of the cross section is obtained by
evaluating the square of the complex amplitudes, following Eqs.~\eqref{eq:elamplitudes},
\eqref{eq:twoamplitudes}, and~\eqref{eq:Cterm}:
\begin{eqnarray}\label{eq:tgen}
\frac{\mathrm{d}\sigma}{\mathrm{d}t} & = & \frac{4\pi\alpha^2(\hbar c)^2}{| t |^2} \times G^4(t) \\ \nonumber
& - &  \; \; \sigmatot \times \frac{\alpha G^2(t)}{|t|}\left[\sin\left(\alpha\phi(t)\right) + \rho \cos\left(\alpha\phi(t)\right) \right] \times \mathrm{e}^{\frac{-B|t| - Ct^{2} - D|t|^{3}}{2}} \\ \nonumber
& + &  \; \; \sigmatot^2 \frac{1+\rho^2}{16\pi(\hbar c)^2} \times \mathrm{e}^{\left({-B|t|-Ct^{2}-D|t|^{3}}\right)}  ,
\end{eqnarray}
where the first term corresponds to the Coulomb interaction, the second to the Coulomb--nuclear interference (CNI),
and the last to the hadronic interaction. This parameterization is used to fit the differential elastic cross
section to extract the physics parameters $\sigmatot$ and $\rho$, and the terms $B$, $C$ and $D$ relevant to the nuclear slope, depending on the model.
This simple parameterization of the differential elastic cross section has been criticized by the authors of Ref.~\cite{TOTEM_1km}. The criticism is based upon the fact that the formula has only been derived for a constant $B$-slope and is thus in principle not valid when the curvature terms $C$ and $D$ are introduced. However, in Ref.~\cite{GrafstromAndRyskin} it is shown that the formula is still valid at a level of $10^{-3}$ in $\rho$ given this kind of $t$-dependence of the slope.
 
The theoretical prediction given by Eq.~\eqref{eq:tgen} also depends on
the Coulomb phase $\phi$ and the form factor $G$.
This analysis uses a conventional dipole parameterization of the proton electric form factor from Ref.~\cite{Cahn}
\begin{equation*}
G(t) = \left(\frac{\Lambda}{\Lambda +|t|}\right)^2 \; ,
\end{equation*}
where $\Lambda=0.71~\GeV^2$.
The uncertainty in the electric form factor is derived by comparing the simple
dipole parameterization with more sophisticated forms~\cite{Simon_FF},
which better describe the high-precision low-energy electron--proton elastic-scattering data~\cite{A1}.
An expression for the Coulomb phase was initially derived in Ref.~\cite{WestAndYennie}
\begin{equation}\label{WestYenniePhase}
\phi(t)=-\ln\frac{B|t|}{2}-\phi_{\textrm{C}}\; ,
\end{equation}
where the model-dependent Coulomb phase shift $\phi_{\textrm{C}}$ was taken to be $\gamma_{\textrm{E}}=0.577$.
Further corrections, primarily from the form factor, were calculated in Ref.~\cite{Cahn} and the
expression nominally used  in this analysis is
\begin{equation}\label{CahnPhase}
\phi(t)=-\left( \gamma_{\textrm{E}} +\ln\frac{B|t|}{2} +\ln\left(1+\frac{8}{B\Lambda} \right) \right) +\frac{4|t|}{\Lambda}\cdot\ln\frac{\Lambda}{4|t|}-\frac{2|t|}{\Lambda}  .
\end{equation}
Uncertainties in the Coulomb phase are estimated by
replacing the parameterization in Eq.~\eqref{CahnPhase} by the simple form in Eq.~\eqref{WestYenniePhase}. This change
has only a minor impact on the cross-section prediction.
Replacing the dipole by other forms also has a negligible impact on the determination of the physics parameters.
 
In the present theoretical framework, a constant nuclear phase is assumed between the real and imaginary parts
of the nuclear amplitude. In Ref.~\cite{GrafstromAndRyskin}, various alternative $t$-dependent nuclear phase models
are discussed which would entail a $t$-dependence of the $\rho$ parameter. The models suggested
in Ref.~\cite{GrafstromAndRyskin}  are considered for systematic uncertainty purposes,
excluding an extreme model featuring a peripheral phase. The model with the largest impact
is a simple model in which the $\rho$-value vanishes at $-t=0.1~\GeV^2$
\begin{equation}\label{eq:simplephase}
\arg f_{\mathrm{N}}(t)=\frac{\pi}{2}-\arctan\left(\rho(0)\left(1+\frac{t}{0.1} \right)\right) .
\end{equation}

\subsection{Simulation model}
\label{sec:ElsA}
Monte Carlo (MC) simulated events are used to calculate acceptance and unfolding corrections.
A fast simulation is used in which the detector resolution is parameterized and tuned to data
and the elastically scattered protons are passed through the LHC lattice by means of the beam transport $2\times2$ matrix according to Eq.~\eqref{eq:transport}. The detector geometry is described by a set of simple requirements.
The fast simulation offers convenient
methods to tune the relevant detector and beam parameters, in terms of resolution and beam divergence,
to measured control observables.
The generation of elastic-scattering events
is performed with a simple `toy' MC simulation. For each event, a $t$-value is randomly drawn
from a probability density function representing the theoretical model.
The nominal model, named $BCD$, includes the small-$|t|$ parameterization in
Eq.~\eqref{eq:tgen} with three slope parameters $B$, $C$ and $D$.
The parameters of the event generation model are summarized in Table~\ref{tab:generation}.
The values of the slope parameters, and also $\sigmatot$ and $\rho$, were iteratively adjusted in the analysis and
are close to the final results.
The validity of this model is limited to relatively small $-t$ values, $|t|< 0.2~\GeV^2$,
which is of relevance for the physics parameter determination in this analysis.
 
\begin{table}
\caption{Event generation parameters used in the elastic-scattering simulation model.}
\label{tab:generation}
\begin{center}
\begin{tabular}{lr}
\hline
Parameter & Value \\
\hline
$\sigmatot$  &  105 mb \\
$B$            &  $21.13~\GeV^{-2}$ \\
$C$            &  $-6.5~\GeV^{-4}$ \\
$D$            &  $17.4~\GeV^{-6}$ \\
$\rho$       &  0.096 \\
$\Lambda$     &  $0.71~\GeV^{2}$  \\
$\phi_{\mathrm{C}}$ & 0.577 \\
\hline
\end{tabular}
\end{center}
\end{table}
The acceptance of the present data set does not cover the dip, but a parameterization based on the model
introduced by Ref.~\cite{BourrelyAndSoffer} to describe this region is used for $|t|> 0.2~\GeV^2$.
 
The simulation model also includes the relevant LHC beam parameters that describe the width of the production
vertex distribution, the intrinsic beam energy spread, and the beam divergence at the IP. The
last is of particular importance, as the angular divergence contributes to the scattering angle resolution.
The divergence is inferred from a combination of measurements of the beam emittance from wire scans performed by the LHC beam instrumentation, from
the beam width measured by the ATLAS inner detector, and in the vertical plane directly from ALFA angular measurements. The resulting beam divergence is implemented per plane, per beam, and per run and, in addition,
the time-dependent emittance growth during the runs is incorporated.
 
A fast parameterization of the detector response is used for the detector simulation, with the
spatial resolution tuned to the measured resolution. The resolution is measured by extrapolating
tracks reconstructed in the inner stations to the outer stations using beam optics matrix-element
ratios and comparing predicted positions with measured positions. Thus, it is a convolution of the
resolutions in the inner and outer stations. The fast simulation is tuned to reproduce this convolved resolution.
A full \GEANT~\cite{GEANT4} simulation is used to set the resolution scale between
detectors at the inner and outer stations, which cannot be determined from the data.

% End of text imported from the .//Sim.tex input file

% The next lines are included from the .//Analysis.tex input file
\section{Data analysis}\label{sec:Analysis}
\subsection{Event selection}\label{sec:Event_selection}
All data used in this analysis were recorded in September 2016 with a beam optics of $\betastar=2.5$~km
in four fills of the LHC resulting in seven ATLAS runs. An event preselection consisting of data quality,
trigger and reconstruction requirements was applied, resulting in a sample of elastic-scattering candidates.
Data quality requirements were applied to ensure
that the ALFA detector was fully operational. Only data recorded outside of the scraping periods are taken into account.
Furthermore, only periods of the data taking where the dead-time
fraction was below $5\%$ are used. The luminosity-weighted average dead-time fraction is below $0.3\%$.
These requirements eliminate less than 5$\%$ of the data.
 
Events are required to pass the trigger conditions for elastic-scattering events,
and have a reconstructed track in all four detectors of the
arm which fired the trigger.
Events with additional tracks in
detectors of the other arm arise from the temporal overlap of halo protons with
elastic-scattering protons and are retained. If halo and elastic-scattering protons overlap in the
same detectors,
which happens typically only on one side,
a track-matching procedure~\cite{STDM-2013-10} between the detectors on each side is applied to identify the elastic
track.
 
Fiducial cuts to ensure good containment inside the detection area are applied to the
vertical coordinate. Tracks must be at least 60~$\upmu$m away from the edge of the detector nearer the beam, where the full detection efficiency
is reached. The cut at the detector edge determines the smallest accessible value of $|t|$.
At large vertical distance, the vertical coordinate must be at least 1~mm away from the shadow
of the beam screen, a protection
element of the quadrupoles, in order to minimize the impact from showers generated in the beam screen.
 
\begin{figure}[h!]
\centering
\includegraphics[width=0.5\textwidth]{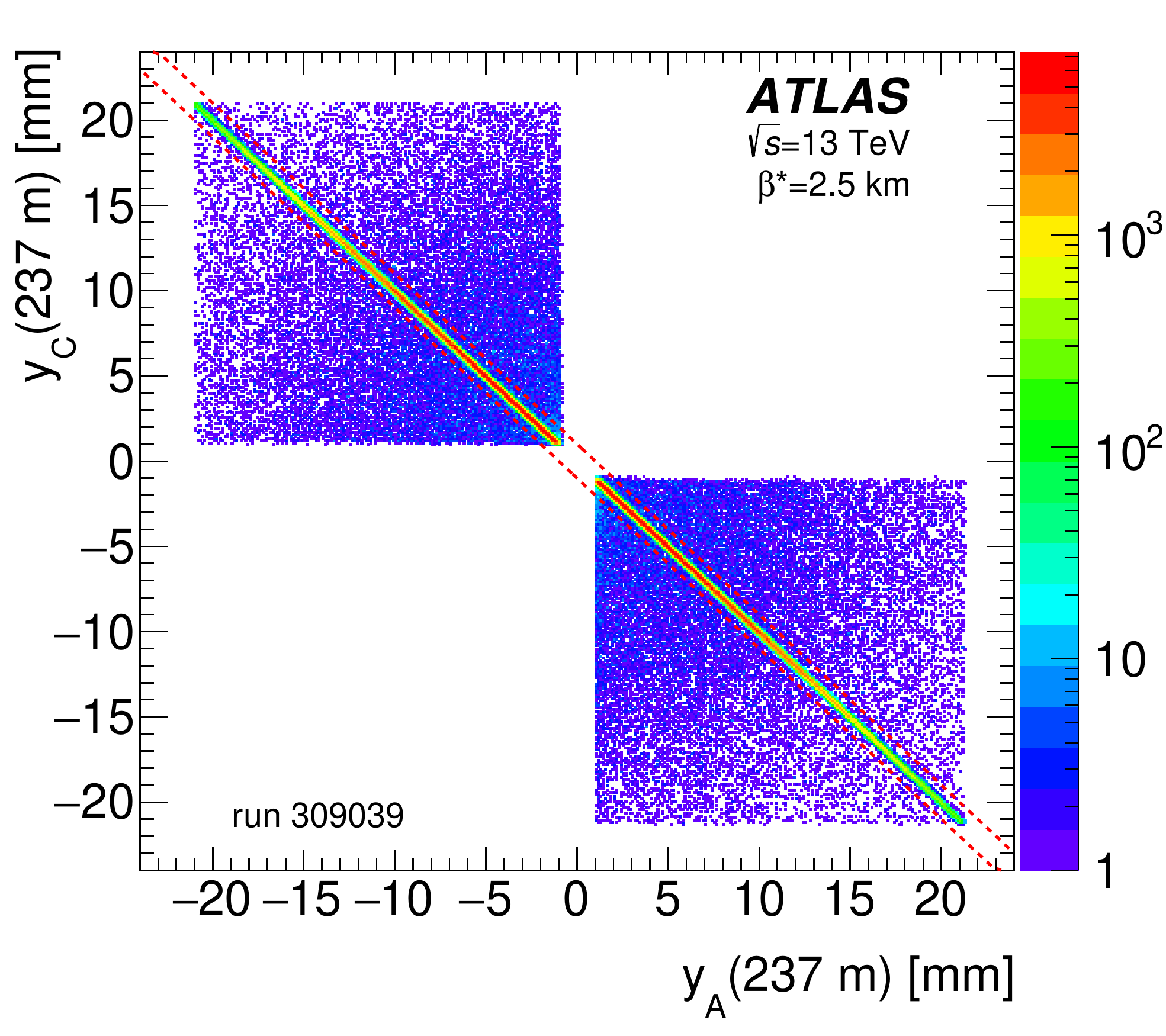}
\caption{The correlation of the $y$ coordinate measured in the inner stations on the A-side and C-side.
Elastic-scattering candidates after preselection and fiducial cuts but before
acceptance and background rejection cuts are shown. Identified elastic events are required to lie between the red lines.
Only data from one run are shown.}
\label{fig:evsel_yinner}
\end{figure}

Further geometrical cuts on the left--right acollinearity are applied, exploiting the back-to-back topology of
elastic-scattering events.
Figure~\ref{fig:evsel_yinner} shows the correlation between left-side (A-side) and right-side (C-side)
positions measured in the vertical plane.
 
\begin{figure}[h!]
\centering
\includegraphics[width=0.5\textwidth]{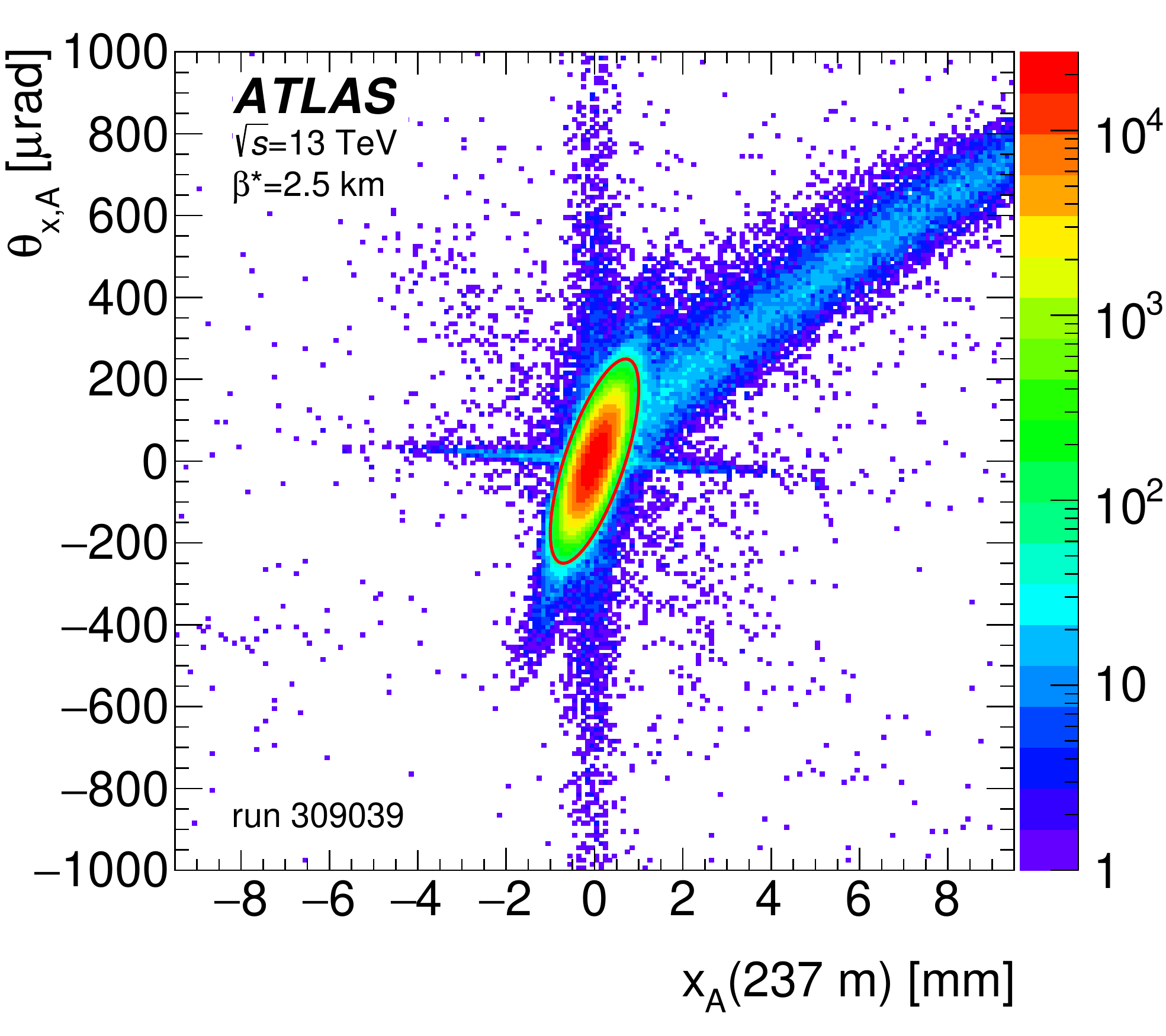}
\caption{The correlation between the coordinate and the local angle in the horizontal plane.
Elastic-scattering candidates after preselection and fiducial cuts but before
acceptance and background rejection cuts are shown. Identified elastic events are required to lie inside an ellipse corresponding to a $3.5\sigma$ selection, represented by a thin red line.
The vertical band corresponds to badly reconstructed tracks, the narrow almost horizontal
band represent the main halo contribution and the structure extending to the upper right quadrant contains off-momentum protons.}
\label{fig:evsel_xtx}
\end{figure}
 
The elastic-scattering candidates are observed in a narrow region along the diagonal,
and events are selected in a band of 2~mm width, as indicated by the red lines in Figure~\ref{fig:evsel_yinner}.
In the horizontal plane the position difference between the left and right sides must be
within $3.5\sigma$
of its resolution determined from simulation. An efficient cut against non-elastic background is obtained from
the correlation of the local angle between two stations and the position in the horizontal plane, as shown in
Figure~\ref{fig:evsel_xtx}, where elastic-scattering events appear inside a broad ellipse with positive slope, whereas
background is concentrated in a narrow ellipse with negative slope or in the band in the top-right quadrant.
A similar selection is also applied in the vertical plane, where elastic events are
selected in a band in the local angle of $40$~$\upmu$rad width. The selection requirements are summarized in Table~\ref{tab:cutflow}, alongside the number of events that meet the requirements at each step of the event selection for the longest run.
\begin{table}[tp]
\begin{center}
\caption{Numbers of events after each stage of the selection for run 309039.
The fractions of reconstructed events with an elastic trigger that survive each event selection cut are shown for each criterion.}
\label{tab:cutflow}
\begin{tabular}{lcccc}
\hline
Selection criterion     &  \multicolumn{4}{c}{Numbers of events} \\
\hline
Preselection    & \multicolumn{4}{c}{$2 \, 558 \, 637$}  \\
& Arm 1 & Fraction & Arm 2 & Fraction \\
Reconstructed tracks                    & $1 \,289 \, 282$ & & $1 \, 269 \, 355$ & \\
Cut on $x$ A vs C ($3.5\sigma$)        &   $1 \, 254 \, 738$ & $97.32\%$ & $1 \, 235 \, 792$ & $97.36\%$\\
Cut on $y$ A vs C ($2$~mm)              &   $1 \, 249 \, 888$ & $96.95\%$ & $1 \, 231 \, 251$ & $96.99\%$\\
Cut on $x$ vs $\theta_x$ ($3.5\sigma$) & $1 \, 248 \, 597$ & $96.84\%$ & $1 \, 230 \, 084$ & $96.91\%$\\
Beam-screen cut                         & $1 \, 243 \, 941$ & $96.48\%$ & $1 \, 225 \, 375$ & $96.53\%$\\
Edge cut                                  & $1 \, 231 \, 848$ & $95.55\%$ & $1 \, 210 \, 759$ & $95.38\%$\\
Cut on $y$ vs $\theta_y$ ($40~\upmu$rad ) & $1 \, 214 \, 717$ & $94.22\%$ & $1 \, 195 \, 251$ & $94.16\%$\\
Total selected & \multicolumn{4}{c}{$2 \, 409 \, 968$} \\ \hline
\end{tabular}
\end{center}
\end{table}
A total of $6.9$~million events were selected.
The selection efficiency relative to the preselection is about $94$\%. The fraction of elastic
pile-up events, where two elastic events from the same bunch crossing are observed in two different arms, is about $0.2\permil$.
The numbers of selected elastic-scattering candidates per run are summarized in Table~\ref{tab:run}.
\begin{table}[tp]
\caption{Fill and run numbers as well as delivered integrated luminosity, number of selected events for the $\beta^{\star}=2.5$~km campaign and reconstruction efficiency per arm (see Section~\ref{sec:RecoEff}).}
\label{tab:run}
\begin{center}
\begin{tabular}{llrrrr}
\hline
Fill & Run & Luminosity [$\upmu$b$^{-1}$] & \multicolumn{1}{c}{Selected elastic} & \multicolumn{2}{c}{Reconstruction efficiency}  \\
&  &  & \multicolumn{1}{c}{event candidates} & Arm 1 [\%] & Arm 2 [\%] \\ \hline
5313 & 308979 & 21.38 & $423 \, 862$       & 84.82 $\pm$ 0.56 & 83.11 $\pm$ 0.87 \\
5313 & 308982 &  6.81 & $136 \, 499$       & 85.84 $\pm$ 0.54 & 84.44 $\pm$ 0.55 \\
5314 & 309010 & 41.27 & $846 \, 581$       & 87.11 $\pm$ 0.51 & 85.00 $\pm$ 0.64 \\
5317 & 309039 & 120.08 & $2 \, 409 \, 968$ & 85.45 $\pm$ 0.49 & 83.23 $\pm$ 0.52 \\
5317 & 309074 & 44.31  & $887 \, 373$      & 85.55 $\pm$ 0.39 & 83.48 $\pm$ 0.48 \\
5321 & 309165 & 55.87  & $1 \, 149 \, 499$ & 87.08 $\pm$ 0.40 & 85.41 $\pm$ 0.44 \\
5321 & 309166 & 50.17 & $1 \, 043 \, 576$  & 88.28 $\pm$ 0.38 & 86.43 $\pm$ 0.45 \\ \hline
\multicolumn{2}{l}{Total} & 339.89 & $6 \, 897 \, 358$ & & \\
\hline
\end{tabular}
\end{center}
\end{table}
 
\subsection{Background determination}\label{sec:Background}
A small fraction of the background is expected to be inside the fiducial volume
defined by the event selection cuts. The background contributions can be clearly observed e.g.\ in
Figure~\ref{fig:evsel_xtx}, where the elastic core populates the interior of the elliptical contour,
whereas distinct patterns seen as narrow or wide bands constitute background from different sources.
A fraction of the background extends into the selected
area and thus constitutes an irreducible component that needs to be subtracted from the reconstructed $t$-spectrum.
The fraction of background events is very small and its subtraction has little impact on the results.
 
Two types of background are considered: non-elastic physics background processes from central diffraction, also known as double-Pomeron exchange (DPE) $pp\rightarrow pp+X$, and accidental coincidences of either halo protons on
both sides or a halo proton on one side and a proton from single diffraction (SD) on the other side.
Other combinations involving double diffraction or protons from non-diffractive processes in conjunction
with a halo proton are not considered, given the overall small level of background.
The strategy for the determination of the background is to use a data-driven template method for the
halo+halo and halo+SD estimation and a MC simulation for DPE.
The DPE simulation is based on the MBR model~\cite{MBR}, which predicts a total DPE cross section
of 0.82~mb at $13~\TeV$.
The event generation for DPE is done with nominal MBR settings as implemented in \PYTHIA[8.303]~\cite{PYTHIA82}.
 
\begin{figure}[htb!]
\centering
\subfloat[Correlation in data.]{
\includegraphics[width=0.49\textwidth]{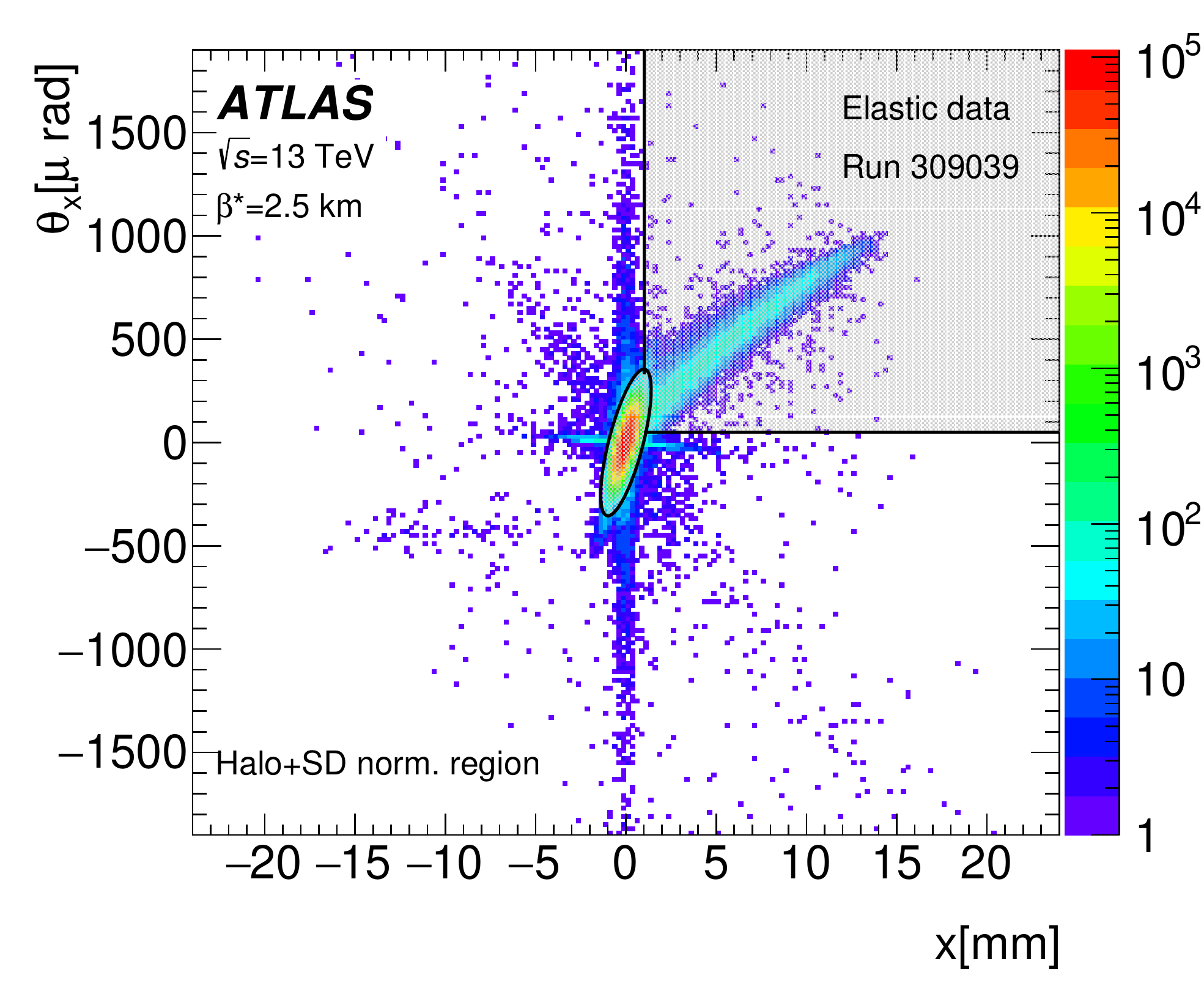}
\label{fig:bg44_xtx_data}
}
\subfloat[Correlation in DPE MC.]{
\includegraphics[width=0.49\textwidth]{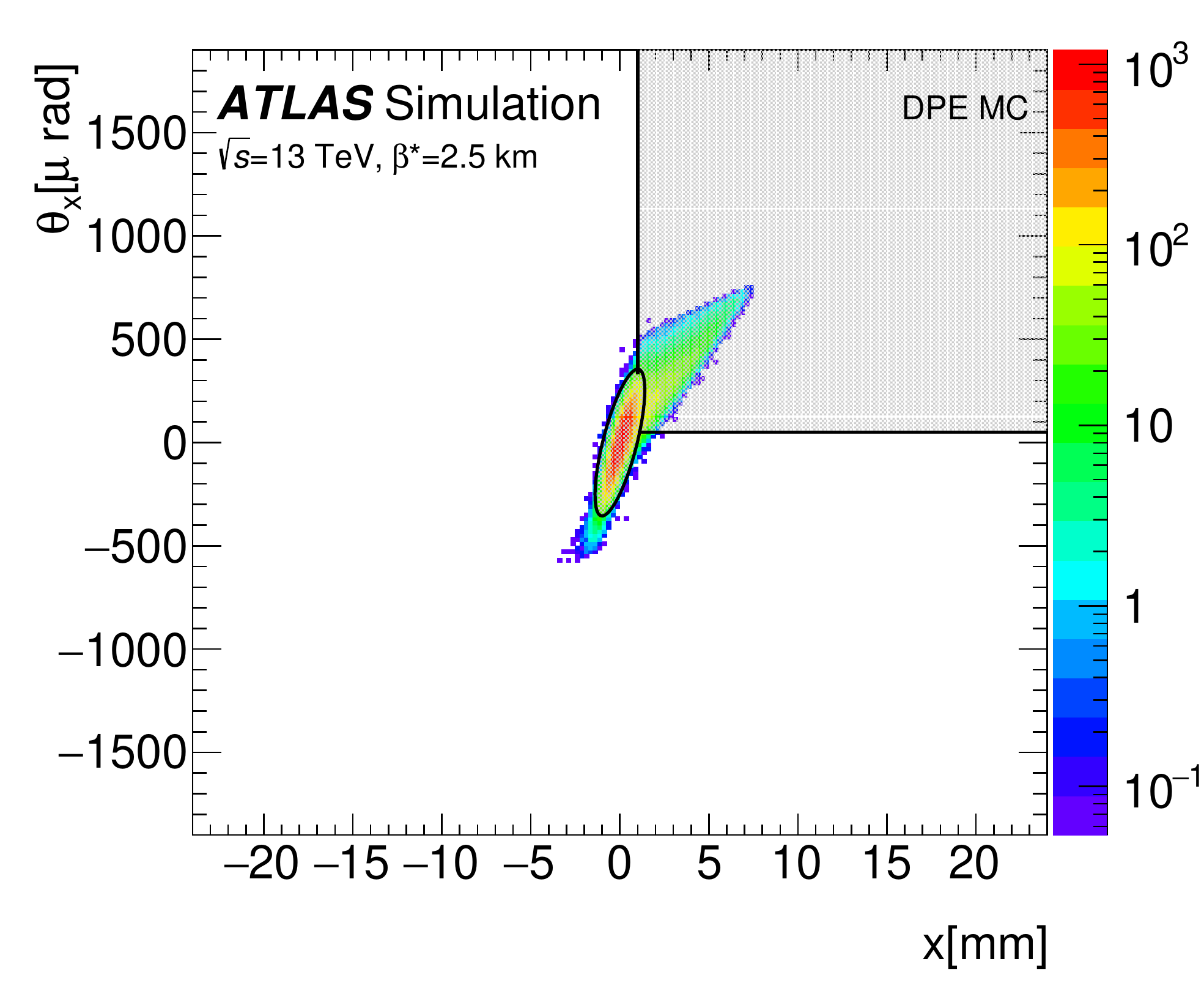}
\label{fig:bg44_xtx_dpe}
}
\newline
\subfloat[Correlation in event-mixing templates.]{
\includegraphics[width=0.5\textwidth]{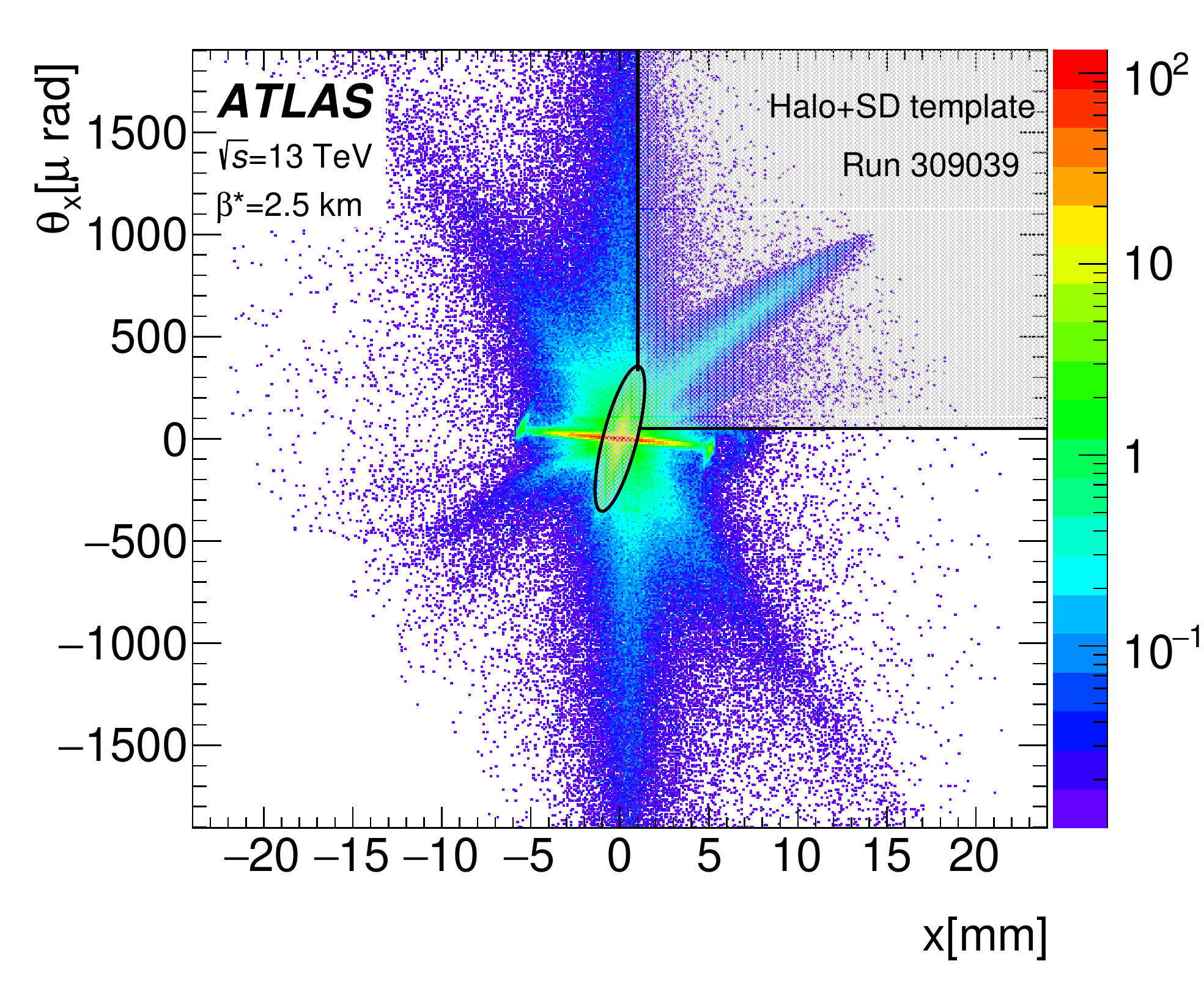}
\label{fig:bg44_xtx_halo}
}
\newline
\caption{Correlation plots before cuts, showing regions used to normalize the event-mixing templates for the background estimate of accidental coincidences between two halo protons or a halo proton and an SD proton. The normalization region is the area in the bottom and left part of the figure.
The shaded region surrounded by the black lines and ellipse is excluded from the normalization. The templates in (b) and (c)
have been normalized to show the expected number of background events present in the data sample.}
\label{fig:bg44}
\end{figure}

In the data-driven method used to calculate the contribution from accidental coincidences, events with tracks
only on one side of the experiment are selected, and those with activity on the opposite side are vetoed.
The veto suppresses contributions from elastic events and the sample composition
is thus dominated by halo and SD events. Then a background
template is constructed by randomly mixing uncorrelated single-sided events from the left side and right side,
corresponding to the background from accidental coincidences.
The absolute normalization of the templates is obtained by comparing the number of events in the data and the template
in control regions, before applying the event selection cuts. The procedure is illustrated in Figure~\ref{fig:bg44},
where the correlation between the horizontal track position and the local angle is shown in data, in DPE simulation and in the templates.
The normalization region is depicted in Figure~\ref{fig:bg44_xtx_data} showing the correlation in data
between the horizontal position and local angle in the horizontal plane. The
events are counted outside of the shaded area in the upper right corner and the elliptical elastic signal region.
The non-shaded normalization region is dominated by background from accidental coincidences of halo protons and free
of elastic events. Simulated DPE events shown in Figure~\ref{fig:bg44_xtx_dpe} contribute
very little to this area, but a large fraction of accidental coincidences from the event-mixing templates shown in Figure~\ref{fig:bg44_xtx_halo} lie
in the normalization region. The templates are scaled such that the number of events in this region equals
the number of events in the data. The irreducible background for selected events is then calculated by applying
the nominal event-selection cuts to the properly scaled template events.
 
Also, the normalization for simulated DPE events is determined from the data by counting events in a region of
the correlation between the vertical coordinate on the A-side and C-side (not shown). The selected normalization
region is mostly populated by DPE events. About $10\%$ of the data events in the normalization area
originate from accidental coincidences, which are subtracted before calculating the DPE scaling factor.
 
\begin{figure}[h!]
\centering
\includegraphics[width=0.6\textwidth]{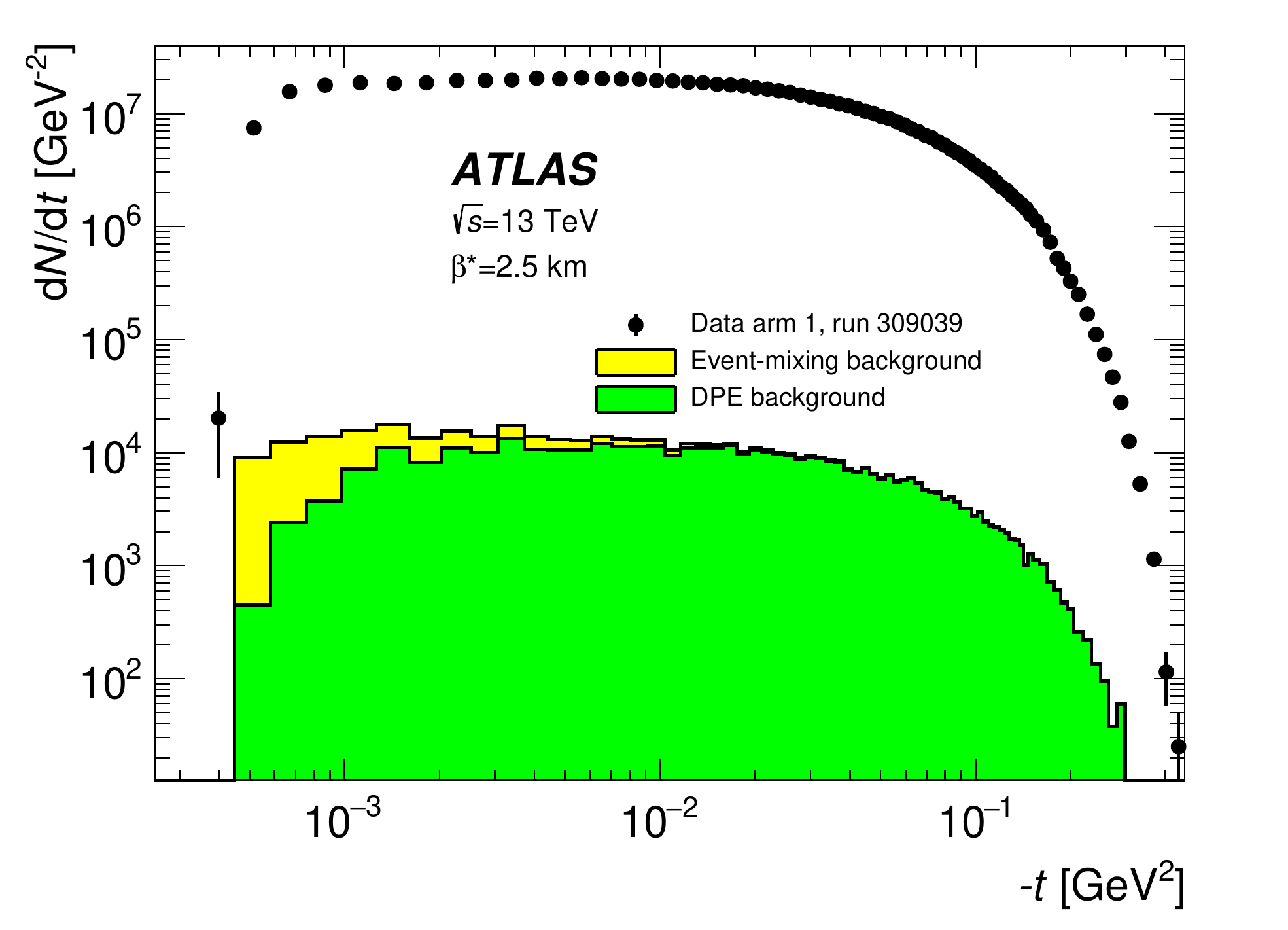}
\caption{The counting rate $\mathrm{d}N/\mathrm{d}t$ in Arm 1,
before corrections, as a function of $t$ compared
to the background from accidental coincidences and DPE.}
\label{fig:background_arm1}
\end{figure}

The resulting background estimate is illustrated in Figure~\ref{fig:background_arm1} for Arm 1, which shows that the
overall level of background is very small and dominated by DPE, except at small $|t|$ where the accidental coincidences prevail.
In previous analyses at $\betastar=90$~m~\cite{STDM-2013-10,STDM-2015-22} a procedure called the `anti-golden' method was
applied, in which events collected in all upper or all lower detectors were used to
estimate the irreducible background in the sample of elastic-scattering candidates.
This method is not used in the present analysis because the detectors are much closer to the beam
and even small vertical beam offsets (up to $80~\upmu$m, see Section~\ref{sec:align}) introduce an
asymmetry between the arms that is not reproduced by the anti-golden method.
 
The resulting background level is $0.75\permil$ on average with very small differences between the different runs.
Systematic uncertainties are estimated by changing the normalization regions, the template composition, and the parameters of the DPE simulation. The nominal normalization region for accidental coincidences shown in Figure~\ref{fig:bg44} in
the plane of $x$ and $\theta_x$ is changed to a region in the plane of $x$ on the A-side and C-side, which exhibits
a different type of correlation. The template composition is varied by imposing a veto on minimum-bias triggers from the central ATLAS detector, thereby depleting the SD content of the templates. The normalization regions for the DPE
background are varied in a similar way. In addition, the DPE background shape is changed by varying the value of the Pomeron
intercept in the simulation from $0.02$ to $0.15$. Varying this MBR parameter was found to have the largest
impact on the DPE shape.
The total background uncertainties are dominated by systematic uncertainties and range from $10.4\%$ to $14.8\%$.
 
\subsection{Reconstruction efficiency}\label{sec:RecoEff}
The reconstruction efficiency accounts for the elastic events which cannot be fully reconstructed
because of the following effects and which therefore need to be excluded from the analysis.
The development
of a hadronic shower in the Roman Pot may lead to configurations where the proton track cannot be reconstructed in
one or more of the four detectors. Also, an overlap with a halo proton or SD proton can lead to
reconstruction failures, particularly if these background protons have initiated a shower either in the
Roman Pot material or upstream as a result of a collimator scattering or beam--gas interaction. The large majority of failed reconstructions are thus related to events with a hit multiplicity that is too large, in which
no meaningful tracks can be determined. The
probability of reconstruction failing because of too few hits is in contrast very low, since for the regular
reconstruction only three good layers out of ten in $u$ and $v$ are required and the single-layer efficiency
is about $90\%$~\cite{ALFA}.

The number of elastically scattered protons that are lost is estimated by a data-driven tag-and-probe method. This method exploits the back-to-back topology of elastic events, allowing
a proton to be tagged on one side of the spectrometer and probe the reconstruction on the other side.
The probability that an elastic event is fully reconstructed is given by the reconstruction efficiency: \\
\begin{equation*}\label{eq:reco_basic}
\varepsilon_{\text{rec}} = N_{\text{reco}} / (N_{\text{reco}}+N_{\text{fail}}) ,
\end{equation*}
 
where $N_{\text{reco}}$ is the number of fully reconstructed elastic-scattering events, which have at least one
reconstructed track in each of the four detectors of a spectrometer arm, and $N_{\text{fail}}$
is the number of not fully reconstructed elastic-scattering events that have reconstructed tracks in
fewer than four detectors.
Events are grouped into several reconstruction cases, for which different selection criteria and corrections are applied,
to determine if an event is from elastically scattered protons, but was not fully reconstructed because of inefficiencies.
 
Both the fully reconstructed and failed events need to have an elastic trigger signal present and need to be inside the acceptance region.
The efficiency is determined separately for the two spectrometer arms.
Based on the number of detectors with at least one reconstructed track, the events are grouped into six reconstruction cases called \textit{topologies} named
4/4, 3/4, 2/4, (1+1)/4, 1/4 and 0/4.
The first number in this notation indicates the number of detectors with at least one reconstructed track.
In the 2/4 case, both detectors with tracks are on one side of the IP and in the (1+1)/4 case they are on different sides.
 
Elastic events for all cases are selected with the event selection described in
Section~\ref{sec:Event_selection}, using the subset of cuts available to each
particular topology.
The tag-and-probe method relies on the very high efficiency of the trigger scintillator
($\varepsilon_{\text{trig}}>99.9\%$) to tag the activity of events in the four detectors of an arm.
The activity is quantified by the number of hits, which must be more than five.
Then selection cuts are applied to detectors with reconstructed tracks to accept events as elastic
or reject them as background. Not all cuts can be applied to every case because limitations
may make them impossible to apply, e.g.\
left--right correlation cuts in the 2/4 case. This results in an overestimation of the elastic event yield and
a reduced background rejection efficiency, which is accounted for by a \textit{phase-space correction} derived
from simulation.
Because of the reduced set of cuts available for topologies with failed reconstructions, the background
rejection is less efficient and more background is present, particularly in the 2/4 topology.
 
\begin{figure}[hbt!]
\centering
\includegraphics[width=0.6\textwidth]{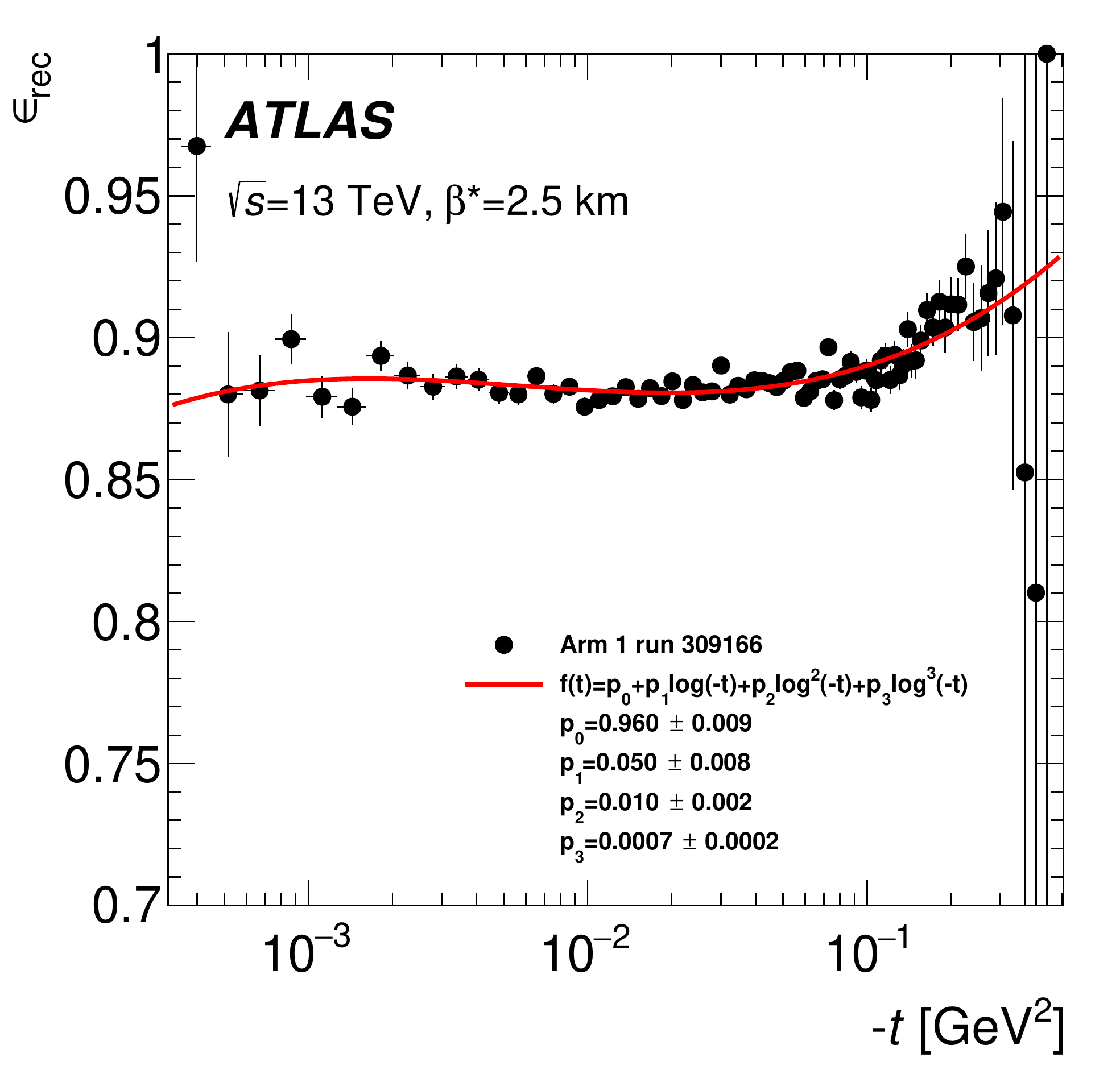}
\caption{The reconstruction efficiency as function of $t$ for
Arm~1. The line corresponds to
a fit of a polynomial function taken as a model of the potential $t$-dependence, which is used in the systematic uncertainty calculation.}
\label{fig:re_tdep}
\end{figure}

The background strategy follows the method applied to `golden' 4/4 events and described in Section~\ref{sec:Background}:
two background contributions from DPE events and accidental halo+SD events are considered, where both are
normalized in control regions specific to the topology.
The single-side templates for the accidental coincidences are extended to also include events where only one
detector has a reconstructed track. For the cases where only two or fewer detectors have reconstructed tracks,
particularly the 2/4 case, the background from accidental coincidences becomes important,
with a similar contribution from the DPE background.
For the cases with only one or no detector with reconstructed tracks, the tracking information is not sufficient for
event classification, and the expected yields are calculated with a probabilistic method~\cite{STDM-2013-10}
based on cases with reconstructed tracks in two or more detectors.
Taking into account background subtraction
and phase-space corrections, the final reconstruction efficiency calculation can be cast
in the following equation:
\begin{equation}\label{eq:recoeff_final}
\varepsilon_{\text{rec}} = \frac{N_{4/4}-B_{4/4}}{\left(N_{4/4}-B_{4/4}\right)+\frac{N_{3/4}-B_{3/4}}{\mathrm{PS}(3/4)} + \frac{N_{2/4}-B_{2/4}}{\mathrm{PS}(2/4)} + \frac{N_{(1+1)/4}-B_{(1+1)/4}}{\mathrm{PS}((1+1)/4)} + N_{\mathrm{lower}}} ,
\end{equation}
where $B_{j/4}$ denotes the background, PS the phase-space correction and $N_{\mathrm{lower}}$ merges
the background-subtracted topologies calculated by the probabilistic method.
\begin{figure}[htb!]
\centering
\includegraphics[width=0.6\textwidth]{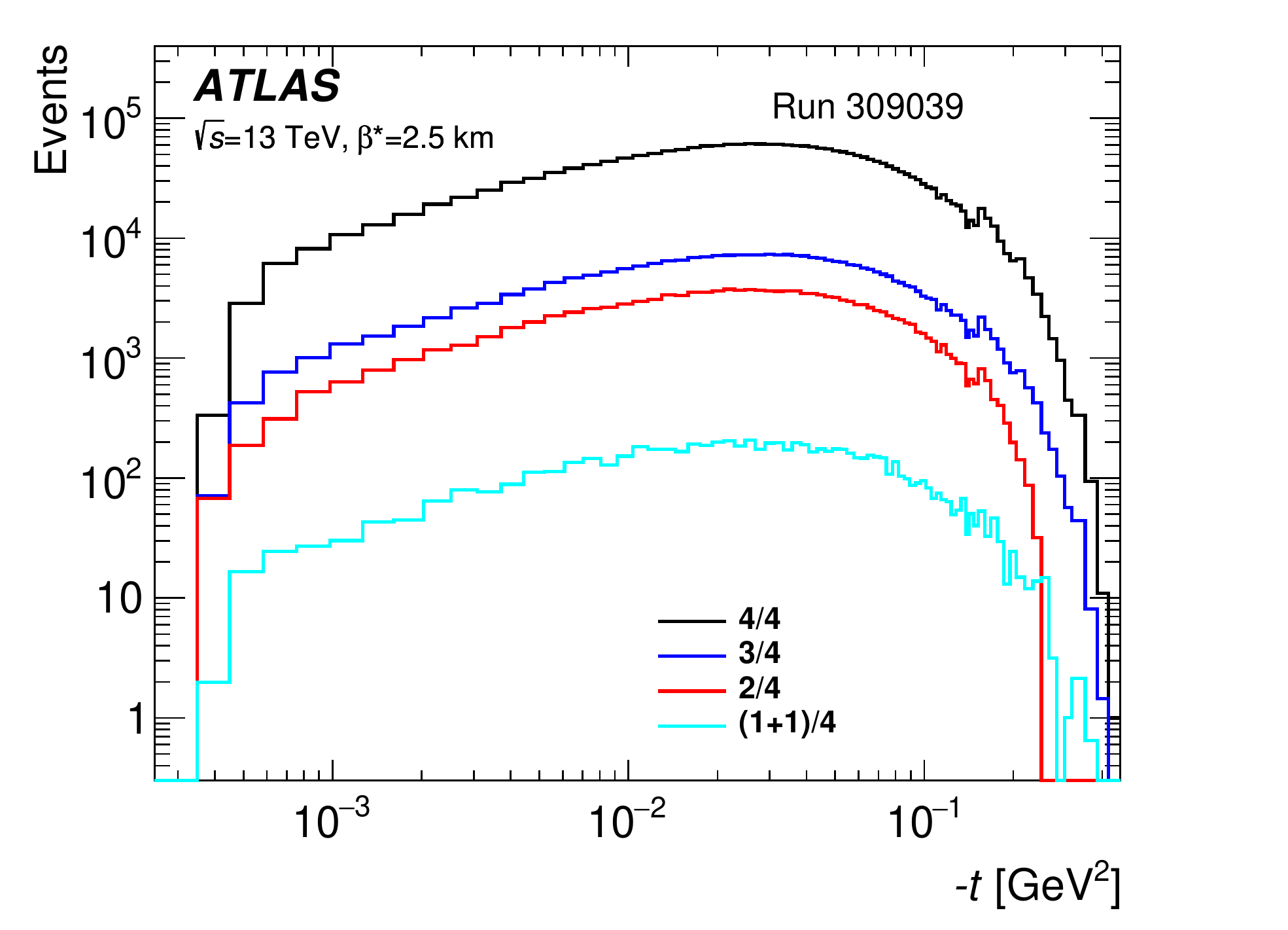}
\caption{$t$-spectra for elastic-scattering candidates with reconstructed tracks in four detectors (4/4) and
for events where in at least one detector the reconstruction failed. Events in the 2/4 topology have two detectors with reconstructed tracks on the same side, those in the (1+1)/4 have one detector with reconstructed tracks per side. Corrections for phase-space
effects are applied and background is subtracted.}
\label{fig:re_topo_cor}
\end{figure}
The $t$-spectra after background subtraction and with phase-space corrections are shown
for different topology classes in Figure~\ref{fig:re_topo_cor}. The shape of the spectra for
topologies with failed reconstruction is in good agreement with the shape of 4/4 cases, except for the 2/4
case where at large $|t|$ values the background, mostly from DPE, is slightly overestimated. This leads to a
bias, which is taken into account in the systematic uncertainty (see Figure~\ref{fig:re_tdep}).
 
\begin{figure}[htb!]
\centering
\includegraphics[width=0.6\textwidth]{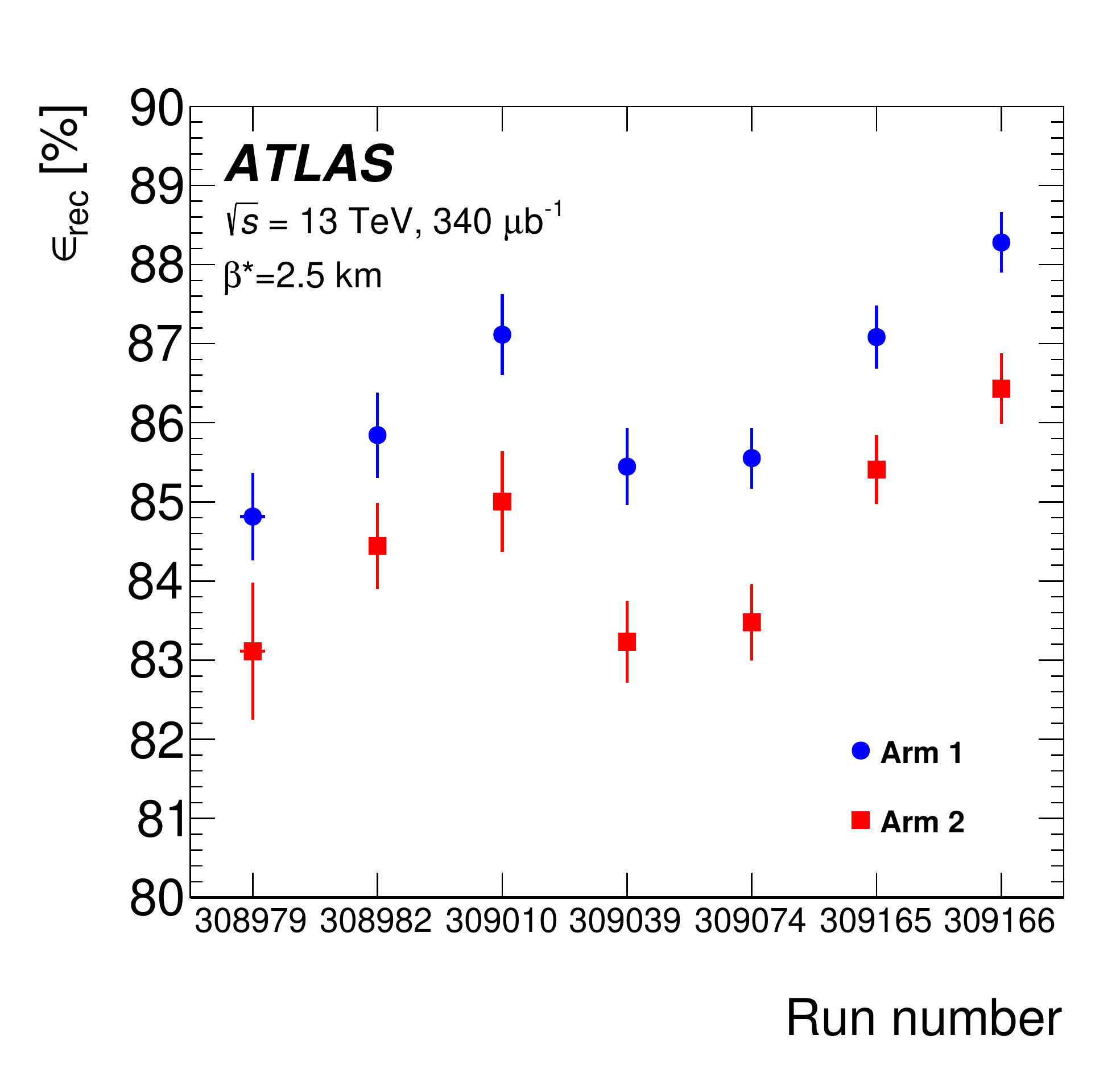}
\caption{Reconstruction efficiency for the two arms as function of the run number. The error bars show the total uncertainty.}
\label{fig:reFinal}
\end{figure}
The final reconstruction efficiency calculated per run and per arm according to Eq.~\eqref{eq:recoeff_final}
is shown in Figure~\ref{fig:reFinal}. The values are around $85\%$, and are typically about 2$\%$ higher for Arm~1 than
for Arm~2 because of a slightly different material distribution~\cite{STDM-2013-10}.
The results are summarized in Table~\ref{tab:run} for both arms. The systematic uncertainties range between $0.4\%$ and $0.9\%$ and are dominated by the composition of the templates for accidental coincidences
and uncertainties in the background subtraction. The former are evaluated by applying different veto conditions
and by increasing and decreasing the SD fraction in the template; the latter are calculated by varying the background normalization regions; and in both cases the procedure described in Section~\ref{sec:Background} is followed.
 
\begin{figure}[htb!]
\centering
\includegraphics[width=0.6\textwidth]{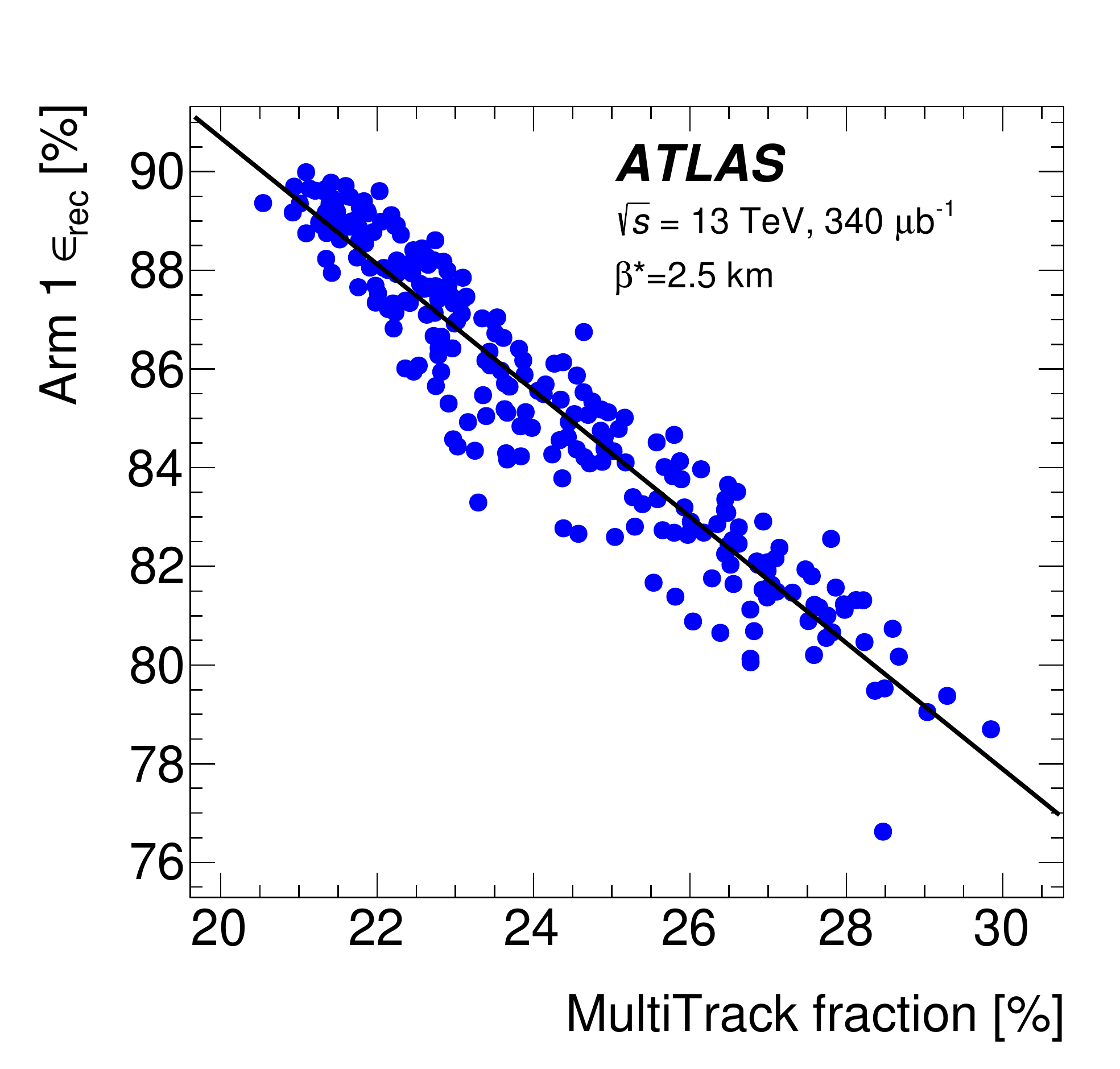}
\caption{The correlation between the reconstruction efficiency in data for Arm~1 and
the fraction of multi-track events. Each point corresponds to a data-recording period
between beam scrapings; data from all runs are combined.}
\label{fig:reMultiTrack}
\end{figure}

The reconstruction efficiency has a time dependence, both within a run and between runs. This time dependence is correlated with the number of halo protons overlapping in time with protons from elastic scattering.
Every proton, be it from elastic
scattering, an SD event or the beam halo, has a certain probability to develop a hadronic shower; thus in
events with overlapping protons the shower probability is approximately twice as high as in events with only one proton per detector. The correlation between the reconstruction efficiency and the number of overlapping halo
events is illustrated in Figure~\ref{fig:reMultiTrack}, where $\varepsilon_{\text{rec}}$ is plotted as a function
of the fraction of multi-track events. Each point in Figure~\ref{fig:reMultiTrack} corresponds to
a data-taking period between beam scrapings. Multi-track events are defined as events with a selected elastically scattered proton in each detector and an
additional reconstructed track in any of the four detectors. Different sources can create additional tracks,
such as cross-talk, but these are time-independent, whereas additional tracks from the halo exhibit a strong time dependence. Therefore, the time-variation of the multi-track fraction is dominated by the halo overlap,
which induces the clear correlation with $\varepsilon_{\text{rec}}$, shown in Figure~\ref{fig:reMultiTrack}.
The reconstruction efficiency is nominally assumed to be independent of the $t$-value because of the uniform
material distribution across the detection surface. However, a possible $t$-dependent bias could be introduced, through deficiencies in the background subtraction method for the 2/4 topology. A model of this possible $t$-dependence,
shown in Figure~\ref{fig:re_tdep}, is used to assess an additional systematic uncertainty.
 
\subsection{Acceptance and unfolding}\label{sec:Acceptance}
\begin{figure}[h!]
\centering
\includegraphics[width=0.6\textwidth]{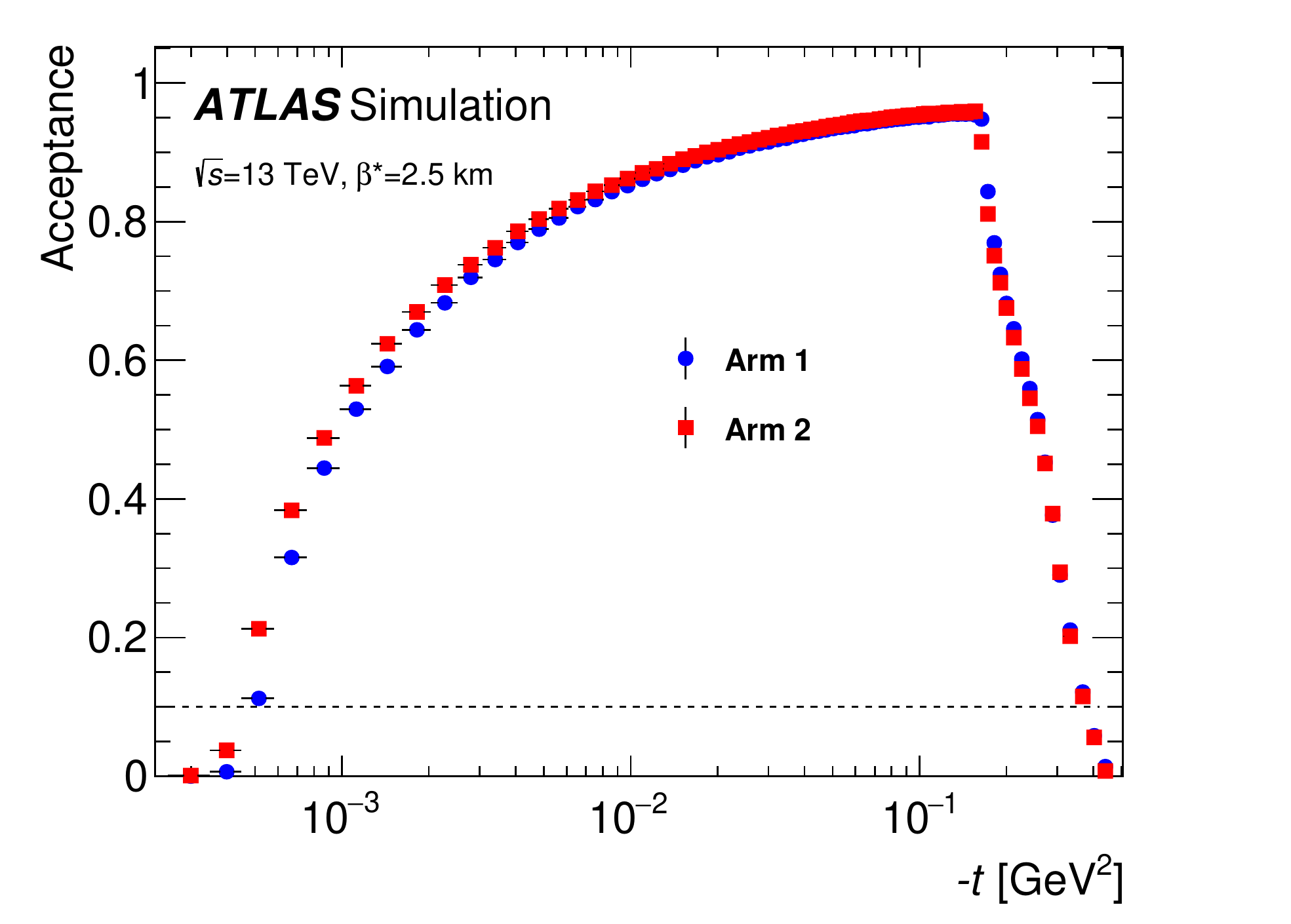}
\caption{The acceptance as a function of the true value of $t$ for each arm. The horizontal line corresponds to an acceptance of $10\%$ required for the extraction of the physics parameters.}
\label{fig:acceptance}
\end{figure}

The acceptance is defined as the ratio of events at particle level passing all geometrical and fiducial acceptance cuts
defined in Section~\ref{sec:Event_selection} to the total number of generated events, and is calculated as a function of $t$. The differential elastic cross section is corrected for acceptance losses.
The calculation is carried out with the simulation model described in Section~\ref{sec:ElsA}.
The acceptance is shown in Figure~\ref{fig:acceptance} for each arm.
The shape of the acceptance curve can be understood from the contributions of the vertical and horizontal
scattering angles to $t$ (see Eq.~\eqref{eq:txy}). The smallest accessible value
of $|t|$ is obtained at the detector edge
and set by the vertical distance of the detector from the beam. Close to the edge, the acceptance is small
because a fraction of the events are lost due to the beam divergence, i.e.\ events inside the acceptance on one
side but outside on the other side. At small $|t|$, up to $-t \sim 0.15~\GeV^2$, vertical and horizontal
scattering angles contribute about equally to a given value of $t$. Larger $|t|$-values imply larger
vertical scattering angles and larger values of $|y|$, and with increasing $|y|$ the fraction of events
lost in the gap between the main detectors decreases.
The maximum acceptance is reached
for events occurring at the largest possible values of $|y|$ within the beam-screen cut.
Beyond that point the acceptance decreases steadily because the events are
required to have larger values of $|x|$, since these $t$-values are dominated by the horizontal scattering
angle component. The difference between the two arms is mainly related to the vertical beam offset,
which brings the detectors of Arm~2 closer to the beam, thereby increasing the acceptance.
Meanwhile, events at large $|y|$ are shadowed by the beam screen, decreasing
the acceptance for Arm~2 at large $|t|$.
 
\begin{figure}[h!]
\centering
\includegraphics[width=0.6\textwidth]{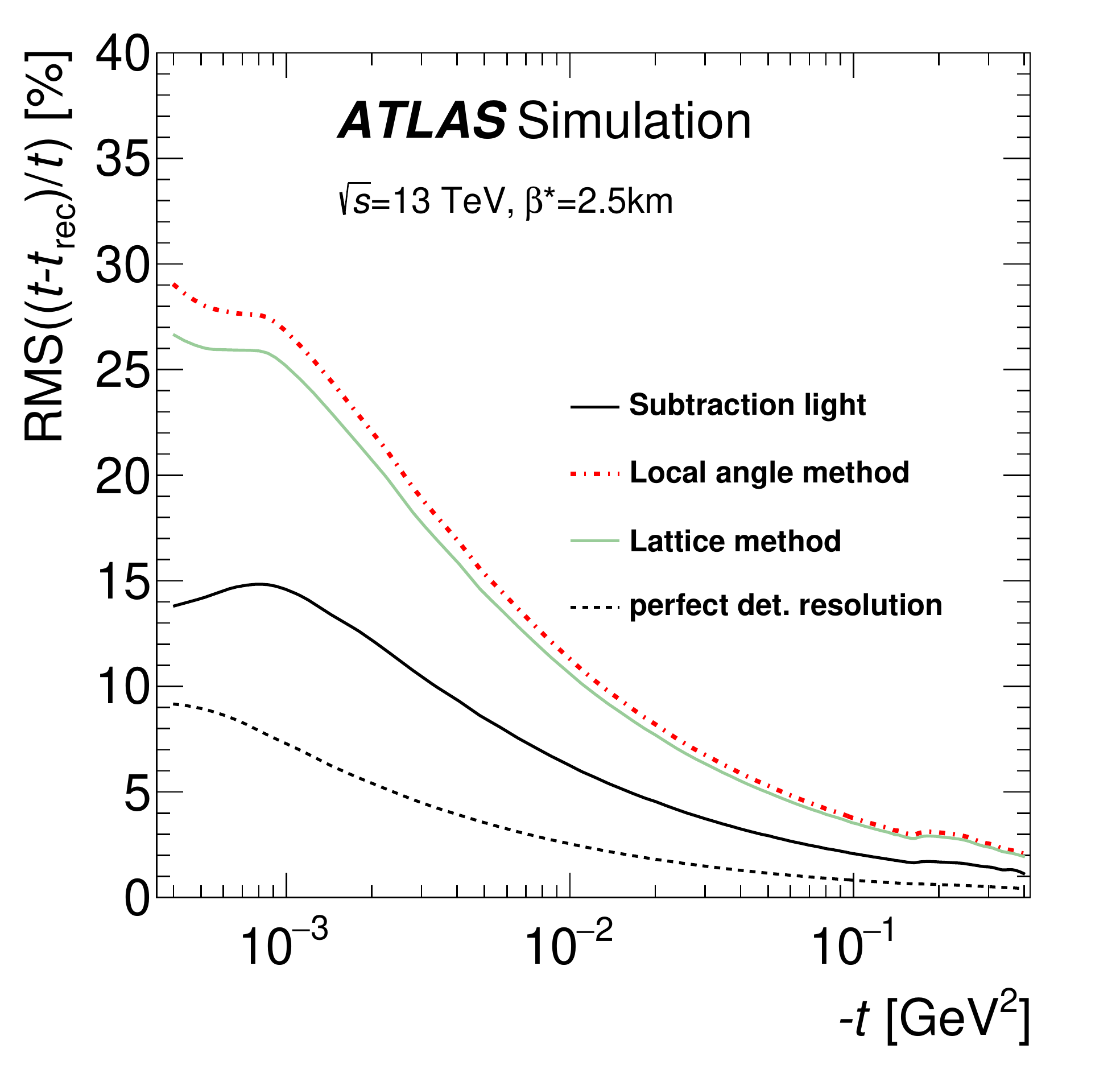}
\caption{The relative resolution $\mathrm{RMS}((t-t_\mathrm{rec})/t)$, where $t_\mathrm{rec}$ is the reconstructed
value of $t$, for different reconstruction methods. The resolution for local
subtraction is not shown as it is practically the same as for the lattice method.
The dotted line shows the resolution without detector resolution, accounting only for
beam effects. It is the same for all methods.}
\label{fig:tres}
\end{figure}

The measured $t$-spectrum is distorted by detector resolution and beam smearing
effects, including angular divergence, vertex smearing and energy smearing.
These effects change the shape of the reconstructed
$t$-spectrum, particularly at small $|t|$ as shown in Figure~\ref{fig:tres},
which also shows how detector resolution effects depend strongly on the reconstruction method.
The magnitude of the matrix elements used in the $t$-reconstruction relative to the detector spatial and
angular resolutions determines the performance of a method.
While the subtraction method receives only small contributions from the detector resolution,
which is good for the space coordinate measurement, all other methods suffer from a sizeable
degradation once detector resolution is included.
The degradation arises from the local angle being poorly measured, given that the 8~m distance between the two
stations is too small to obtain good precision.
 
\begin{figure}[hb!]
\centering
\includegraphics[width=1.0\textwidth]{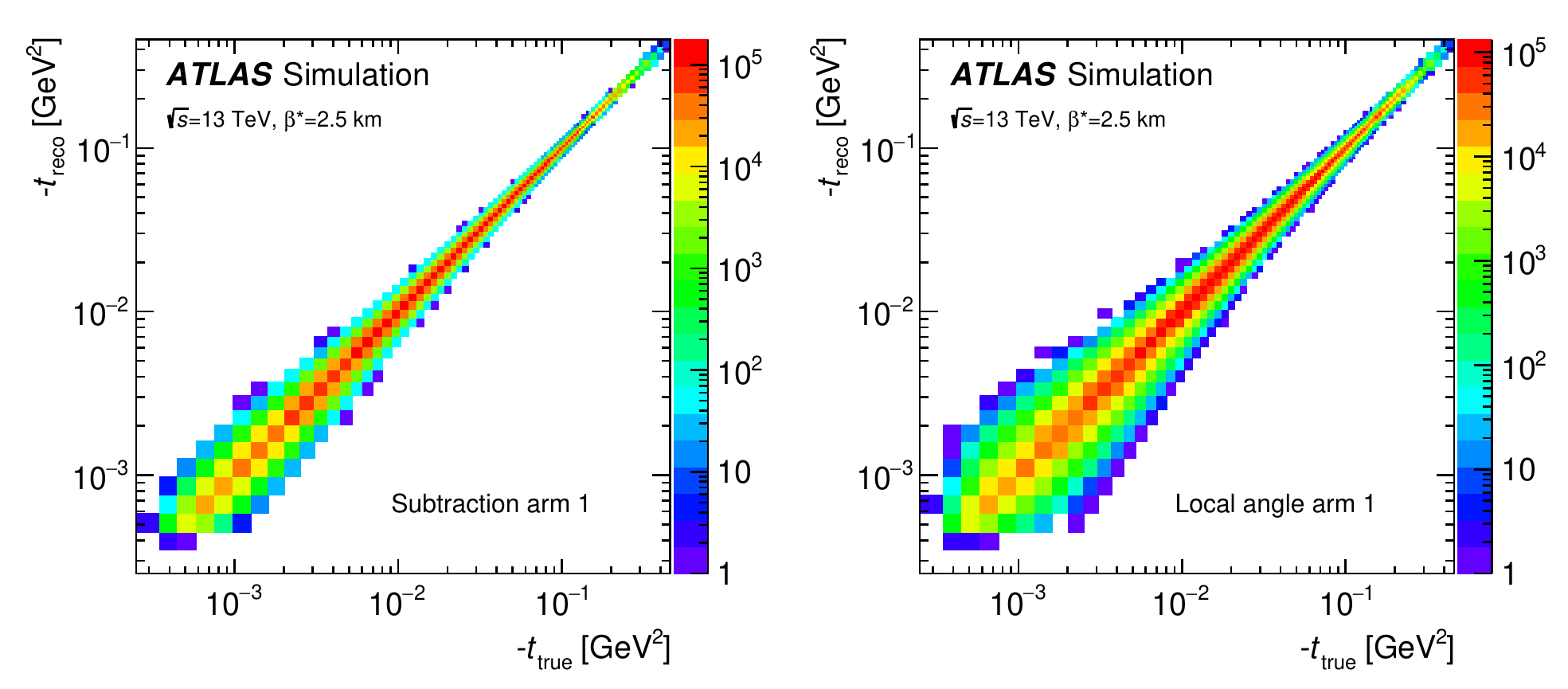}
\caption{The migration matrix of the $t$-value at reconstruction and generation level in Arm~1 for
the subtraction and
local angle methods.}
\label{fig:migration}
\end{figure}

After background subtraction, the measured $t$-spectrum in each arm is corrected for migration
effects using an iterative, dynamically stabilized unfolding method~\cite{IDS}.
MC simulation is used to obtain the migration matrix shown in Figure~\ref{fig:migration}
that is used in the unfolding. The superior resolution of the subtraction method means the migration
almost ends only one or two bins away from the diagonal, whereas a few more
non-diagonal elements are populated for the local angle method.
For the subtraction method the impact on the $t$-spectrum is very small and confined to the first few bins,
whereas for the local angle method the corrections are slightly larger and flat for $-t>10^{-3}~\GeV^{2}$.
Overall, the impact of the resolution on the reconstructed $t$-spectrum is negligible, except for very
small $|t|$ at the detector edge, where the beam divergence is important.
The very first bin at smallest $|t|$ is dominated by divergence effects. The events that contribute to this bin have a small vertical
scattering angle and would, in the absence of divergence, be outside of the acceptance
defined by the vertical cut close
to the detector edge. The divergence-induced angular smearing folded with the steeply falling scattering
angle distribution causes many more events to migrate into that bin than out of it.
 
The results are cross-checked using an unfolding method based on the singular value decomposition method~\cite{SVD}.
The unfolding procedure is applied to the distribution obtained using all selected events, after background
subtraction in each elastic arm.
A data-driven closure test is used to evaluate any bias in the unfolded data spectrum shape due
to mis-modelling of the reconstruction-level spectrum shape in
the simulation. The simulation is reweighted,
according to a polynomial function parameterizing the data/MC difference,
at particle level such that the reconstruction-level spectrum in simulation matches the data.
The modified reconstruction-level simulation is unfolded
using the original migration matrix, and the result is compared with the modified particle-level spectrum.
The resulting bias is considered as systematic uncertainty; it is very small.
 
The unfolding procedure introduces a statistical correlation between the bins of the $t$-spectrum,
which is incorporated in a covariance matrix included when fitting the data.
 
\subsection{Beam optics}\label{sec:Optics}
The reconstruction of the $t$-value requires knowledge of the elements of the transport matrix.
The transport matrix can be calculated from the design of the 2.5~km beam optics,
which consists of a sequence of the beam elements including the alignment parameters of the magnets,
the magnet currents and the field calibrations.
This initial set of matrix elements is referred to as the `design optics'.
Small corrections, allowed within the range of the systematic uncertainties,
need to be applied to the design optics for the measurement of the physics parameters,
following a procedure developed in Refs.~\cite{STDM-2013-10,STDM-2015-22}.
Constraints on beam optics parameters
are derived from the ALFA data, exploiting the fact that the reconstructed scattering angle
must be the same for different reconstruction methods using different transport matrix elements.
The beam optics parameters are determined from a global fit, using these constraints, with the design optics as
a starting value. The tuned parameters are the field strengths of the quadrupoles, quantified
by the k-values, located between the IP and
ALFA, and named Q1--Q6 (see Figure~\ref{fig:alfa_layout}), in beam~1 and beam~2.
For the 2.5~km beam optics,
the following k-values are tuned to fulfil the ALFA constraints: a single k-value correction for Q1 and Q3, because they are connected to a common power line, and independent k-values for Q5 and Q6.
In the fit, the k-values for beam~1 and beam~2 are
treated as independent parameters, resulting in a total of six fitted parameters.
 
Reconstructed tracks from elastic-scattering events are used to derive two classes of data-driven constraints
on the beam optics:
\begin{itemize}
\item Correlations between A-side and C-side measurements of positions or angles, and between positions or angles
measured in the inner and outer stations of ALFA on the same side. These are used to infer the ratio of matrix elements in the beam transport matrix.
The resulting constraints are independent of any optics input.
\item Correlations between the reconstructed scattering angles. These are calculated using different methods
to derive further constraints on matrix elements as scaling factors. These factors
indicate the amount of scaling needed for a given matrix element ratio
to equalize the measurements of the scattering angle. These constraints depend on the given optics model.
The design beam optics with quadrupole currents measured during the run is used as a reference to calculate the constraints.
\end{itemize}
For each constraint, the bias induced by the measurement method because of resolution or acceptance
effects is estimated by evaluating the constraint in simulation and comparing
the value with that from the design optics. An additive correction is then applied to the constraint values
obtained from data before comparison with beam optics calculations. These corrections are generally small
and at the level of a few per mille.
 
\begin{figure}[hb!]
\centering
\includegraphics[width= 0.6\textwidth]{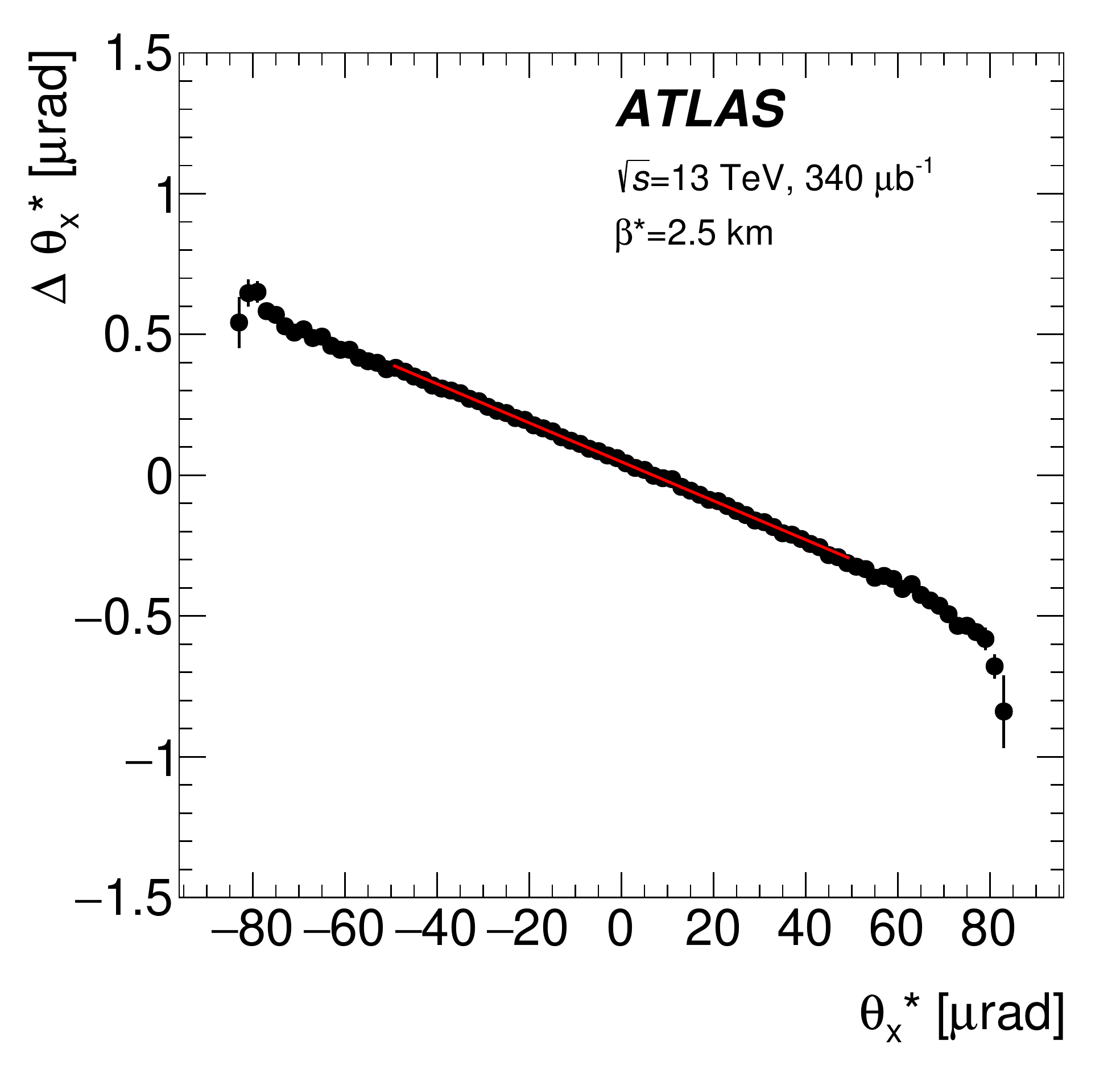}
\caption{The difference in reconstructed scattering angle $\Delta\theta^\star_x$ between the subtraction and local angle methods as a function of the scattering angle using the subtraction method for the outer detectors. In each bin of the scattering angle the points show the mean value of $\Delta\theta^\star_x$ and the error bar represents the RMS. The red line represents a fit used to determine the value of the slope.}
\label{fig:optics_perx}
\end{figure}

A stringent constraint from the second class is illustrated in Figure~\ref{fig:optics_perx}.
This example shows the comparison
of the scattering angle in the horizontal plane reconstructed with the subtraction method, Eq.~\eqref{eq:subtraction},
which is based on the position and $M_{12,x}$,
and the local angle method Eq.~\eqref{eq:localangle}, which is based on the local angle and $M_{22,x}$.
Figure~\ref{fig:optics_perx} shows the difference in scattering angle between the two methods as a function of the scattering angle determined with the subtraction method. The slope is extracted using a linear fit in the central region (red line). This slope is used as the scaling factor for the matrix element ratio $M_{12}/M_{22}$.
Its value of about $1\%$ indicates that the design optics ratio $M_{12,x}/M_{22,x}$ needs to
be increased by $1\%$ to obtain the same scattering angle, on average, from both methods using data.
 
\begin{figure}[hb!]
\centering
\includegraphics[width=\textwidth]{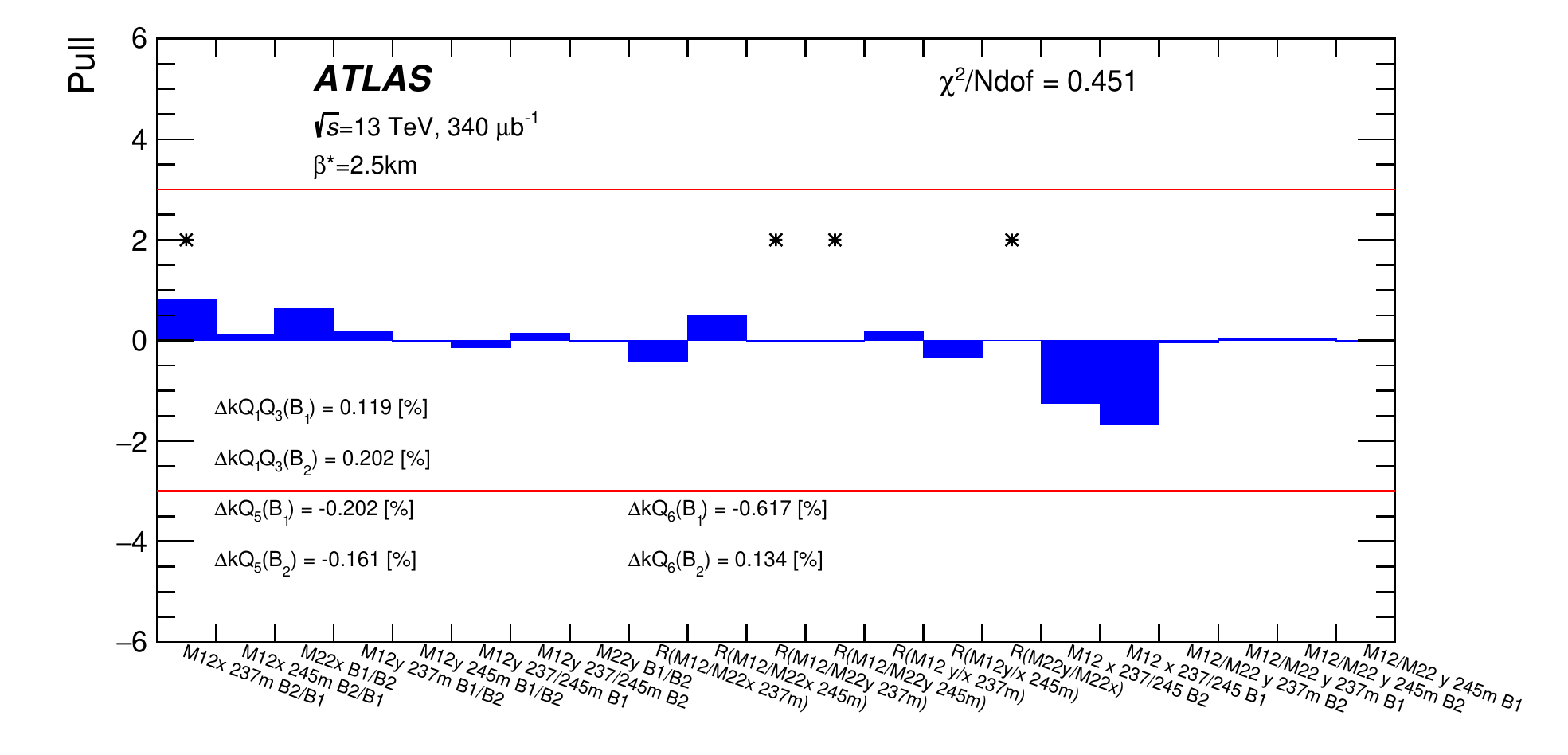}
\caption{Pulls from the effective beam optics fit of Q1, Q3, Q5 and Q6 to the ALFA constraints. The constraints marked with an asterisk are excluded from the calculation of the $\chi^{2}$. The red lines indicate pull values of $\pm 3$.}
\label{fig:optics_hfit_Q5Q6Q13}
\end{figure}

A total of 21 constraints were determined from the ALFA data.
A sub-set of 17 constraints with the best precision are
normally included in the $\chi^2$ calculation of the fit, four constraints with larger experimental
uncertainties and less constraining power are not included but kept for control.
All ALFA constraints are treated as uncorrelated in the fit, but coherent changes in the constraints
observed under experimental variations of the analysis are
taken into account in the systematic uncertainty.
In the minimization procedure, the beam
optics calculation program MadX~\cite{madx} is used to extract the optics parameters and
to calculate the matrix element ratios for a given set of magnet strengths.
The ALFA system provides precise constraints on the matrix element ratios,
but cannot probe the deviations of single magnets. There are, therefore, several sets of optics parameters
which minimize the $\chi^2$, arising from different
combinations of magnet strengths.
The chosen configuration, called the effective optics, is one solution among many.
Different sets of quadrupoles are included in the fit, and a variation of the choice of set
represents the main contribution to the systematic uncertainty.
Figure~\ref{fig:optics_hfit_Q5Q6Q13} shows the pull of the ALFA constraints after the minimization, which resulted in
magnet strength corrections ranging from 0.1\% to 0.6\%, with rather large differences between beam~1 and beam~2,
most notably for Q6 with a negative correction of $-0.62\%$ in beam~1 and a positive correction of $0.13\%$ in beam~2.
The $\chi^{2}$ of the fit includes the systematic uncertainties of the constraints and is of good quality
with $\chi^{2}/N_{\textrm{dof}}=0.45$.
The resulting change in the transport matrix elements most important for $t$-reconstruction is up to 1\%.
 
Systematic uncertainties for the k-values and matrix elements include the coherent change of constraints under
experimental variations. A total of 13 variations, including
the nominal constraints obtained from the average of the arms and
the constraints from individual arms, are taken into account and the standard deviation of the resulting
distribution is assigned as an uncertainty. The dominant systematic uncertainty with the largest impact on
the $t$-reconstruction is obtained when treating only the k-values of Q5 and Q6 as free parameters in the fit,
which still results in an acceptable fit quality.
 
\subsection{Luminosity}\label{sec:Luminosity}
The general methods for luminosity determination in ATLAS are described in Ref.~\cite{lumi13TeV}. However, a dedicated measurement is required due to the special $\betastar = 2.5$~km optics resulting in different conditions shown in  Table~\ref{tab:parameters}.  The ATLAS general strategy to provide a reliable luminosity determination and to properly assess the systematic
uncertainties is to compare the measurements of different detectors and algorithms. This section describes the luminosity determination for this run and its systematic uncertainty using LUCID  (LUminosity measurement with a Cherenkov Integrating Detector)~\cite{lucid} as the
baseline alongside approaches using the Beam Conditions Monitor (BCM)~\cite{bcm}
and the Inner Detector (ID)~\cite{atlas1} of ATLAS.
Each detector and each algorithm is calibrated in special van der Meer (vdM) scans~\cite{svdm},
except the track counting algorithm, which is cross-calibrated to LUCID in parts of the vdM scan run with head-on collisions.
Table~\ref{tab:parameters} summarizes the differences between high-$\betastar$ runs, high-luminosity runs and vdM scans.
Such differences are due to the number of colliding bunches and the average numbers of interactions per bunch crossing
(pile-up parameter $\mu$), leading to an instantaneous luminosity in the high $\betastar$ regime
which is up to seven orders of magnitude lower than in high-luminosity running, and three orders of magnitude lower than in the vdM scans.
\begin{table}[tp]
\centering
\caption{Main parameters for high-$\betastar$ runs, vdM scans and high-luminosity runs~\cite{lumi13TeV}.}
\begin{tabular}{c c c c c c c}
\hline
Parameters & 2016 high-$\betastar$ runs & 2016 vdM scan  & High-luminosity runs\\
\hline
Number of colliding bunches & 4--5 & 32 & 2208 \\
Average pile-up parameter $\mu$ & 0.002--0.006 & 0.5 & 41 \\
Instantaneous luminosity (10$^{27}$ cm$^{-2}$ s$^{-1}$) & $\sim$ 1.4--4 & 2.6 $\cdot$ 10$^{3}$ & 13 $\cdot$ 10$^{6}$ \\
$\betastar$ (m) & 2500 & 19 & 0.4 \\
\hline
\end{tabular}
\label{tab:parameters}
\end{table}
 
The  algorithms listed in Table~\ref{tab:algo} were used in the final analysis.
They were chosen as having the sensitivity needed for high-$\betastar$ runs and at the same time exhibiting the most
favourable background conditions. Both LUCID and the BCM have detectors on both the A-side and C-side of the ATLAS IP.
The A-side detectors were not used in OR-mode due to  high background on the A-side.
The track-counting algorithm using ID data is independent of LUCID and the BCM. The track-counting
algorithm obtains the per-bunch visible interaction value $\mu_{\textrm{vis}}$ from the mean number of reconstructed tracks per bunch crossing
averaged over a luminosity block, which corresponds to a period of about one minute with approximately constant luminosity. The track measurements used in this analysis are based on the
TightModLumi\footnote{The TightModLumi working point is named {\em 2017 selection} in Ref.~\cite{lumi13TeV}.} working point~\cite{lumi13TeV}; they are available for only two of the ALFA runs (308979 and 309165).
 
\begin{table}[tp]
\centering
\caption{The luminosity determination algorithms used for the high-$\betastar$ analysis. For LUCID
the activity is defined by at least one hit in a tube. BI and BI2 correspond to two different sets of photomultipliers used by LUCID. For the BCM the activity is defined by at least one hit in one of the four sensors. The total integrated luminosity and relative deviation from the nominal LUCID\_EventORC\_BI algorithm is given
for the algorithms available for all runs. Only statistical uncertainties are shown.}
\begin{tabular}{l l c c}
\hline
Algorithm & Definition & $L_{\textrm{int}}~[\upmu \textrm{b}^{-1}]$ & Deviation [\%]\\
\hline
LUCID\_EventORC\_BI  &  Activity on the C-side in LUCID & $339.9 \pm 0.1$ & -- \\
LUCID\_EventAND\_BI  &  Activity on  both sides in LUCID & -- & -- \\
LUCID\_EventORC\_BI2 &  Activity on the C-side in LUCID & $341.1 \pm 0.1$ & $+0.4$  \\
LUCID\_EventAND\_BI2 &  Activity both sides in LUCID  & $341.2 \pm 0.2$ & $+0.4$ \\
BCM\_T\_EventAND     &  Activity on both sides in the BCM & $337.2 \pm 0.2$ & $-1.8$ \\
TightModLumi         &  Track counting using the Inner Detector & -- & -- \\
\hline
\end{tabular}
\label{tab:algo}
\end{table}
 
\begin{figure}
\centering
\includegraphics[width=0.8\textwidth]{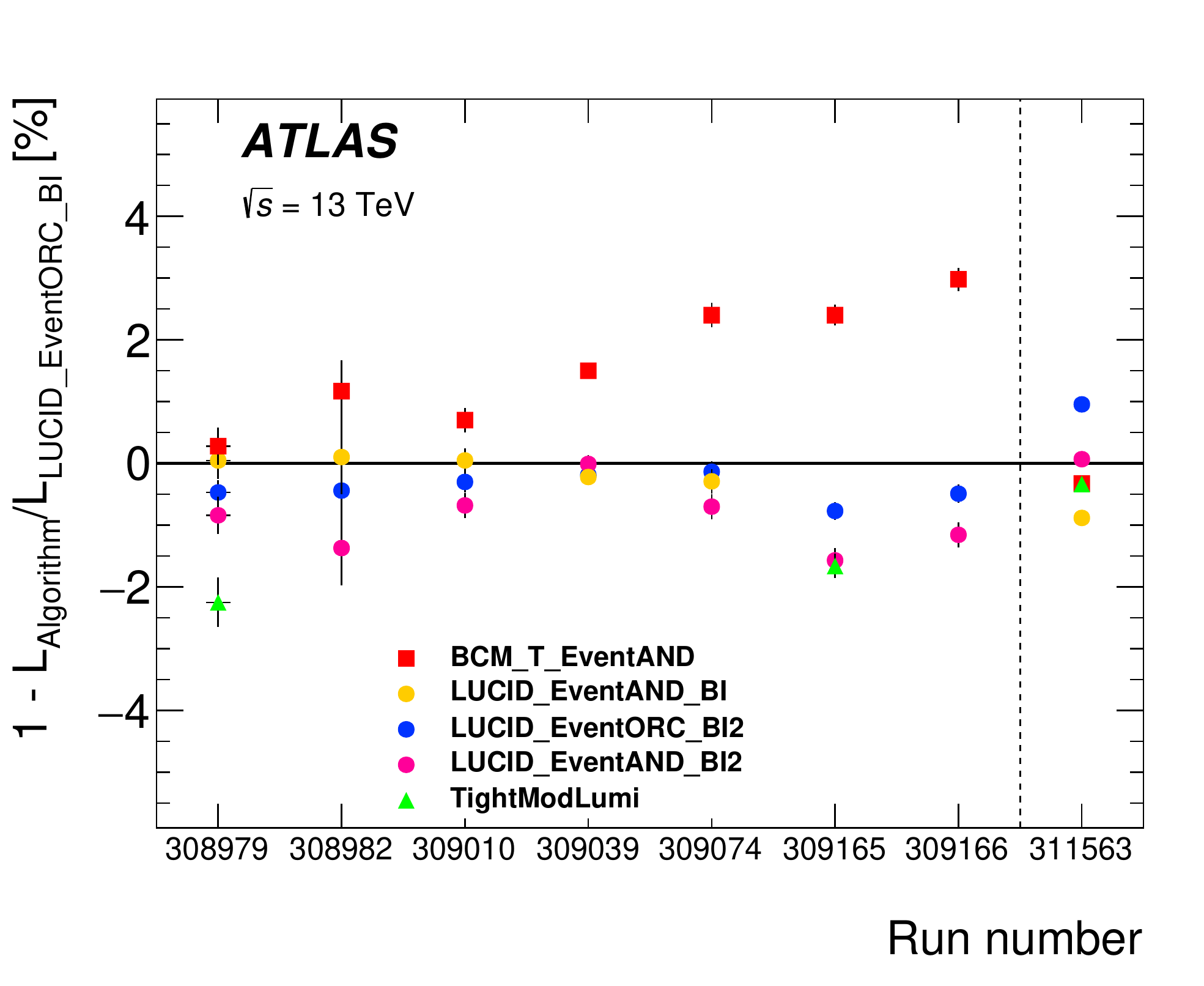}
\caption{Percentage deviation of the integrated luminosity with respect to the reference algorithm LUCID\_EventORC\_BI as a function of the run number. The very last point refers to the MD1814 run. All the algorithms were calibrated in the vdM run.}
\label{fig:lumi_ratios}
\end{figure}

LUCID\_EventORC\_BI, being the most stable algorithm with favourable background conditions, was chosen as the reference algorithm and the other algorithms are used to evaluate the systematic uncertainties.
The percentage differences with respect to the reference algorithm are shown in Figure~\ref{fig:lumi_ratios} for all runs as a
function of the run number. This is a key plot in the context of evaluating the systematic uncertainties.
The main sources of systematic uncertainty are the vdM calibration uncertainty, calibration transfer uncertainty, long-term stability uncertainty and background uncertainty.
 
\textbf{vdM calibration uncertainty:}
This source of uncertainty is the same as for the standard luminosity
analysis, and in 2016 it was evaluated to be 1.1\%~\cite{lumi13TeV}, while the central value
of the calibration was updated for this analysis, applying the techniques described in Ref.~\cite{lumi13TeV}.
 
\textbf{Calibration transfer uncertainty:}
This arises because the vdM scans are performed at a luminosity $10^3$~times higher than the luminosity
during the ALFA data taking (see Table~\ref{tab:parameters}).
Any non-linearity between the vdM-scan and ALFA-run luminosity
regimes would appear as systematic deviation between algorithms in the stability plot shown in Figure~\ref{fig:lumi_ratios}.
It is therefore included in the
deviations of the various independent algorithms from the reference one.
 
\textbf{Long-term stability uncertainty:}
Unlike the standard data taking, the period with high-$\betastar$ runs was very short,
lasting only a few days. Nevertheless, these runs were acquired four months
after the vdM-scan session and about one month before a vdM-like scan, the so-called MD1814 scan.
The stability of the detectors and algorithms from the time of the vdM-scan session was
evaluated through the comparison between these two scans, and it
is reflected in the stability plot of Figure~\ref{fig:lumi_ratios}. Thus, no additional uncertainty is needed to
account for the long-term stability.
 
\textbf{Background uncertainty}:
The single-beam background to the luminosity signal, as estimated from unpaired bunches, appears negligible for all
LUCID and BCM algorithms used in this analysis. There is no sign of collision background for the chosen algorithms.
A more quantitative constraint on collision background can be inferred from the internal consistency, over the
entire ALFA running period, of measurements reported by five independent luminosity algorithms with very different intrinsic
background sensitivity. Therefore, the internal consistency of the measurements displayed in Figure~\ref{fig:lumi_ratios}  implicitly sets an upper limit on the possible impact of collision backgrounds on the reported integrated luminosity.
 
The total systematic uncertainty of the
luminosity measurement can be obtained from the absolute
calibration uncertainty of the vdM scan and from the stability and consistency among the
various algorithms.
Figure~\ref{fig:lumi_ratios} shows the deviation of the luminosity from the value given by the reference algorithm run by run.
However, the run-integrated luminosity is most relevant for the determination of $\sigmatot$. The run-integrated luminosity measured by the various algorithms is compared with the reference LUCID\_EventORC\_BI value. This provides the uncertainty related to all listed effects:  calibration transfer, long-term stability and background.
The biggest deviation in the run-integrated luminosity is found to be $-1.8\%$ for the
BCM\_T\_EventAND algorithm. Unfortunately, the TightModLumi track-counting algorithm was only available in two runs.
The TightModLumi run-integrated luminosity in these two runs deviates from the value given by
reference LUCID algorithm by $1.85\%$.
This number is therefore taken as the stability and consistency uncertainty.
 
Adding in quadrature the uncertainty in the absolute calibration (1.1\%) and
that associated with the stability and consistency of the available
independent luminosity measurements (1.85\%), a total systematic
uncertainty of 2.15\% is obtained for the integrated luminosity
delivered during the seven high-$\betastar$ runs.
 
The final value of the integrated luminosity is therefore:
\begin{equation*}
L_{\textrm{int}} = 339.9 \pm 0.1 ~\mbox{(stat.)} \pm 7.3 ~\mbox{(syst.)}~\upmu \textrm{b}^{-1} .
\end{equation*}

% End of text imported from the .//Analysis.tex input file

% The next lines are included from the .//Results.tex input file
\section{Results}\label{sec:Results}
\subsection{The differential elastic cross section}\label{sec:Telastic}
Several corrections are applied to calculate the differential elastic cross section.
The corrections are made individually per run and per arm before combining all runs and the
two arms in the final differential elastic cross section.
In a given bin $t_i$, the cross section is calculated according to the following formula:
\begin{equation*}\label{eq:cross-section}
\frac{\mathrm{d}\sigma}{\mathrm{d}t_i} = \frac{1}{\Delta t_i}\times \frac{{\cal M}^{-1}[N_i - B_i]}{A_i  \times \epsilon_{\mathrm{rec}} \times \epsilon_{\mathrm{trig}} \times \epsilon_{\mathrm{DAQ}}  \times L_{\mathrm{int}} } ,
\end{equation*}
where $\Delta t_i$ is the bin width, ${\cal M}^{-1}$ represents the unfolding procedure applied to the
background-subtracted number of events $N_i - B_i$, $A_i$ is the acceptance,
$\epsilon_{\mathrm{rec}}$ is the event reconstruction efficiency, $\epsilon_{\mathrm{trig}}$ is the trigger efficiency,
$\epsilon_{\mathrm{DAQ}}$ is the dead-time correction and $L_{\mathrm{int}}$ is the integrated luminosity used for this
analysis.
The binning in $t$ is chosen to be appropriate for the experimental resolution and statistical uncertainty. At small $|t|$ the selected bin width is two times the resolution.
At larger
$|t|$ the bin width is increased to compensate for the lower number of events from the exponentially falling distribution.
The  resulting differential elastic cross section obtained using the subtraction method is shown in Figure~\ref{fig:tfit_nuisance_subtraction} and numerical values
with uncertainties are summarized in Tables~\ref{tab:tval1} and~\ref{tab:tval2} in the Appendix.
\subsubsection*{Experimental systematic uncertainties}
\begin{figure}[htbp]
\centering
\subfloat[Alignment and effective optics]{
\includegraphics[width=0.45\textwidth]{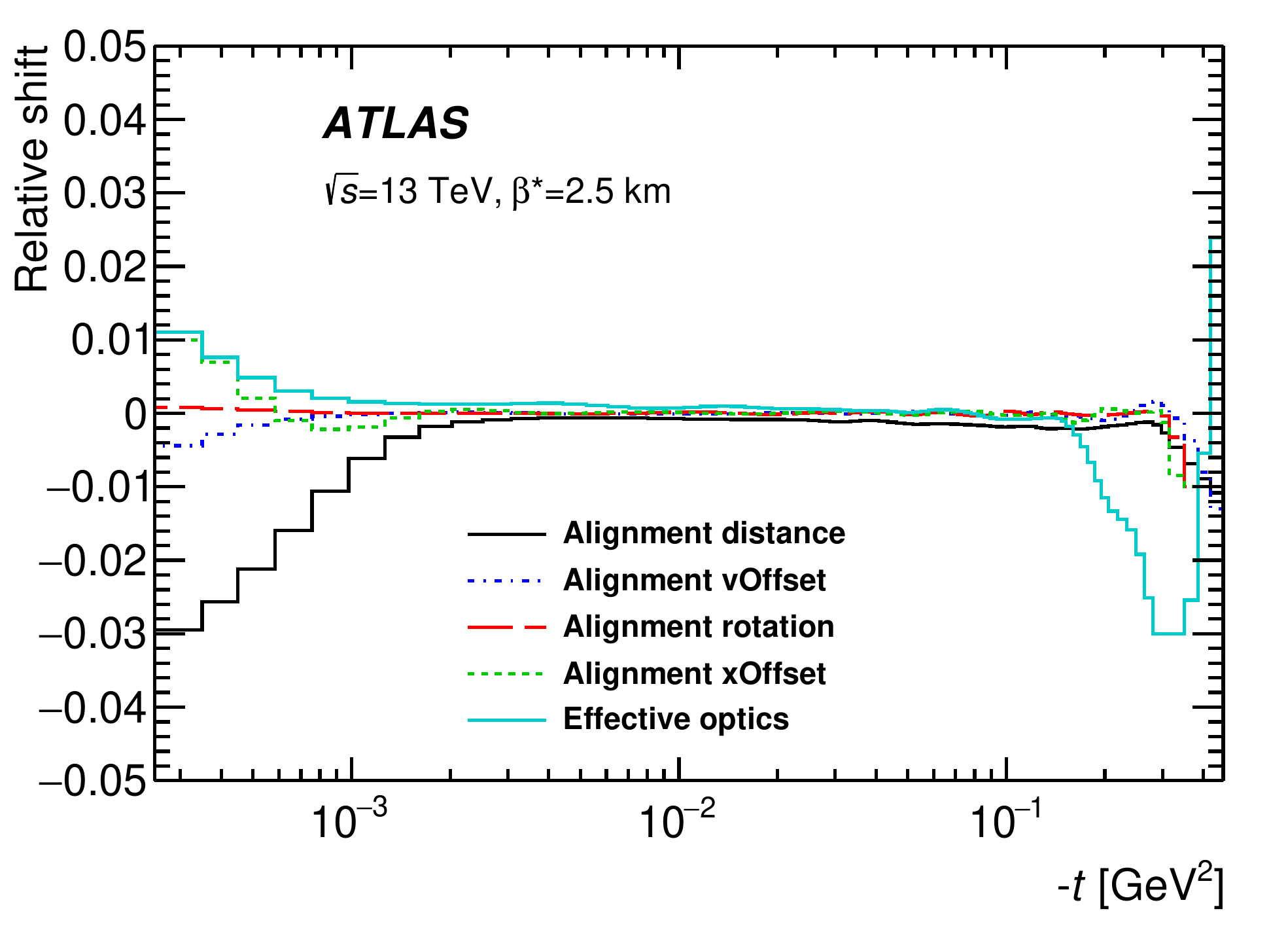}
\label{fig:tshift1}
}
\subfloat[Beam properties]{
\includegraphics[width= 0.45\textwidth]{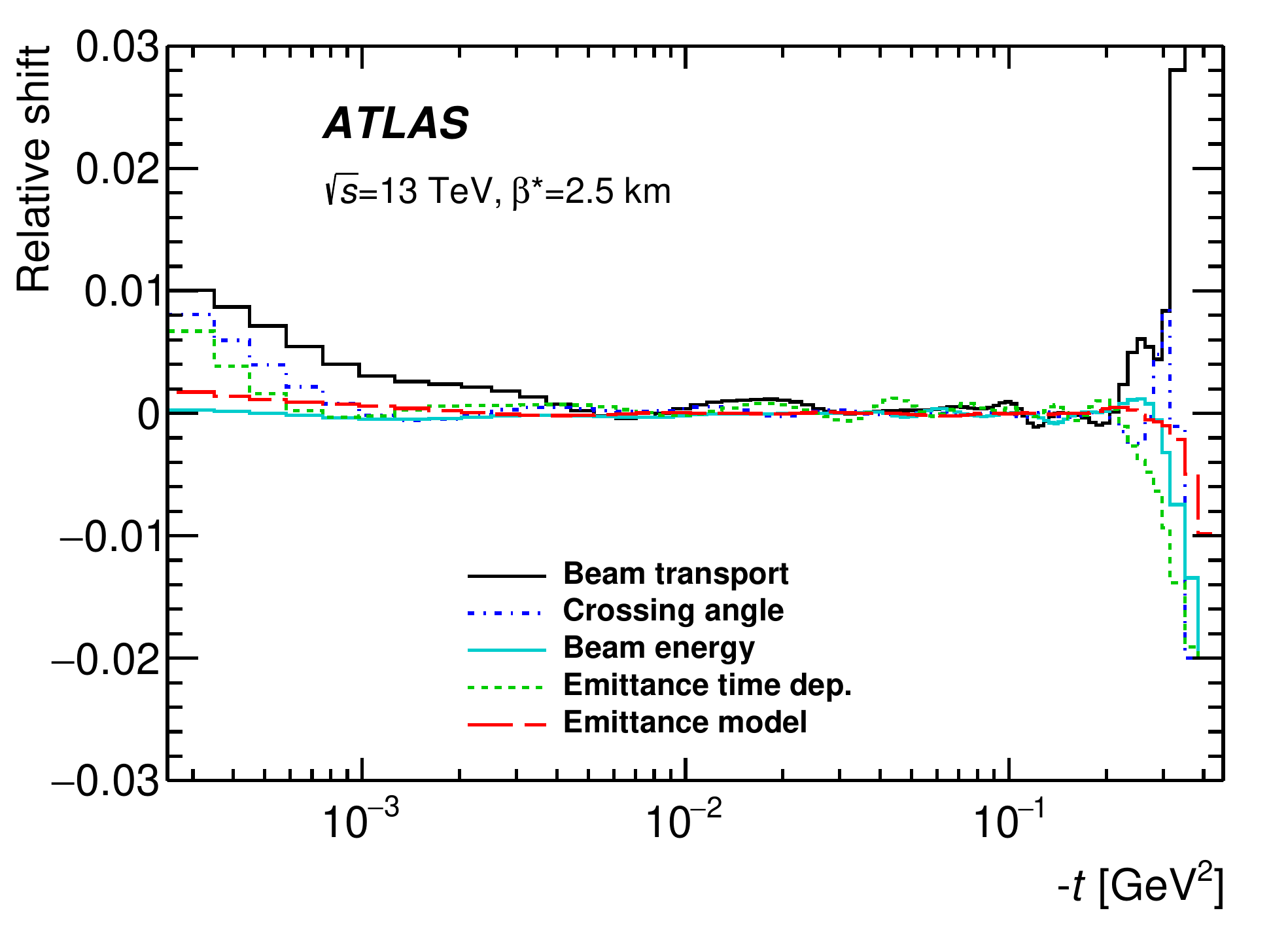}
\label{fig:tshift2}
}
\caption{Relative systematic shifts of $\mathrm{d}\sigma/\mathrm{d}t$ as a function of $t$ resulting from uncertainties in the
\protect\subref{fig:tshift1} alignment and effective optics and
\protect\subref{fig:tshift2} beam properties. For cases where two-sided uncertainties are available the larger one was retained and assigned as a symmetric uncertainty.}
\label{fig:tshift12}
\end{figure}

\begin{figure}[htbp]
\centering
\subfloat[Simulation, detector resolution and background]{
\includegraphics[width=0.45\textwidth]{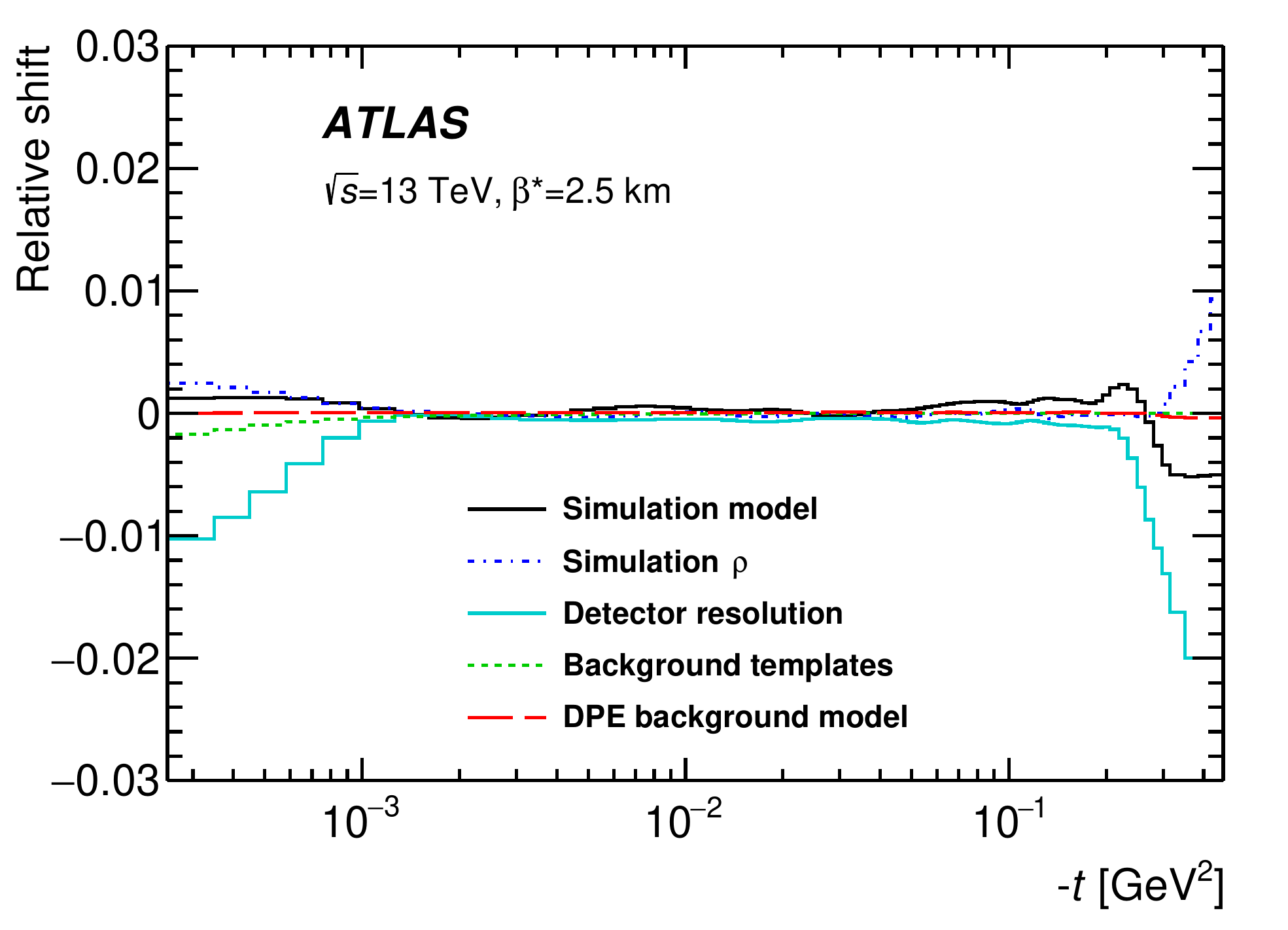}
\label{fig:tshift3}
}
\subfloat[Reconstruction efficiency, luminosity and tracking]{
\includegraphics[width= 0.45\textwidth]{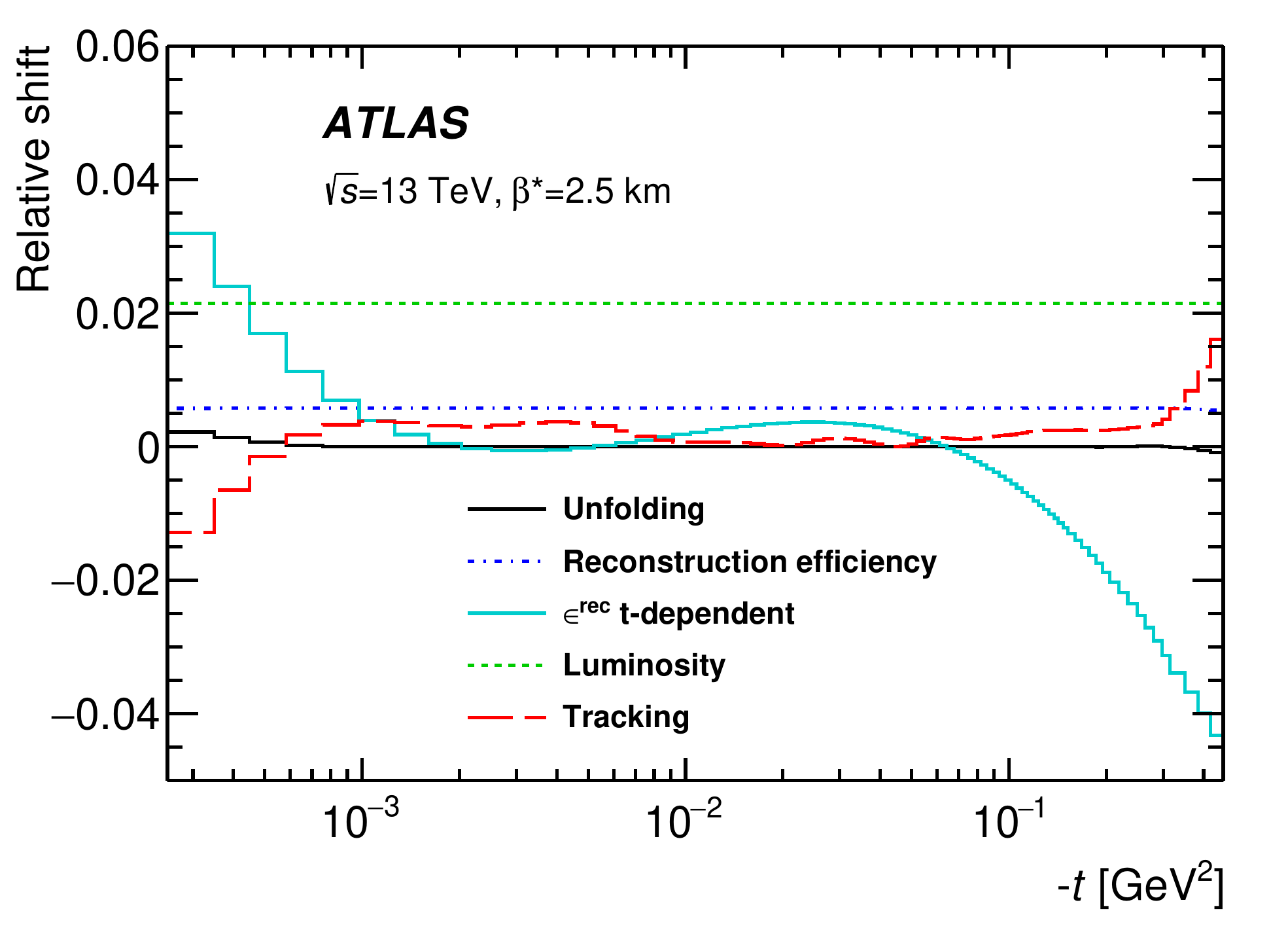}
\label{fig:tshift4}
}
\caption{Relative systematic shifts of $\mathrm{d}\sigma/\mathrm{d}t$ as a function of $t$ resulting from uncertainties in the
\protect\subref{fig:tshift3} simulation, detector resolution, background and
\protect\subref{fig:tshift4} reconstruction efficiency, luminosity and tracking. For cases where two-sided uncertainties are available the larger one was retained and assigned as a symmetric uncertainty.}
\label{fig:tshift34}
\end{figure}

Several variations of the analysis are performed under different experimental conditions in
order to assess the systematic uncertainties.
For certain sources, several variations potentially probing the same effect are considered. From these variations, the
largest deviation from the nominal cross section is retained as the systematic uncertainty.
A total of 20
sources of uncertainty are propagated to the differential elastic cross section as follows:
\begin{itemize}
\item For the alignment, separate uncertainties are calculated for different components of the alignment procedure.
The dominant uncertainty is related to the global correction to the distance parameter determined by the procedure
described in Section~\ref{sec:align}, and the correction is varied by its total uncertainty. For the vertical beam offset, the uncertainties
are assumed to be fully correlated between inner and outer stations but anti-correlated between the left and right sides, and thus
systematic shifts preserve the internal consistency of the data.
Uncertainties related to the horizontal
offset and to the rotation are propagated directly to the differential cross section,
where both the full-correlation and anti-correlation assumptions were tested. The resulting uncertainties are rather small and similar for the two correlation assumptions, and the
larger one is retained as the nominal uncertainty.
\item The nominal effective optics model with corrections to the strength of Q1, Q3, Q5 and Q6 is replaced by a simplified model in which only Q5 and Q6 are tuned (see Section~\ref{sec:Optics}). This alternative optics model yields by far the largest cross-section difference
with respect to the nominal beam optics. Several other uncertainties related to the input optics constraints, to the inclusion of Q2 and Q4 in the fit and a variation of their strength by its uncertainty, to the quadrupole alignment and to the
propagation of the intrinsic effective optics fit are considered as a stability test. These gave rise to very small changes
in the cross section which were not included in the systematic uncertainty.
\item The beam transport in the simulation nominally achieved with the transport matrix is replaced by the symplectic PTC transport module in MadX~\cite{madx}, where the quadrupole strengths of the effective optics were fed into a MadX strength file.
The difference between PTC-tracking and matrix transport also accounts for the simplification in the effective optics fit where a beamline set-up was used, as opposed to the multi-turn ring set-up nominally used in MadX for LHC simulations.
\item The nominal crossing angle for the 2.5~km campaign was zero, but the measurement of the crossing angle by the beam position monitors around
IP1 has a limited precision estimated to be $10~\upmu$rad. A residual non-compensated
crossing angle in the vertical plane with parallel-to-point focusing beam optics would cause a shift of the beam
position at ALFA. Such a shift would trivially be corrected by the alignment procedure.
In the horizontal plane, a crossing angle
would also shift the beam centre and, in addition, broaden the beam distribution. The latter effect is assessed by applying
a crossing angle in the MadX beam transport simulation and then used for acceptance and unfolding corrections in the analysis.
\item The nominal beam energy value of $p=6500~\GeV$ is varied by its uncertainty of $0.1\%$
\cite{LHC_Energy} simultaneously in the data for $t$-reconstruction according to Eq.~\eqref{eq:t-basic},
and in the simulation both at the generator level and again in the $t$-reconstruction.
\item The emittance model used to calculate the angular divergence in the simulation is varied by its uncertainty
of the order of $10\%$, and the time dependence of the emittance is replaced by a luminosity-weighted average.
\item In the simulation the nominal model for the $t$-spectrum with three slope parameters $B$, $C$, $D$
(see Section~\ref{sec:Theory}) is replaced by a model with a constant exponential slope $B$.
\item The value of the $\rho$-parameter in simulation is varied by $\pm 0.01$, corresponding to its uncertainty (see Section~\ref{sec:Tfit}.
\item The detector resolution in the simulation is varied, replacing the resolution determined from the collision data with values
3--4~$\upmu$m smaller, as predicted from the \GEANT simulation, and 4--5~$\upmu$m larger, as
measured using test-beam data. Additionally, a small $y$-dependence (10\%) of the resolution observed in the data
is parameterized and used instead of the nominal constant value.
\item The uncertainty in the background from accidental coincidences is calculated from a variation of the normalization
regions and template composition. For DPE the normalization region is also varied and the shape is changed in the
simulation, as explained in Section~\ref{sec:Background}.
\item The intrinsic unfolding uncertainty is determined from the data-driven closure test explained in Section~\ref{sec:Acceptance}.
\item The event reconstruction efficiency is varied by its uncertainty discussed in Section~\ref{sec:RecoEff}. Furthermore, a possible $t$-dependence of the reconstruction efficiency is taken into account.
\item The luminosity uncertainty of 2.15$\%$ is propagated to the cross section (see Section~\ref{sec:Luminosity}).
\item For the tracking efficiency, the minimum number of fibre layers in both $u$ and $v$ required to reconstruct a track is varied between three and four (see Section~\ref{sec:align}).
\end{itemize}
The systematic uncertainties related to the sources above are calculated using the
offset method. In this method, the nominal value of a certain parameter in the analysis chain is varied according to the
assigned uncertainty. The shift in bin $t_i$ for systematic uncertainty source $k$,
$\delta_k(t_i)=\mathrm{d}\sigma_k(t_i)/\mathrm{d}t - \mathrm{d}\sigma_{\mathrm{nominal}}(t_i)/\mathrm{d}t$
is recorded, keeping track of the sign and thereby accounting
for correlations across the $t$-spectrum.
Systematic shifts are included in the fit used to determine the physics parameters, as outlined in Section~\ref{sec:Tfit}.
The systematic $t$-dependent shifts for the differential elastic cross section are shown in Figures~\ref{fig:tshift1},
\ref{fig:tshift2}, \ref{fig:tshift3}, and~\ref{fig:tshift4} for the entire range of the $t$-spectrum.
The dominant systematic uncertainties at small $|t|$ are related to the vertical alignment, detector resolution and
possible $t$-dependent reconstruction efficiency; the second and third effects are also important at large $|t|$.
 
\subsection{Fitting procedure}\label{sec:Tfit}
The physics parameters, total cross section, $\rho$-parameter and the slope parameters are obtained from a fit of the theoretical spectrum (Eq.~\eqref{eq:tgen}),
including the CNI term, to the measured differential cross section.
Both the statistical and systematic uncertainties as well as their correlations are taken into account in the fit.
The statistical correlations are included in the covariance matrix calculated in
the unfolding procedure. The correlations of systematic uncertainties are taken into account by using a
profile minimization procedure~\cite{profile}, where nuisance parameters corresponding to all 20 systematic shifts are included,
and the $\chi^2$ is given by:
\begin{eqnarray*}\label{eq:profile_fit}
\chi^2 & = & \sum_{i,j}\left[\left(D(i)-\left(1+ \sum_{l=1}^{2}\alpha_l\right) \times T(i) - \sum_{k=1}^{18} \beta_k \times \delta_k(i)\right) \times V^{-1}(i,j)\right. \\ \nonumber
& &  \left.\times \left(D(j)-\left(1+ \sum_{l=1}^{2}\alpha_l\right) \times T(j) - \sum_{k=1}^{18} \beta_k \times \delta_k(j)\right)\right]
+\sum_{k=1}^{18}\beta_k^2 + \sum_{l=1}^{2}\frac{\alpha_l^2}{\epsilon_l^2} ,
\end{eqnarray*}
where $D(i)$ is the measured value of the elastic cross section in bin $i$, $T(i)$ the theoretical
prediction and $V(i,j)$ the statistical covariance matrix. For each systematic uncertainty that changes the shape
of the $t$-spectrum, a nuisance
parameter $\beta_k$ multiplying the corresponding shift $\delta_k$ is fitted as a free parameter, and
a penalty term $\sum_k\beta_k^2$ is added to the $\chi^2$.
Two scale parameters, $\alpha_l$, are used to describe the rescaling of the
normalization of the theoretical prediction due to the $t$-independent uncertainties in the luminosity and
the reconstruction efficiency $\epsilon_l$. The sum in quadrature of these two scale factors divided by their
uncertainties results in a second penalty term, $\sum_{l=1}^{2}\alpha_l^2/\epsilon_l^2$.
The uncertainties in the physics parameter values returned by the profile fit account for both the experimental
systematic and statistical uncertainties. In the case of the time-dependent emittance model
and the $t$-dependent reconstruction efficiency, the fit could somewhat constrain the nuisance parameters, resulting in slightly reduced systematic uncertainties.
 
\begin{figure}[htb!]
\centering
\includegraphics[width=0.7\textwidth]{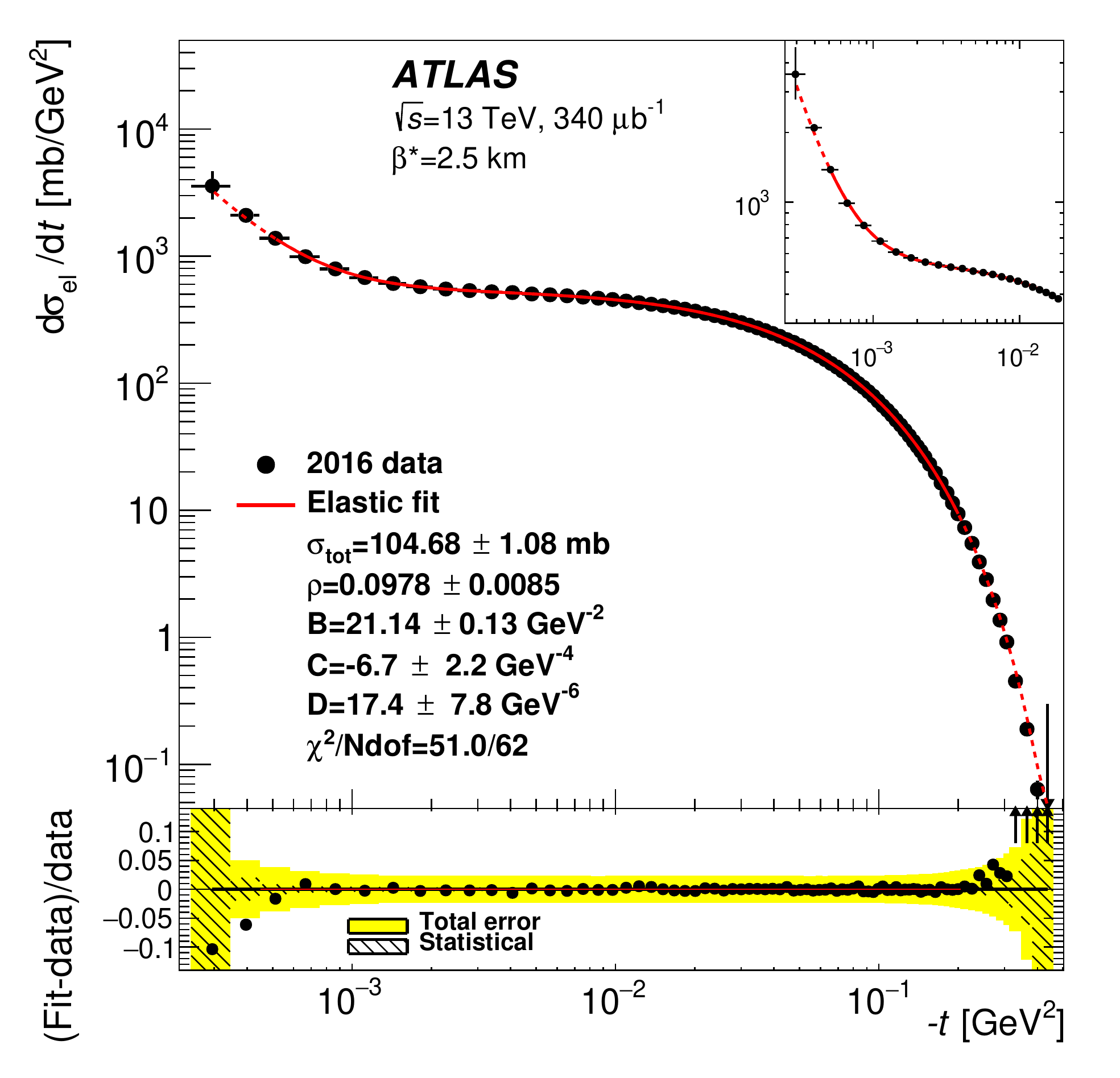}
\caption{A fit of the $BCD$-model with $\sigmatot$, $\rho$, $B$, $C$ and $D$ as free parameters to the
differential elastic cross section reconstructed with the subtraction method. The red full line represents the result
of the fit inside the fit range, the dotted red line shows the extrapolation of the fit outside, including a
special model for large $|t|$ matched to the fitted function at $-t=0.2~\GeV^2$.
In the lower panel, the points represent the deviation from unity of the ratio of the fit to the data,
the yellow area represents the total experimental uncertainty and the hatched area the statistical component. The arrows indicate points out of the scale. The inset plot shows a zoom in the small-$|t|$ region.}
\label{fig:tfit_nuisance_subtraction}
\end{figure}
The differential cross section and the fitted theoretical prediction are shown in
Figure~\ref{fig:tfit_nuisance_subtraction}.
The fit range is chosen from $t = -4.5\cdot10^{-4}~\GeV^2$ to $t = -0.205~\GeV^2$.
The lower $|t|$ value is chosen to be as close as possible
to $t=0$ to maximize the coverage of the CNI region while having an acceptance above 10\%.
The value chosen for the upper limit is motivated by theoretical considerations not to extend the fit into
the region where the shape of the differential cross section is influenced by the approach of the dip.
The main fit results are $\sigmatot=104.7 \pm 1.1$~mb and $\rho = 0.098 \pm 0.011$; further
details are summarized in Table~\ref{tab:profile_fit}.
\begin{table}[tp]
\begin{center}
\caption{Results of the profile fit to the differential elastic cross section.}
\begin{tabular}{l|ccccc}
\hline
& $\sigmatot~[\mbox{mb}]$ & $\rho$ & $ B~[\GeV^{-2}]$ & $C~[\GeV^{-4}]$ & $D~[\GeV^{-6}]$\\
\hline
Central value  & 104.68~~~~ & 0.0978 & 21.14~~ & $-6.7$~~\, & 17.4~~ \\ \hline
Statistical error    & 0.22 & 0.0043 & 0.07 & 1.1 & 3.8 \\
Experimental error    & 1.06 & 0.0073 & 0.11 & 1.9 & 6.8 \\
Theoretical error      & 0.12 & 0.0064 & 0.01 & ~~0.04 & ~~0.15 \\ \hline
Total error    & 1.09 & 0.0106 & 0.13 & 2.3 & 7.8 \\ \hline
\end{tabular}
\label{tab:profile_fit}
\end{center}
\end{table}
Here, the experimental uncertainty is obtained from the profile fit, the statistical component is calculated
from a bootstrap method~\cite{bootstrap}, and the theoretical uncertainties are discussed below.
 
\subsection{Theoretical uncertainties}\label{sec:Syst}
The theoretical systematic uncertainty is calculated by changing the assumptions of the model used
in the profile fit (see Section~\ref{sec:Theory}). The parameterization of the nuclear amplitude, nominally using
three slope parameters $B$, $C$ and $D$ in the exponential term, is reduced to two parameters $B$ and $C$ and
to a single constant slope $B$. In each case the upper end of the fit range is progressively
reduced in order to obtain a good description of the data. Further variations include changes in the
parameterization of the electric form factor of the proton and of the Coulomb phase, as well as the
inclusion of a term related to the magnetic moment
of the proton, which induces a spin dependence in the Coulomb amplitude~\cite{Bourrely_Spin}.
For the nuclear phase, several assumptions about the possible $t$-dependence were investigated.
The systematic uncertainty is calculated with a simple model of a $t$-dependence according to Eq.~\eqref{eq:simplephase}. Other nuclear phase models suggested in Ref.~\cite{GrafstromAndRyskin} were
also evaluated, but were not taken into account in the uncertainty calculation. Among these, the largest
change is observed when using a peripheral phase, but according to Ref.~\cite{GrafstromAndRyskin}
this phase model represents an extreme case with a large change increasing
with $|t|$.
The resulting theoretical uncertainties are summarized in Table~\ref{tab:model_syst}.
 
\begin{figure}[hbt!]
\centering
\includegraphics[width=0.65\textwidth]{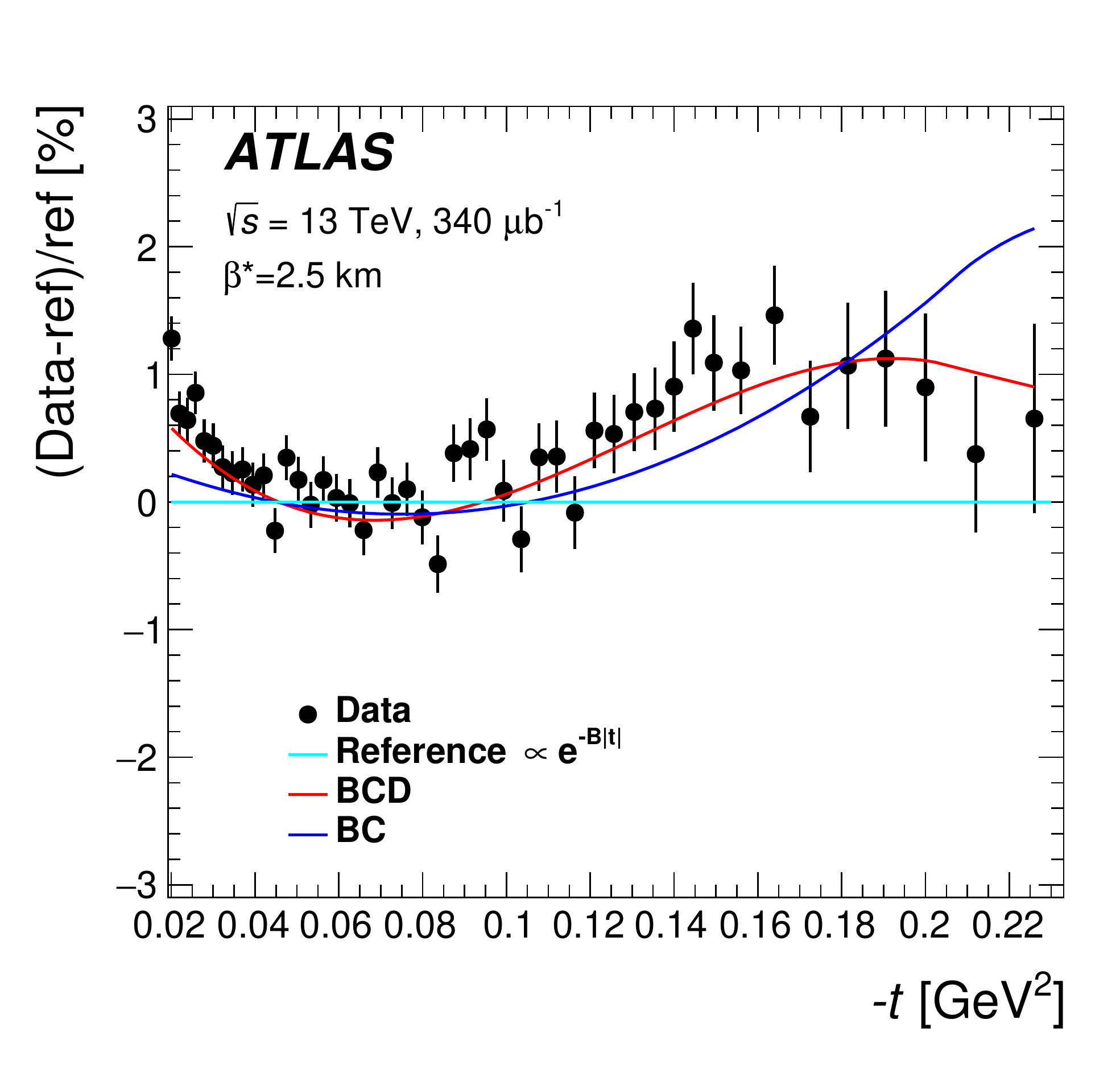}
\caption{The data normalized to a reference exponential function compared to models of the nuclear amplitude with a $t$-dependent
exponential slope. The error bars include the statistical uncertainty only.}
\label{fig:curvature_plot}
\end{figure}

\begin{table}[tp]
\caption{Systematic uncertainties originating from the choice of model fitted to the data. Details of the models for the nuclear phase can be found in Ref.~\cite{GrafstromAndRyskin}. The simple phase model is included in the systematic uncertainty calculation, and the other nuclear phase model results are given for reference.}
\label{tab:model_syst}
\begin{center}
\begin{tabular}{l|ccccc}
\hline
Variation   & $\Delta\sigmatot$ [mb] & $\Delta\rho$   & $\Delta B$ [$\GeV^{-2}$] & $\Delta C$ [$\GeV^{-4}$] & $\Delta D$ [$\GeV^{-6}$] \\ \hline
$BC$              & 0.04     & 0.0018 & -- & -- & -- \\
$B$               & 0.09     & 0.0047 & -- & -- & -- \\
Form factor       & 0.01     & 0.0022        & 0.01         & $< 0.1$~~~\, & 0.1 \\
Spin dependence   & $< 0.01$~~~\, & 0.0013   & $< 0.01$~~~\ & $< 0.1$~~~\, & $< 0.1$~~~\, \\
Coulomb phase     & 0.02     & 0.0030        & $< 0.01$~~~\ & $< 0.1$~~~\, & $< 0.1$~~~\, \\
Simple phase      & 0.05     & $< 0.0001$~~\,& 0.01         & $< 0.1$~~~\, & 0.1 \\ \hline
Total theoretical & 0.12     & 0.0064        & 0.01         & ~~0.04  & ~~0.15 \\ \hline
Peripheral phase  & 0.21     & $-0.0060$~~\, & 0.80         & 7.3     & 12.2~~ \\
Durand-Ha phase   & 0.02     & $< 0.0001$~~\,& 0.01         & $< 0.1$~~~\, & 0.2 \\
Standard phase    & 0.02     & $< 0.0001$~~\,& $< 0.01$~~~\ & $< 0.1$~~~\, & 0.1 \\
Bailly phase      & 0.01     & $< 0.0001$~~\,& $< 0.01$~~~\ & $< 0.1$~~~\, & $< 0.1$~~~\, \\
\hline
\end{tabular}
\end{center}
\end{table}
The presence of a $t$-dependent slope in the data requiring an extension of the simple exponential
model~$\sim \exp{(-B|t|})$
used in previous analyses \cite{STDM-2013-10,STDM-2015-22} with less constraining data is illustrated in
Figure~\ref{fig:curvature_plot}.
The plot shows the differential cross-section data normalized to a reference exponential fit in the range from $-t=0.04~\GeV^2$ to $-t=0.2~\GeV^2$ where
the effect is most visible. Other models fitted in the same range but with a $t$-dependent slope are normalized
to the same reference and found to describe the data better, with one additional parameter $C$ giving a satisfactory
description up to $-t=0.1~\GeV^2$ and
the full $BCD$-model up to $-t=0.2~\GeV^2$. The regime beyond $-t=0.2~\GeV^2$ is characterized by the onset of diffractive interference
phenomena leading to the dip, which requires a different type of parameterization~\cite{PhillipsAndBarger,BourrelyAndSoffer}.
 
\subsection{Stability checks}\label{sec:Stability}
\begin{figure}[htb!]
\centering
\subfloat[$\sigmatot$]{
\includegraphics[width=0.49\textwidth]{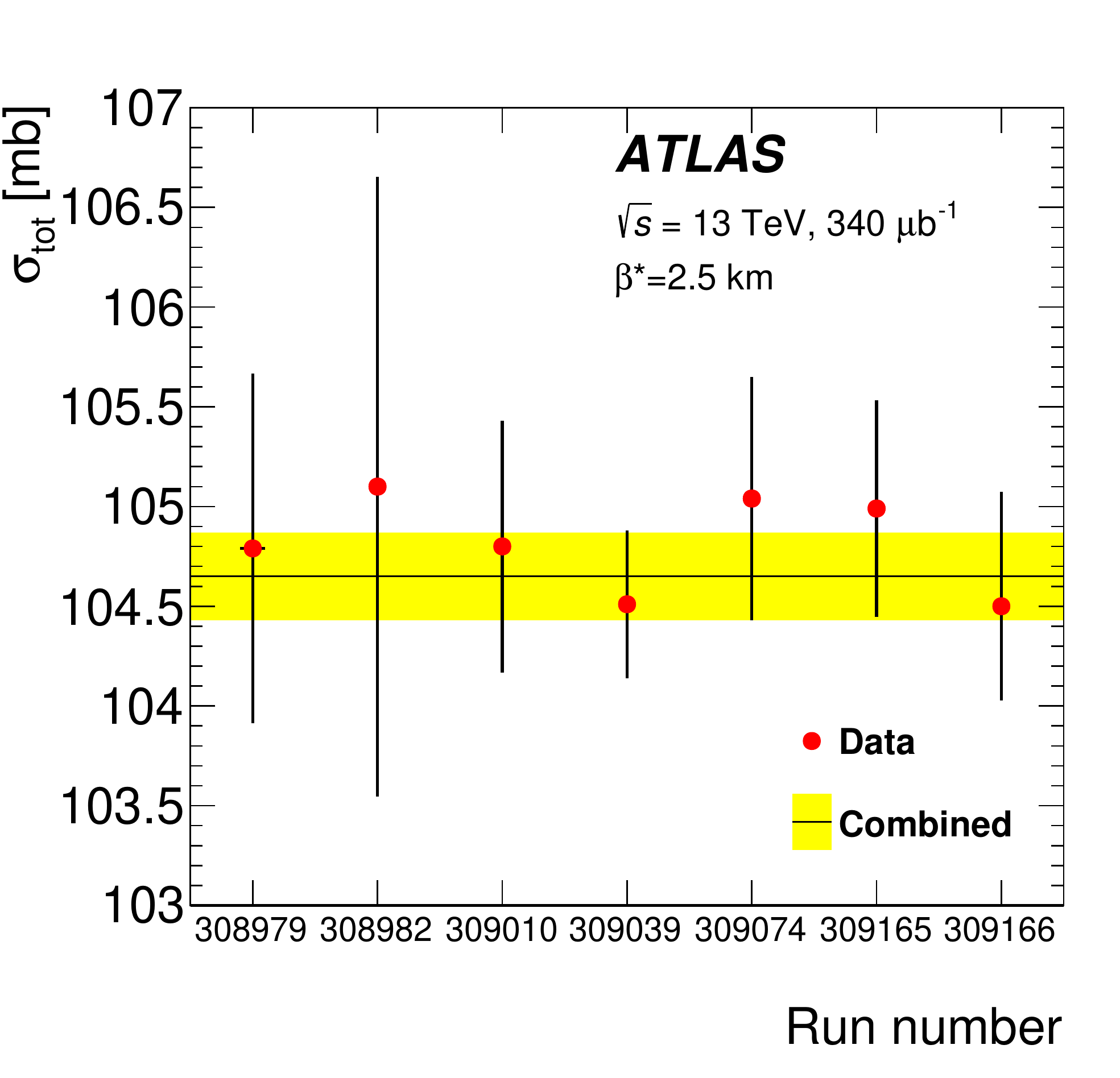}
\label{fig:run7_stot}
}
\subfloat[$\rho$]{
\includegraphics[width=0.49\textwidth]{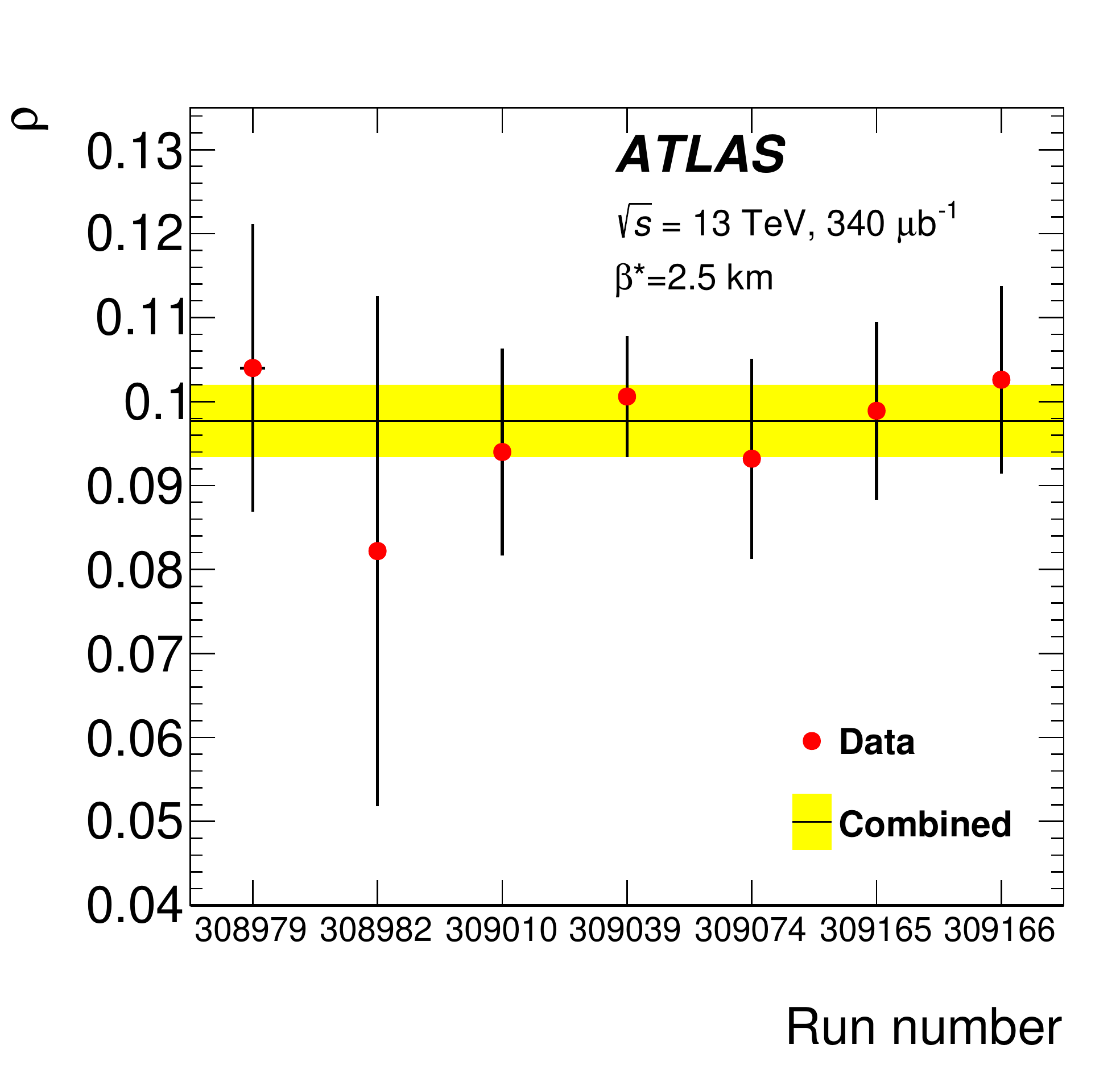}
\label{fig:run7_rho}
}
\caption{Dependence of the fitted values of \protect\subref{fig:run7_stot} $\sigmatot$ and
\protect\subref{fig:run7_rho} $\rho$ on the run number. The error bars show statistical uncertainties only.}
\label{fig:run7}
\end{figure}

Several cross-checks of the analysis are performed in order to verify the stability of the results.
A test for possible time dependence is performed by repeating the analysis separately for each run in the campaign. The
results shown in Figure~\ref{fig:run7} demonstrate that the physics parameters determined per run are  compatible
with one another and with results from the time-integrated combined data set.
A very basic test probing the sensitivity to different fitting options consists of a simple fit that uses only the statistical covariance matrix
in the definition of $\chi^2$.
\begin{table}[tp]
\begin{center}
\caption{Results of a simple fit using only statistical uncertainties.}
\label{tab:cov_fit}
\begin{tabular}{l|cccccc}
\hline
& $\sigmatot~[\mbox{mb}]$ & $\rho$ & $ B~[\GeV^{-2}]$ & $C~[\GeV^{-4}]$ & $D~[\GeV^{-6}]$  & $\chi^2/N_{\textrm{dof}}$\\
\hline
Central value  & 104.73~~ & 0.0926 & 21.16 & $-7.6$~~\, & 20.9~~ & 55.5/62\\
Uncertainty    & ~~0.07 & 0.0035 & ~~0.08 & 1.1 & 4.3 \\ \hline
\end{tabular}
\end{center}
\end{table}
The result given in Table~\ref{tab:cov_fit} shows good agreement with the profile fit. There are small shifts of the central
values which originate from different
weights of the bins inside the fit range when systematic uncertainties are omitted, the latter having a different $t$-shape than statistical
uncertainties.
 
Independent measurements are performed in each arm of ALFA,
and are combined before fitting the physics parameters.
The fits can, however, also be done
per arm for the subsample collected in that arm, in which case the acceptance is reduced roughly by a factor of two.
In order to test the consistency of the measurements in the two arms, the same standard fit range used in the nominal
combined analysis is applied, although the acceptance per arm is below 10$\%$ in the first bin. This test is carried
out with a fit including only statistical uncertainties. The difference between
the arms is found to be consistent with the arm-dependent vertical offset and reconstruction efficiency uncertainties.
 
As outlined in Section~\ref{sec:measure}, the $t$-value can be reconstructed with four different methods, with the subtraction
method being the nominal method. Different methods use different combinations of the transport matrix in the reconstruction.
Table~\ref{tab:profile_methods} reveals good
consistency among the methods, with the small differences being within the beam optics uncertainty.
\begin{table}[tp]
\begin{center}
\caption{Results of the profile fit to the differential elastic cross section
measured with different $t$-reconstruction methods. The quoted error includes both the statistical and
experimental systematic uncertainties, but not the theoretical uncertainty.
}
\label{tab:profile_methods}
\begin{tabular}{l|cccccc}
\hline
Method  & $\sigmatot~[\mbox{mb}]$ & $\rho$ & $ B~[\GeV^{-2}]$ & $C~[\GeV^{-4}]$ & $D~[\GeV^{-6}]$ & $\chi^2/N_{\textrm{dof}}$\\
\hline
Subtraction & $104.68 \pm 1.08$ & $0.0978\pm0.0085$ & $21.14\pm0.13$ & $-6.7 \pm 2.2$ & $17.4 \pm 7.8$ & 51/62 \\
Local angle & $104.50 \pm 1.05$ & $0.1029\pm0.0081$ & $21.16\pm0.14$ & $-6.9 \pm 2.3$ & $19.3 \pm 8.0$ & 82/62\\
Lattice & $104.87 \pm 1.03$ & $0.0956\pm0.0085$ & $21.02 \pm 0.13$ & $-4.9\pm 2.2$ & $12.5 \pm 7.5$ & 83.1/62\\
Local subtraction & $104.52 \pm 1.04$ & $0.0910\pm0.0091$ & $20.99\pm0.15$ & $-5.1 \pm 2.4$ & $12.3 \pm 8.1$ & 76.6/62  \\ \hline
\end{tabular}
\end{center}
\end{table}
 
Finally the dependence of the fit results on the choice of fit range was investigated. Both the upper and lower
ends of the fit range were varied. The largest change was found when the lower end of the fit range was moved to
$-t_{\textrm{min}}=1.26\cdot10^{-3}~\GeV^2$ where a resulting shift of 0.2~mb for $\sigmatot$ and of 0.0017 for $\rho$ was observed with
negligible changes in the slope parameters. Moving the upper end of the fit range to
$-t_{\textrm{max}}=0.15~\GeV^2$ had little impact on $\sigmatot$ and $\rho$, but changed the slope
parameter $C$ by $1.2~\GeV^{-4}$ and the slope parameter $D$ by $4.7~\GeV^{-6}$. These changes are significantly smaller than the quoted
experimental uncertainties.
 
\subsection{Elastic and inelastic cross sections, and $\sigmael/\sigmatot$}\label{sec:Phys_elastic}
The total elastic cross section measured inside the fiducial volume is obtained by
integrating the observed differential elastic cross section
in the available $t$ range from $-t=2.5 \cdot 10^{-4}~\GeV^2$ to $-t=0.46~\GeV^2$. The result is
\begin{displaymath}
\sigmael^{\textrm{obs}} = 27.80~\pm~0.09~\mbox{(stat.)}~\pm~0.64~\mbox{(syst.)}~\mbox{mb} .
\end{displaymath}
This measurement accounts for both the strong interaction and electromagnetic interaction in
the CNI region.
For the extrapolation to the full phase space, a common approach is to use only the nuclear part of the fit, which
in the past typically consisted of only a single exponential term $\sim\exp(-B|t|)$, and to integrate from zero to infinity. The Coulomb interaction
needs to be disregarded in this approach because only the hadronic part of elastic scattering is to be integrated.
In this analysis, the nuclear amplitude parameterization is extended to include the curvature parameters $C$ and $D$ and the integration is carried out numerically from
zero to $1000~\GeV^2$, neglecting the part of the cross section at higher $|t|$. The result is
\begin{displaymath}
\sigmael^{\textrm{extr}} = 27.27 \pm 1.10 ~\mbox{(exp.)} \pm 0.30 ~\mbox{(th.)} ~\mbox{mb} ,
\end{displaymath}
where the first uncertainty is propagated from the profile fit and thus includes experimental systematic and statistical uncertainties
and the second uncertainty accounts for the model dependence.
The same integration can also be restricted to the measurement range and reveals that a contribution
of about 0.75~mb can be attributed to electromagnetic interactions inside the fiducial volume.
The fact that the observed cross section in the fiducial volume is larger than the cross section extrapolated
to full phase space is related to the large contribution from Coulomb scattering in the small-$|t|$ region of the fiducial volume.
 
The total integrated (hadronic) inelastic cross section is obtained by subtracting the extrapolated elastic cross section from the total cross section. The result is
\begin{displaymath}
\sigmainel = 77.41 \pm 1.07 ~\mbox{(exp.)} \pm 0.18 ~\mbox{(th.)} ~\mbox{mb} ,
\end{displaymath}
where again the uncertainty, taking into account correlations, is propagated from the profile fit to the total and extrapolated elastic cross sections
in the first error, and the second error represents the theoretical uncertainties. This result
from ALFA can be compared with the ATLAS measurement of the inelastic cross section using the MBTS~\cite{ATLAS_inel13}, which is
$\sigmainel = 78.1 \pm 2.8$ mb, where the uncertainty is dominated by the extrapolation to full phase space
covering the diffractive region with protons of momentum loss $\Delta p / p < 5 \cdot 10^{-6}$. The two measurements are compatible within the uncertainties.
 
Another derived quantity is the ratio of the elastic cross section to the total cross section,
\begin{displaymath}
\frac{\sigmael}{\sigmatot} = 0.257 \pm 0.008 ~\mbox{(exp.)} \pm 0.009 ~\mbox{(th.)} .
\end{displaymath}
The normalization uncertainty between $\sigmael$ and $\sigmatot$ is correlated and partly cancels out in the ratio.
The value of this ratio should asymptotically approach $1/2$ in the black-disk limit (see Section~\ref{sec:el_tot}),
a value that the present measurement is far below. A study
of the ratio's energy evolution may reveal additional aspects of strong interactions in low-momentum transfer reactions.
 
\subsection{Absolute luminosity calibration}\label{sec:Lumical}
The initial goal for ALFA was to provide an absolute calibration of the
luminosity~\cite{ATLAS-TDR-18}.
The method requires covering the CNI region of the $t$-spectrum to gain sensitivity to the Coulomb interaction,
which is independent of $\sigmatot$. In the nuclear regime the normalization is fully correlated with the
total cross section and an independent measurement of luminosity and $\sigmatot$ is impossible.
The good CNI-coverage of the present data set allows an evaluation of this method, as demonstrated recently in Ref.~\cite{TOTEM_2p5km}.
Because the differential elastic cross section normalized by the luminosity from LUCID was used to
constrain the alignment, this absolute calibration test can only be used to assess the achievable precision.
 
Different fitting strategies were considered for luminosity calibration. In the actual implementation, one
more parameter $x_L$ is introduced as a multiplicative factor to the nominal luminosity and is fitted together with
the five physics parameters over the nominal fit range. For this study a fit including only statistical uncertainties is
used and the result for $x_L$ is
\begin{displaymath}
x_L=1.008 \pm 0.020 ~\mbox{(stat.)}\; \pm 0.059 ~\mbox{(syst.)} ,
\end{displaymath}
where the first uncertainty is purely statistical and the second accounts for the experimental systematic uncertainties. For the other physics parameters, values compatible with the nominal values are found. The experimental systematic uncertainties are obtained by repeating the fit under the same variations of the analysis that were taken
into account for the nominal profile fit, except for the luminosity uncertainty. The systematic uncertainty of $x_L$ is
dominated by the distance uncertainty (0.038) and the possible $t$-dependence of the reconstruction efficiency
(0.039), whereas all other uncertainties are much smaller and when added in quadrature amount to 0.027.
The dependence on the fit range was also investigated and a variation of its lower end induces a change of 0.035 in $x_L$ and
the upper-end variation induces a change of 0.014. These changes are well covered by the quoted uncertainty.
 
Another fit option consists of a separation of the fits in two steps. In the first step, the slope parameters are fitted at
large $|t|$ between $-t=0.01~\GeV^2$ and $-t=0.2~\GeV^2$ with nominal luminosity in a region where the data have a negligible sensitivity
to $\rho$. The actual value of $\rho$ used in the fit was verified to have no impact on the slope parameters $B$, $C$ and $D$.
In the second step, the values of the parameters $B$, $C$ and $D$ determined in the first step are fixed and a fit is performed in the complementary small-$|t|$ region from $-t=4.5 \cdot 10^{-4}~\GeV^2$ to $-t=0.01~\GeV^2$ with $\sigmatot$, $\rho$ and $x_L$ as free parameters.
The result from this fit set-up is
\begin{displaymath}
x_L=1.017 \pm 0.021 ~\mbox{(stat.)} \pm 0.061 ~\mbox{(syst.)} .
\end{displaymath}
The results are thus rather similar to those from the six-parameter simultaneous fit, with marginally larger uncertainties.
Also, the dependence on the fit range is similar, with the exception of the variation of the lower end of the fit range,
which induces a slightly larger change in $x_L$.
The alternative absolute luminosity calibration with ALFA revealed the expected consistency
with the nominal luminosity, presented in Section~\ref{sec:Luminosity}, based on the LUCID detector with vdM calibration,
but the $2.15\%$ precision of the latter is far better.

% End of text imported from the .//Results.tex input file

% The next lines are included from the .//Interpretation.tex input file
\section{Interpretation}\label{sec:Interpretation}
\subsection{Energy evolution of $\sigmatot$ and $\rho$}
Elastic scattering is a low-$\pT$ process, and a perturbative expansion cannot be applied.
Therefore, $\sigmatot$ and the $\rho $-parameter cannot be calculated from first principles in QCD.
Nonetheless, the fundamental principles of analyticity of the elastic-scattering amplitude, crossing symmetry,
and unitarity must be respected, which are captured by dispersion relations.
Dispersion relations connect the energy evolution of the $\rho $-parameter with the
energy evolution of $\sigmatot$.
 
The COMPETE Collaboration~\cite{compete} provides global fits to elastic-scattering measurements in $pp$, $\ppbar$
and other beam-particle collisons from low energies of around $5~\GeV$ to LHC energies and above.
The $\pp$ case at high energies is of interest here. At high energy, they use a crossing-even
amplitude ($C^{+}$), which implies that the $\ppbar$ and $\pp$ cross sections converge asymptotically.
This amplitude is based upon the Pomeron concept and a double and triple pole is assumed, leading to an
energy evolution of $\sigmatot$ combining $\ln{s}$ and $\ln^{2}{s}$ terms. COMPETE does not use a crossing-odd
amplitude ($C^{-}$) corresponding to the so-called Odderon. Recently, the $t$-spectrum measured in $\pp$ collisions at
the LHC by TOTEM was extrapolated to the Tevatron energy and compared with the direct measurement
of the $\ppbar$ spectrum by D0 to claim evidence for the Odderon~\cite{TOTEM_D0}.
 
\begin{figure}[h!]
\centering
\includegraphics[width=0.8\textwidth]{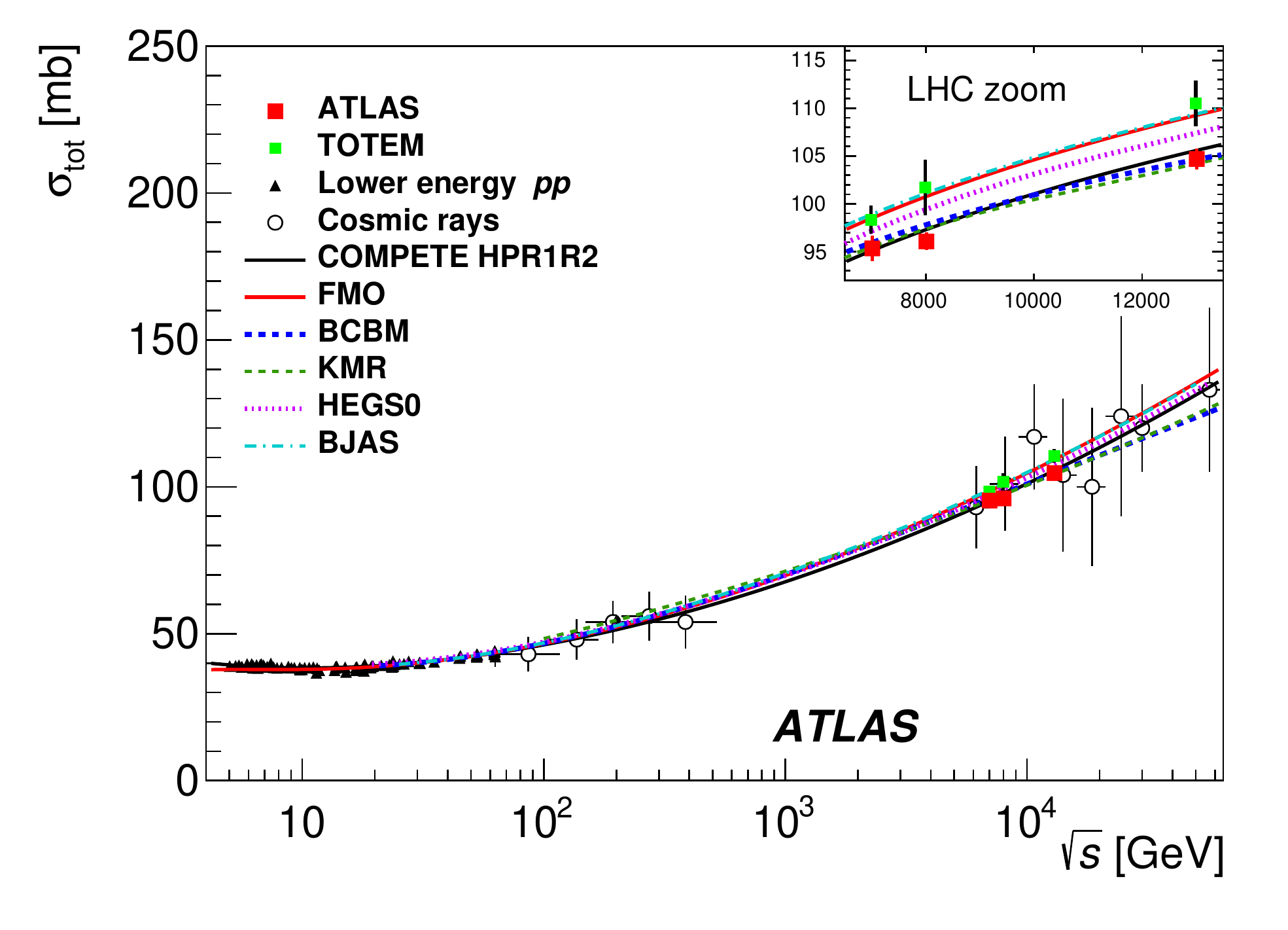}
\caption{Energy evolution of the total cross section compared to different model predictions.}
\label{fig:running_models_stot}
\end{figure}

\begin{figure}[h!]
\centering
\includegraphics[width=0.8\textwidth]{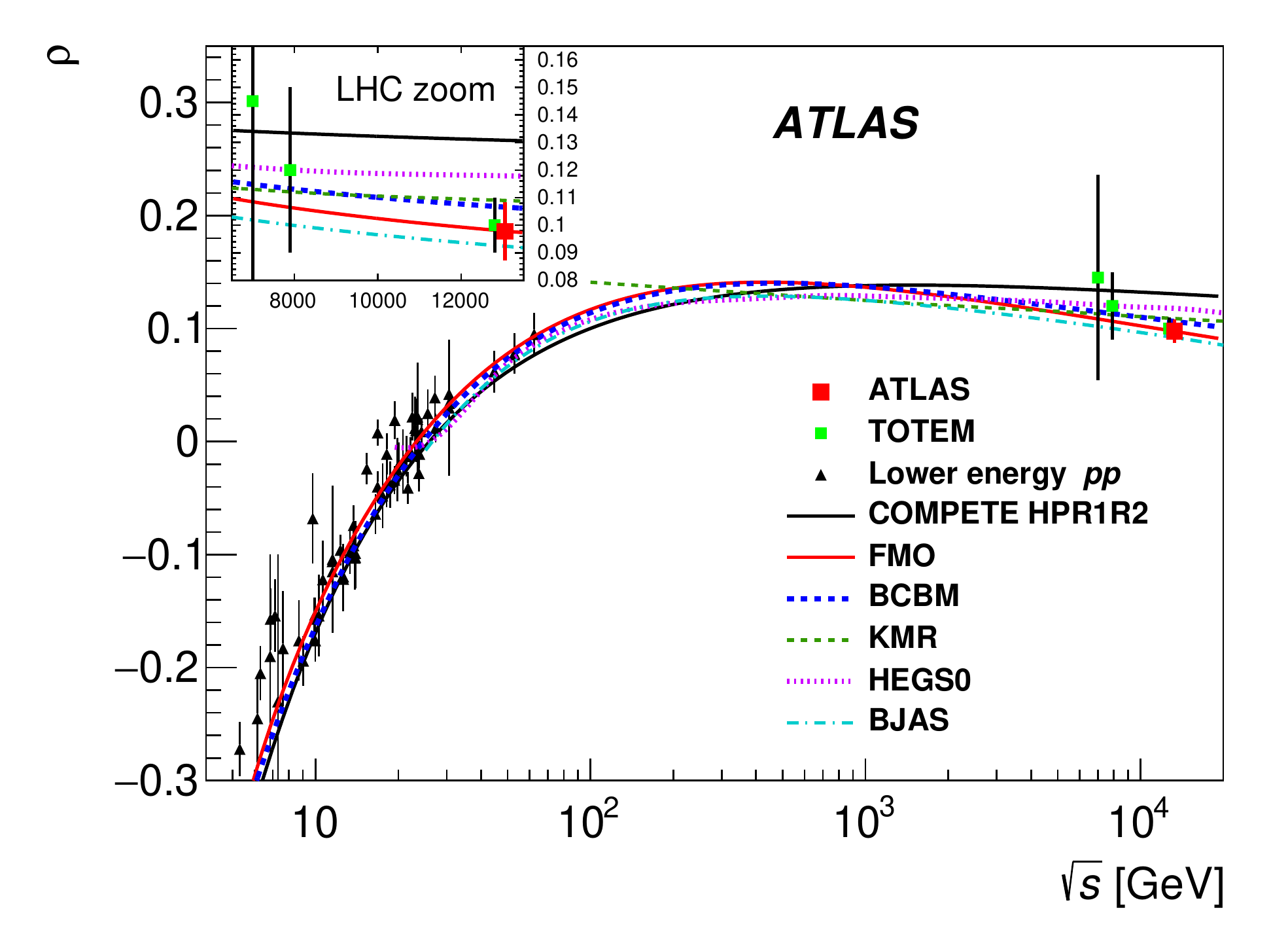}
\caption{Measurements of the $\rho$-parameter at different centre-of-mass energies compared to different model predictions.}
\label{fig:running_models_rho}
\end{figure}

Figures~\ref{fig:running_models_stot} and~\ref{fig:running_models_rho} show the world data
for $\sigmatot$ and $\rho $
together with the new result from ALFA. Also shown in these figures are the predictions
of the COMPETE model (black lines), taken from the PDG~\cite{PDG_2020}, and of other models discussed below.
The COMPETE prediction corresponds to their latest global fit for $\sigmatot$ and $\rho$, which includes
all data available up to $7~\TeV$ and also the TOTEM point at $8~\TeV$.
 
The new ALFA measurement of $\rho$ is compatible within uncertainties with the recent TOTEM measurement~\cite{TOTEM_2p5km},
but the TOTEM value of the total cross section is about 5.8~mb higher than the ALFA measurement.
The difference between ALFA and TOTEM corresponds to a discrepancy of about $2.2\sigma$ assuming uncorrelated
uncertainties. This trend was already observed at 7 and $8~\TeV$ and is confirmed at $13~\TeV$.
The difference between the ALFA and TOTEM measurements of the differential elastic cross
section is essentially confined to the normalization, whereas the shapes are mostly in good agreement.
 
ATLAS and TOTEM use quite different methods to obtain the absolute normalization.
ATLAS uses the method outlined in Section~\ref{sec:Luminosity} for a precise luminosity measurement,
while TOTEM uses the so-called luminosity-independent method. The luminosity-independent
method requires a simultaneous measurement of the total inelastic rate including a MC estimation
of the non-measurable contribution from low-mass diffraction, which for TOTEM covers masses below $4.6~\GeV$.
 
The COMPETE prediction of $\sigmatot$ is in good agreement with the ALFA data at $13~\TeV$ but it exceeds
the $\rho$ measurements of ALFA and TOTEM by more than $3 \sigma$.
One possible explanation for a lower $\rho $-value could be that $\sigmatot$ asymptotically
grows slightly slower than the $\ln^{2}{s}$ evolution assumed by COMPETE.
In the 1980s, Block and Cahn~\cite{Block_and_Cahn_damping} and
Bourrely and Martin (BCBM)~\cite{Bourrely_and_Martin} suggested the possibility of a
damped $\ln^{2}{s}$ amplitude which gives an energy dependence of the form $\ln^{2}{s}/(1+\alpha \ln^{2}{s})$,
where $\alpha$ is the damping factor.
The proposed amplitude modifies the high-energy behaviour of $\rho$ and $\sigmatot$ and the latter
will asymptotically approach a constant value.
Our global fit to the latest ALFA and TOTEM data together with lower-energy $pp$ data using this amplitude and
applying dispersion relations is shown by the blue dashed lines in Figures~\ref{fig:running_models_stot} and~\ref{fig:running_models_rho}.
A fair description of the ALFA data is found for a damping factor $\alpha=0.0014$.
The damped-amplitude model was suggested in the context of putting limits on predictions of
$\sigmatot$ at energies beyond the ISR and the amplitude was not derived from an underlying physics model.
The damping factor was thus regarded as a measure of the energy scale where a possible deviation from
$\ln^{2}{s}$ evolution might set in. Later, there were attempts to explain a slower rise of $\sigmatot$ in relation
to Color Glass Condensate models~\cite{GotsmanCGC}.
 
An alternative way to generate a low $\rho$-value is to assume the existence of a
crossing-odd amplitude. This would imply that differences between anti-particles and particles
would prevail asymptotically. Such an amplitude generates a lower $\rho$-value for $\pp$ scattering
and a higher value for $\ppbar$ scattering relative to the crossing-even amplitude.
This type of amplitude was first proposed in 1973, e.g.\ in Ref.~\cite{Odderon1973}. Much later,
the $C^{-}$ amplitude was associated with a three-gluon state in QCD, in contrast to the two-gluon state of the Pomeron.
 
The original proposal from 1973 was extreme in the sense that
the strong interaction was assumed to be as strong as allowed by axiomatic field theory.
This implies that not only the imaginary part of the
scattering amplitude increases as $\ln^{2}{s}$ but also the real part.
The corresponding model was named
FMO~\cite{FMO}  (the Froissaron Maximal Odderon) and is displayed in
Figures~\ref{fig:running_models_stot} and~\ref{fig:running_models_rho} (red lines).
This model was tuned to the TOTEM data, which are well described,
whereas the ALFA cross-section data at 7 and $8~\TeV$ were discarded from the model
tuning and are thus not well described.
The FMO model generates a shift $\Delta\rho\backsim-0.04$ at $13~\TeV$, whereas in Ref.~\cite{RyskinOdderon}
the Odderon contribution was estimated to be $O(1~\mbox{mb})$ and thus \(\Delta\rho\backsim-0.01\).
 
Figures~\ref{fig:running_models_stot} and~\ref{fig:running_models_rho} show the result of the model
constructed by the KMR group~\cite{KMR_inelastic} (green dashed lines). This is a two-channel eikonal model with few parameters and
it uses all available  high-energy data for $\rho$ and $\sigmatot$, as well as the
corresponding differential elastic cross sections, and also all available measurements
of low-mass diffraction.
The latest TOTEM data at $13~\TeV$ are also included as a constraint.
For the calculation shown, they have only used a $C^{+}$ amplitude and yet they get a reasonable
description of \(\rho\).
They also estimate the contribution of the Odderon (not included in the figure) and find its contribution
is only $\Delta\rho\backsim -0.005$.
 
Figures~\ref{fig:running_models_stot} and~\ref{fig:running_models_rho} also show the results of two other models.
The HEGS (High Energy General Structure) model~\cite{HEGS0} (purple dotted lines) is based on reggenized gluon exchange and
generalized parton distributions. Their estimate of the Odderon contribution is small~\cite{HEGS_Odderon}.
The description of \(\rho\) overshoots the experimental values of TOTEM and ALFA, but not as much as the
COMPETE prediction.
In their fit to the data, they have included all data up to $7~\TeV$ and the point at $8~\TeV$ from TOTEM.
 
The BJAS model (from the initials of the authors of Ref.~\cite{Broniowski} and shown by the light blue dotted-dashed lines) uses both
a $ C^{+}$ amplitude and a  $ C^{-}$ amplitude. The $ C^{-}$ amplitude resembles the Maximal Odderon and thus describes the \(\rho\) data points very well at $13~\TeV$.
This model has a distinct feature that the inelastic profile as a function of the impact parameter $b$
has a small dip at $b=0$, which can be interpreted as \textit{hollowness}.
All data available up to $13~\TeV$ were used
to determine the parameters of this model.
 
To better quantify the level of agreement between the models and the data, a $\chi^2$-comparison is presented
in Table~\ref{tab:model_chi},
using either all available data from $\pp$ collisions or the partial contributions from ALFA and TOTEM.
COMPETE provides a good description of the data up to $8~\TeV$, but it fails to describe the 13 TeV $\rho$ data.
The FMO model is overall in poorest
agreement with the $\pp$ data and disagrees
with the ALFA data, but it yields the best description of the TOTEM data, to which it was tuned.
The BCBM model achieves the best overall description and is also in good agreement with the ALFA data but,
given the discrepancy in the total cross-section data, is not in good agreement with TOTEM.
For the other models,
only discrete values are provided, and thus the comparison is restricted to the LHC range.
 
\begin{table}[tp]
\caption{Fit quality of predictions for $\sigmatot$ and $\rho$ considering the entire data set of $\pp$ collisions and the
individual contributions from ALFA and TOTEM data on $\sigmatot$ and $\rho$ to the $\chi^2$.
The last column indicates the LHC data included in the tuning of the models (A for ALFA, T for TOTEM, with the energy in TeV).}
\label{tab:model_chi}
\begin{center}
\begin{tabular}{l|ccc|l}
\hline
Model & Global $\chi^2/N_\textrm{dof}$ & ALFA & TOTEM  & LHC data included \\
& & partial $\chi^2/N_\textrm{dof}$ & partial $\chi^2/N_\textrm{dof}$ & in model tuning \\ \hline
COMPETE HPR1R2 & 1.42 & 3.00  & 3.50 & A 7; T 7, 8  \\
FMO            & 1.61 & 9.50 & 0.13   & T 7, 8, 13 \\
BCBM           & 1.03 & 0.81 & 2.04 & all \\ \hline
KMR            &      & 0.85 & 2.29 & A 7, 8; T 7, 8, 13 \\
HEGS           &      & 8.10  & 0.83 & A 7; T 7, 8 \\
BJAS           &      & 11.90~~ & 0.29 & A 7; T 7, 8, 13 \\ \hline
\end{tabular}
\end{center}
\end{table}
The conclusion drawn from these model comparisons is that it is difficult to separate possible effects from a
flatter energy dependence of $\sigmatot$ and contributions from a possible $C^{-}$ contribution
in the form of an Odderon.
The fact that the size of the Odderon contribution is subject to large uncertainties
complicates the matter further.
 
\subsection{Evolution of the ratio $\sigmael/\sigmatot$}\label{sec:el_tot}
\begin{figure}[h!]
\centering
\includegraphics[width=0.8\textwidth]{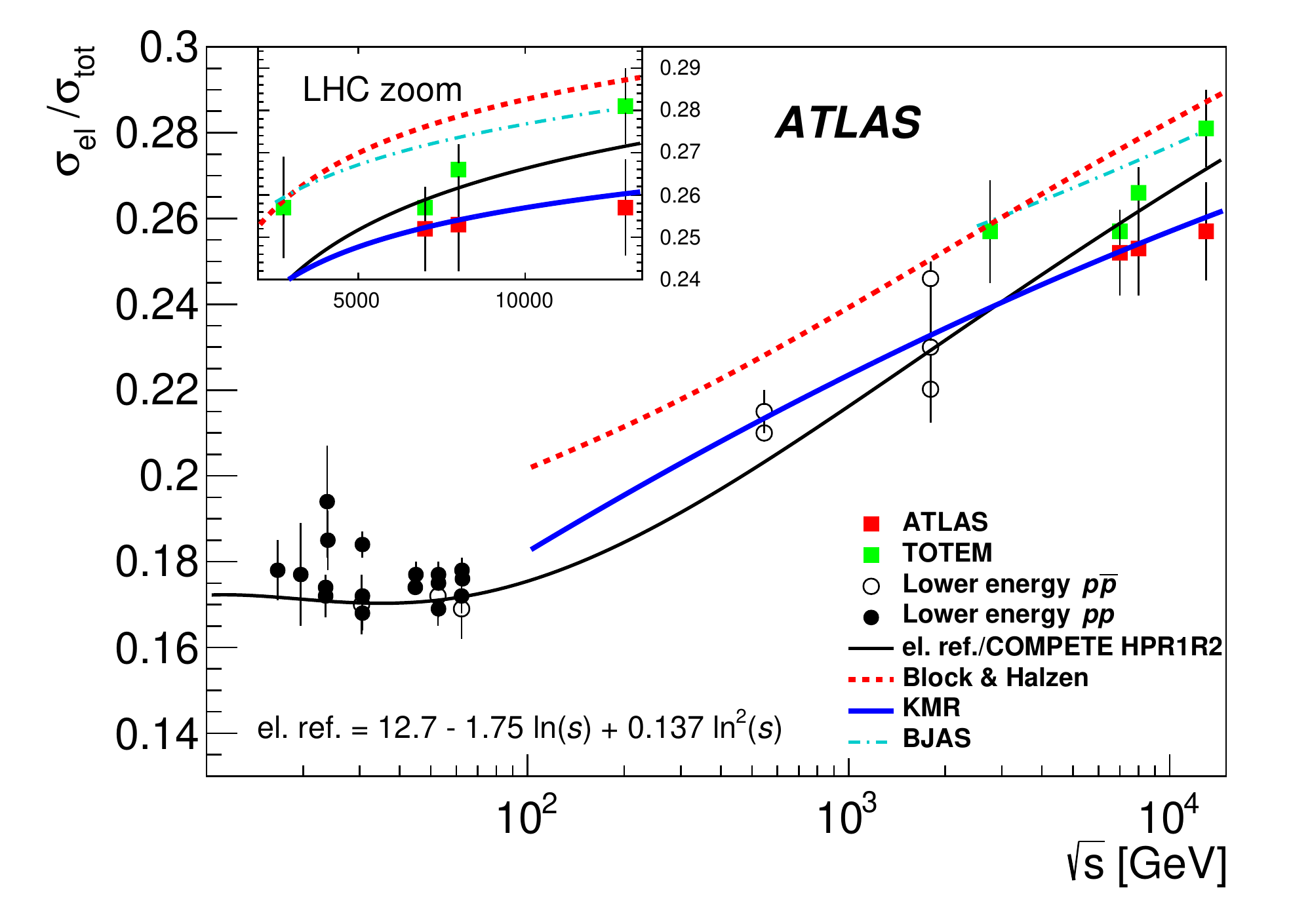}
\caption{Measurements of the ratio $\sigmael/\sigmatot$ at different centre-of-mass energies compared to model  predictions and for illustrative purposes a conventional parameterization of the elastic cross section divided by the COMPETE prediction of $\sigmatot$.}
\label{fig:selstot}
\end{figure}

At the highest energies hadron--hadron scattering cross sections may asymptotically approach the
geometrical size of the
hadrons, a configuration which is referred to as the \textit{black-disc limit}.
In this case, the elastic and inelastic
cross sections become equal, and thus the ratio $\sigmael/\sigmatot$ is  $1/2$.
At present energies, this ratio is
closer to $1/4$ but an increase from the ISR to LHC energies is clearly
observed.
The energy evolution of $\sigmael/\sigmatot$ is investigated in Figure~\ref{fig:selstot}, where the new
measurement at 13~\TeV\ is compared with the values from TOTEM and lower energies.
In order to guide the eye, the data are
compared with a simple parameterization of $\sigmael$~\cite{STDM-2013-10,STDM-2015-22} normalized to the COMPETE prediction of $\sigmatot$.
A set of consistent calculations of the ratio using $\sigmael$ and $\sigmatot$ calculated
in the same framework are also shown in Figure~\ref{fig:selstot}.
The prediction from Block and Halzen~\cite{BlockHalzen} was derived before the LHC data was available and
clearly overshoots the data as well as the BJAS prediction. The Block and Halzen model is only valid for data
above $100~\GeV$. This is also true for the KMR model~\cite{KMR_inelastic}. However, this model is in good agreement with the ATLAS data and also lower-energy data in the range of applicability.
 
\subsection{Inelastic cross section}
\begin{figure}[h!]
\centering
\includegraphics[width=0.8\textwidth]{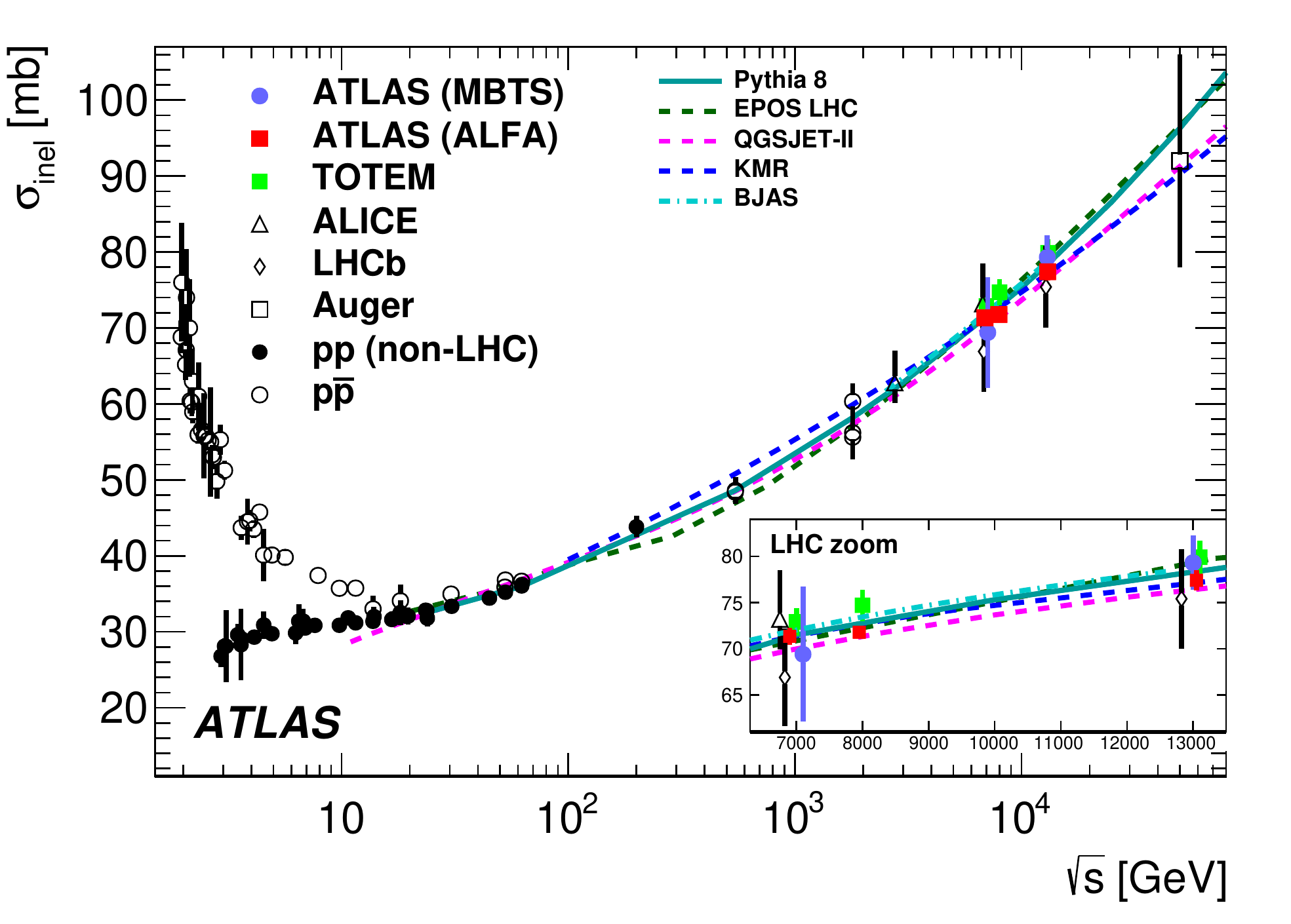}
\caption{Comparison of inelastic cross-section measurements presented here with other
published measurements and model predictions as a function of the centre-of-mass energy.}
\label{fig:inelastic}
\end{figure}

The measurement of the inelastic cross section determined here is compared with the ALFA measurements
in Run~1, direct measurements from ATLAS~\cite{ATLAS_inel,ATLAS_inel13} and other experiments~\cite{PDG_2020,TOTEM_lumindep,TOTEM_8TeV,ALICE_inel,LHCb_inel,Auger,STAR},
together with generator predictions~\cite{QGSJETII,PYTHIA,EPOS-LHC} and  model calculations by KMR and BJAS in Figure~\ref{fig:inelastic}.
The most precise measurement in the LHC energy range are the ALFA measurement followed by
the TOTEM  measurements. The other measurements, not using Roman Pots, have limited precision due
to the fact that the
full inelastic phase space is not covered and they have to rely upon theoretical models to extrapolate
beyond the fiducial volume.
 
\subsection{Nuclear slope $B$}
In the context of Regge field theory, the energy evolution of the nuclear slope at small $|t|$ is related
to the slope of the Pomeron Regge trajectory $\alpha^\prime$. Taking $B$ as the leading slope term at
small $|t|$, its evolution with $s$ is expected to be
\begin{displaymath}
B=B_0+2\cdot\alpha^\prime \ln\left(\frac{s}{s_0}\right) .
\end{displaymath}
The increase of the $B$ slope corresponds to a reduction of the emission cone of elastically
scattered particles,
a phenomenon known as the \textit{shrinkage of the forward cone}.
It was verified that this linear relation is quite accurate at lower energy, where the scattering process is dominated by a single Pomeron
exchange but the energy is still high enough for secondary Regge trajectories to be unimportant.
However, at higher energies, in a situation with multi-Pomeron exchange, the increase of $B$ is expected
to be faster than linear and an
additional term in $\ln^2\left(s/s_0\right)$ may be needed to describe the evolution.
This may be the case at the LHC, as pointed out in Ref.~\cite{Scheg}.
 
In the present analysis the $t$-dependence of the nuclear slope is taken into account, but in order to
compare it with previous measurements at lower energy the leading slope parameter
$B$ is determined in a dedicated
fit at small $|t|$, where the slope is approximately constant, without the curvature parameters $C$ and $D$.
 
\begin{figure}[hb!]
\centering
\includegraphics[width=0.8\textwidth]{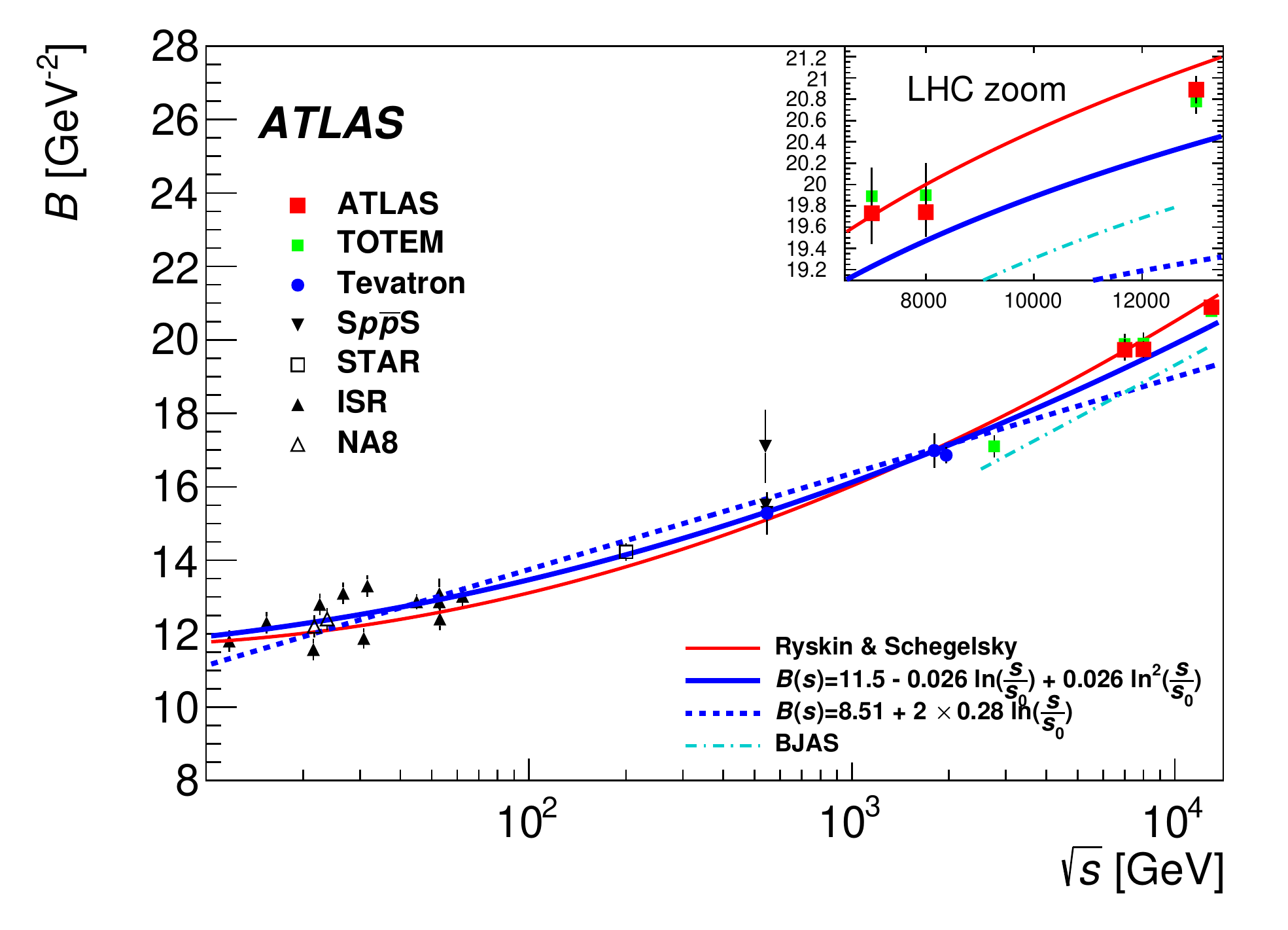}
\caption{Measurements of the nuclear slope parameter $B$ determined at small $|t|$ compared to different predictions of
its energy evolution.}
\label{fig:BslopeS}
\end{figure}

In Figure~\ref{fig:BslopeS} the LHC data are compared with lower-energy data
and with the model in Ref.~\cite{Scheg} (red line), which included only TOTEM data at $7~\TeV$ for the LHC
and a restricted selection of lower-energy data, particularly ISR data.
Also shown in Figure~\ref{fig:BslopeS} is an attempt to
repeat the fit with the same parameterization as in Ref.~\cite{Scheg},
but including all data (blue solid line). For comparison,
a simple fit is also shown using only the linear term in $\ln\left(s/s_0\right)$ (dashed blue line).
 
While a quadratic energy
evolution appears to be favoured by the data, it must also be concluded that the selection of
lower-energy data has
a significant impact on the shape of the evolution, and the fit including all data does not give a perfect
description of the LHC data.  A good description
of the 7 and $8~\TeV$ data is obtained with the selection in Ref.~\cite{Scheg}, but the prediction at $13~\TeV$ is higher than the measurements.
 
The prediction of the BJAS model (light blue dashed-dotted line) does not provide a good description of the nuclear slope's energy evolution,
as can be seen in Figure~\ref{fig:BslopeS}.
 
\FloatBarrier

% End of text imported from the .//Interpretation.tex input file

\FloatBarrier

% The next lines are included from the .//Conclusion.tex input file
\section{Conclusion}\label{sec:Conclusion}
This paper presents measurements of the total cross section, $\rho$-parameter and nuclear slope parameters
using elastic $pp$ scattering data at $\rts=13~\TeV$ recorded by the ALFA subdetector of ATLAS
in 2016 in a special LHC run with $\betastar=2.5$~km optics, corresponding to an integrated luminosity of $340~\upmu$b$^{-1}$.
The analysis introduces dedicated data-driven methods to determine the relevant
beam optics parameters, event reconstruction efficiency, simulation tuning and vertical Roman Pot alignment.
The alignment precision achieved, about $22~\upmu$m,  is required to determine the $\rho$-parameter,
because the main sensitivity is at small $|t|$ in the CNI region where the alignment has the largest impact.
 
From a fit to the differential elastic cross section in the range from $-t = 4.5 \cdot 10^{-4}~\GeV^{2}$ to $-t = 0.2~\GeV^{2}$,
the total cross section and $\rho$-parameter are determined to be:
\begin{eqnarray*}
\sigmatot(pp\rightarrow X) & = &  \mbox{104.68} \; \pm \mbox{1.08} \; (\mbox{exp.})  \pm \mbox{0.12} \; (\mbox{th.})  \; \mbox{mb}, \\
\rho & = & \mbox{0.0978} \; \pm \mbox{0.0085} \; (\mbox{exp.})  \pm \mbox{0.0064} \; (\mbox{th.}) ,
\end{eqnarray*}
where the first error accounts for all experimental systematic uncertainties and includes the statistical component, and the second
is related to the model uncertainties.
The experimental systematic uncertainty is dominated by the uncertainty in the luminosity and the
alignment.
This analysis included a dedicated luminosity determination for the $\betastar=2.5$~km run, which is directly used
to normalize the cross-section measurements.
This new value of the total cross section is about~5.8 mb lower than the measurement from the TOTEM Collaboration,
corresponding approximately to a $2.2\sigma$ tension, assuming uncorrelated uncertainties. A similar difference was already
observed at 7 and $8~\TeV$. The main difference is traced back to the normalization of the differential
elastic cross section measured by ATLAS and TOTEM.
 
A study of the form of the $t$-spectrum revealed the need to introduce a $t$-dependent parameterization
of the exponential nuclear slope (see Eq.~\eqref{eq:Cterm}), assumed to be constant at 7 and $8~\TeV$. It is found in this analysis that with two more parameters $C$ and $D$ in addition to $B$ used in previous measurements
a satisfactory description is achieved; they are measured to be:
\begin{eqnarray*}
B & = &  \mbox{21.14} \; \pm \mbox{0.13}~\GeV^{-2}  , \\
C & = &  \mbox{$-6.7$} \; \pm \mbox{2.2}~\GeV^{-4}  , \\
D & = &  \mbox{17.4} \; \pm \mbox{7.8}~\GeV^{-6}  .
\end{eqnarray*}
 
The new data for $\sigmatot$ and $\rho$ are compared with lower-energy data, and the energy evolution of these data is analysed
in the context of model studies of the evolution. The widely used COMPETE model yields a good description of the ATLAS and
lower-energy total cross-section data, but fails to describe the $\rho$ measurements at $13~\TeV$.
The FMO model, tuned to the TOTEM data, featuring a maximal Odderon contribution in the nuclear amplitude, was also investigated and found to be in good agreement with the $\rho$ measurements but it exceeds the ATLAS total cross-section measurement.
A simultaneous fit exploiting dispersion relations was performed to $\sigmatot$ and $\rho$ data, using the BCBM
parameterization of a purely even but damped amplitude,
and was found to give a good description of both $\sigmatot$ and $\rho$.
This study shows that the commonly accepted energy evolution as implemented in the COMPETE model is in tension with
the $13~\TeV$ elastic-scattering data.
Nonetheless, further research is needed to understand whether the low value of $\rho$ can be attributed to the Odderon
or other effects in strong interactions.
 
This analysis also measures the inelastic cross section, and its value is the most precise of the five available LHC measurements.
 
The ratio $\sigmael/\sigmatot$, a measure of the opaqueness of the proton, continues to grow slowly with energy, and
its evolution is well described by the KMR model.
The measurement remains far from probing the black-disc limit, i.e.\ a totally opaque proton.
 
The study of the evolution of the leading nuclear slope $B$ confirms the tendency already found at 7 and $8~\TeV$:
in the energy range of the LHC, the shrinkage of the forward cone starts to accelerate relative to the linear dependence
found previously at energies below the LHC range, as expected from the onset of multi-Pomeron exchange.

% End of text imported from the .//Conclusion.tex input file

\section*{Acknowledgements}

% The next lines are included from the .//acknowledgements/Acknowledgements.tex input file

We thank CERN for the very successful operation of the LHC, as well as the
support staff from our institutions without whom ATLAS could not be
operated efficiently.
 
We acknowledge the support of
ANPCyT, Argentina;
YerPhI, Armenia;
ARC, Australia;
BMWFW and FWF, Austria;
ANAS, Azerbaijan;
CNPq and FAPESP, Brazil;
NSERC, NRC and CFI, Canada;
CERN;
ANID, Chile;
CAS, MOST and NSFC, China;
Minciencias, Colombia;
MEYS CR, Czech Republic;
DNRF and DNSRC, Denmark;
IN2P3-CNRS and CEA-DRF/IRFU, France;
SRNSFG, Georgia;
BMBF, HGF and MPG, Germany;
GSRI, Greece;
RGC and Hong Kong SAR, China;
ISF and Benoziyo Center, Israel;
INFN, Italy;
MEXT and JSPS, Japan;
CNRST, Morocco;
NWO, Netherlands;
RCN, Norway;
MEiN, Poland;
FCT, Portugal;
MNE/IFA, Romania;
MESTD, Serbia;
MSSR, Slovakia;
ARRS and MIZ\v{S}, Slovenia;
DSI/NRF, South Africa;
MICINN, Spain;
SRC and Wallenberg Foundation, Sweden;
SERI, SNSF and Cantons of Bern and Geneva, Switzerland;
MOST, Taiwan;
TENMAK, T\"urkiye;
STFC, United Kingdom;
DOE and NSF, United States of America.
In addition, individual groups and members have received support from
BCKDF, CANARIE, Compute Canada and CRC, Canada;
PRIMUS 21/SCI/017 and UNCE SCI/013, Czech Republic;
COST, ERC, ERDF, Horizon 2020 and Marie Sk{\l}odowska-Curie Actions, European Union;
Investissements d'Avenir Labex, Investissements d'Avenir Idex and ANR, France;
DFG and AvH Foundation, Germany;
Herakleitos, Thales and Aristeia programmes co-financed by EU-ESF and the Greek NSRF, Greece;
BSF-NSF and MINERVA, Israel;
Norwegian Financial Mechanism 2014-2021, Norway;
NCN and NAWA, Poland;
La Caixa Banking Foundation, CERCA Programme Generalitat de Catalunya and PROMETEO and GenT Programmes Generalitat Valenciana, Spain;
G\"{o}ran Gustafssons Stiftelse, Sweden;
The Royal Society and Leverhulme Trust, United Kingdom.
 
The crucial computing support from all WLCG partners is acknowledged gratefully, in particular from CERN, the ATLAS Tier-1 facilities at TRIUMF (Canada), NDGF (Denmark, Norway, Sweden), CC-IN2P3 (France), KIT/GridKA (Germany), INFN-CNAF (Italy), NL-T1 (Netherlands), PIC (Spain), ASGC (Taiwan), RAL (UK) and BNL (USA), the Tier-2 facilities worldwide and large non-WLCG resource providers. Major contributors of computing resources are listed in Ref.~\cite{ATL-SOFT-PUB-2021-003}.

% End of text imported from the .//acknowledgements/Acknowledgements.tex input file

\clearpage
\appendix
\part*{Appendix}\label{sec:Appendix}
\addcontentsline{toc}{part}{Appendix}

% The next lines are included from the .//Appendix.tex input file
Numerical values of the differential elastic cross section are given in
the range from $-t= 2.5 \cdot 10^{-4}~\GeV^2$ to $-t=0.05785~\GeV^2$ in Table~\ref{tab:tval1} and
from $-t=0.05785~\GeV^2$ to $-t=0.46~\GeV^2$ in Table~\ref{tab:tval2}.
\begin{table}[tp]
\caption{The measured values of the differential elastic cross section with statistical and systematic uncertainties. The central $t$-values in each bin are calculated from simulation.
This table gives values from $-t=0.00025~\GeV^2$ to $-t=0.05785~\GeV^2$.}
\label{tab:tval1}
\begin{center}
\hspace*{-0.1cm}
\begin{tabular}{ccccccc}
\hline \hline
Low $|t|$ edge  & High $|t|$ edge & Central $|t|$ & $\mathrm{d}\sigma_{\mathrm{el}}/\mathrm{d}t$ & Stat. uncert.  & Syst. uncert. & Total uncert.\\
$[$\GeV$^2$]   & [\GeV$^{2}$] &  [\GeV$^{2}$] & [mb/\GeV$^2$] & [mb/\GeV$^2$] & [mb/\GeV$^2$] & [mb/\GeV$^2$] \\ \hline
0.00025 & 0.00035 & 0.00029 & 3662 & 944 & 207 & 966 \\
0.00035 & 0.00045 & 0.00040 & 2136 & 42 & 98 & 107 \\
0.00045 & 0.00058 & 0.00051 & 1401 & 12 & 52 & 53 \\
0.00058 & 0.00076 & 0.00067 & 998 & 6 & 31 & 31 \\
0.00076 & 0.00098 & 0.00086 & 797 & 4 & 21 & 21 \\
0.00098 & 0.00126 & 0.00112 & 680.1 & 2.8 & 16.4 & 16.6 \\
0.00126 & 0.00161 & 0.00143 & 610.6 & 2.3 & 14.1 & 14.3 \\
0.00161 & 0.00203 & 0.00182 & 576.1 & 2.0 & 13.1 & 13.3 \\
0.00203 & 0.00252 & 0.00227 & 552.5 & 1.7 & 12.5 & 12.6 \\
0.00252 & 0.00308 & 0.00279 & 537.2 & 1.6 & 12.2 & 12.3 \\
0.00308 & 0.00371 & 0.00339 & 525.4 & 1.4 & 11.9 & 12.0 \\
0.00371 & 0.00442 & 0.00407 & 518.0 & 1.3 & 11.7 & 11.8 \\
0.00442 & 0.00521 & 0.00482 & 504.8 & 1.2 & 11.4 & 11.5 \\
0.00521 & 0.00608 & 0.00564 & 497.2 & 1.1 & 11.2 & 11.3 \\
0.00608 & 0.00703 & 0.00655 & 487.9 & 1.0 & 11.0 & 11.0 \\
0.00703 & 0.00805 & 0.00754 & 476.4 & 1.0 & 10.7 & 10.7 \\
0.00805 & 0.00916 & 0.00860 & 466.8 & 0.9 & 10.4 & 10.5 \\
0.00916 & 0.01035 & 0.00975 & 456.0 & 0.9 & 10.2 & 10.3 \\
0.01035 & 0.01162 & 0.01098 & 443.0 & 0.9 & 9.9 & 10.0 \\
0.01162 & 0.01297 & 0.01229 & 430.3 & 0.8 & 9.7 & 9.7 \\
0.01297 & 0.01441 & 0.01369 & 418.7 & 0.8 & 9.4 & 9.5 \\
0.01441 & 0.01592 & 0.01516 & 407.9 & 0.7 & 9.2 & 9.2 \\
0.01592 & 0.01752 & 0.01672 & 395.9 & 0.7 & 9.0 & 9.0 \\
0.01752 & 0.01921 & 0.01836 & 383.0 & 0.7 & 8.7 & 8.7 \\
0.01921 & 0.02099 & 0.02009 & 369.5 & 0.6 & 8.4 & 8.4 \\
0.02099 & 0.02285 & 0.02191 & 354.1 & 0.6 & 8.0 & 8.0 \\
0.02285 & 0.02479 & 0.02381 & 340.6 & 0.6 & 7.7 & 7.7 \\
0.02479 & 0.02681 & 0.02579 & 327.9 & 0.6 & 7.4 & 7.4 \\
0.02681 & 0.02892 & 0.02786 & 313.3 & 0.5 & 7.1 & 7.1 \\
0.02892 & 0.03112 & 0.03002 & 299.7 & 0.5 & 6.8 & 6.8 \\
0.03112 & 0.03340 & 0.03225 & 285.9 & 0.5 & 6.5 & 6.5 \\
0.03340 & 0.03577 & 0.03458 & 272.6 & 0.5 & 6.2 & 6.2 \\
0.03577 & 0.03823 & 0.03699 & 259.6 & 0.5 & 5.9 & 5.9 \\
0.03823 & 0.04078 & 0.03949 & 246.5 & 0.4 & 5.5 & 5.6 \\
0.04078 & 0.04341 & 0.04208 & 234.0 & 0.4 & 5.3 & 5.3 \\
0.04341 & 0.04612 & 0.04475 & 220.6 & 0.4 & 5.0 & 5.0 \\
0.04612 & 0.04892 & 0.04751 & 209.8 & 0.4 & 4.7 & 4.7 \\
0.04892 & 0.05181 & 0.05035 & 197.6 & 0.4 & 4.4 & 4.4 \\
0.05181 & 0.05478 & 0.05328 & 185.81 & 0.34 & 4.16 & 4.17 \\
0.05478 & 0.05785 & 0.05630 & 175.05 & 0.32 & 3.92 & 3.93 \\
 
\hline
\end{tabular}
\end{center}
\end{table}
\begin{table}[tp]
\begin{center}
\caption{The measured values of the differential elastic cross section with statistical and systematic uncertainties. The central $t$-values in each bin are calculated from simulation.
This table gives values from $-t=0.05785~\GeV^2$ to $-t=0.42~\GeV^2$.}
\label{tab:tval2}
\hspace*{-0.1cm}
\begin{tabular}{ccccccc}
\hline \hline
Low $|t|$ edge  & High $|t|$ edge & Central $|t|$ & $\mathrm{d}\sigma_{\mathrm{el}}/\mathrm{d}t$ & Stat. uncert.  & Syst. uncert. & Total uncert.\\
$[$\GeV$^2$]   & [\GeV$^{2}$] &  [\GeV$^{2}$] & [mb/\GeV$^2$] & [mb/\GeV$^2$] & [mb/\GeV$^2$] & [mb/\GeV$^2$] \\ \hline
0.05785 & 0.06100 & 0.05941 & 164.08 & 0.31 & 3.67 & 3.69 \\
0.06100 & 0.06424 & 0.06260 & 153.67 & 0.29 & 3.44 & 3.45 \\
0.06424 & 0.06757 & 0.06588 & 143.41 & 0.28 & 3.21 & 3.22 \\
0.06757 & 0.07098 & 0.06925 & 134.49 & 0.26 & 3.01 & 3.02 \\
0.07098 & 0.07448 & 0.07271 & 125.04 & 0.25 & 2.80 & 2.81 \\
0.07448 & 0.07807 & 0.07625 & 116.44 & 0.25 & 2.62 & 2.63 \\
0.07807 & 0.08174 & 0.07988 & 107.89 & 0.23 & 2.43 & 2.44 \\
0.08174 & 0.08550 & 0.08360 & 99.64 & 0.22 & 2.25 & 2.26 \\
0.08550 & 0.08936 & 0.08741 & 93.00 & 0.21 & 2.11 & 2.12 \\
0.08936 & 0.09329 & 0.09130 & 85.92 & 0.21 & 1.96 & 1.97 \\
0.09329 & 0.09731 & 0.09527 & 79.35 & 0.19 & 1.82 & 1.83 \\
0.09731 & 0.10140 & 0.09933 & 72.70 & 0.18 & 1.67 & 1.68 \\
0.10140 & 0.10560 & 0.10347 & 66.55 & 0.17 & 1.54 & 1.55 \\
0.10560 & 0.11000 & 0.10776 & 61.35 & 0.16 & 1.43 & 1.44 \\
0.11000 & 0.11400 & 0.11198 & 56.31 & 0.16 & 1.32 & 1.33 \\
0.11400 & 0.11870 & 0.11631 & 51.31 & 0.15 & 1.22 & 1.23 \\
0.11870 & 0.12330 & 0.12096 & 46.96 & 0.14 & 1.12 & 1.13 \\
0.12330 & 0.12800 & 0.12561 & 42.70 & 0.13 & 1.03 & 1.04 \\
0.12800 & 0.13300 & 0.13045 & 38.74 & 0.12 & 0.95 & 0.95 \\
0.13300 & 0.13800 & 0.13545 & 34.99 & 0.11 & 0.87 & 0.87 \\
0.13800 & 0.14200 & 0.13998 & 31.97 & 0.11 & 0.80 & 0.81 \\
0.14200 & 0.14700 & 0.14446 & 29.30 & 0.10 & 0.74 & 0.75 \\
0.14700 & 0.15200 & 0.14946 & 26.39 & 0.10 & 0.68 & 0.68 \\
0.15200 & 0.16000 & 0.15589 & 23.11 & 0.08 & 0.60 & 0.61 \\
0.16000 & 0.16800 & 0.16389 & 19.71 & 0.08 & 0.53 & 0.53 \\
0.16800 & 0.17700 & 0.17236 & 16.45 & 0.07 & 0.46 & 0.46 \\
0.17700 & 0.18600 & 0.18136 & 13.74 & 0.07 & 0.39 & 0.40 \\
0.18600 & 0.19500 & 0.19037 & 11.44 & 0.06 & 0.34 & 0.35 \\
0.19500 & 0.20500 & 0.19982 & 9.40 & 0.05 & 0.30 & 0.30 \\
0.20500 & 0.21900 & 0.21168 & 7.33 & 0.05 & 0.24 & 0.25 \\
0.21900 & 0.23300 & 0.22569 & 5.51 & 0.04 & 0.19 & 0.20 \\
0.23300 & 0.24900 & 0.24055 & 3.938 & 0.034 & 0.146 & 0.149 \\
0.24900 & 0.26400 & 0.25606 & 2.866 & 0.030 & 0.115 & 0.119 \\
0.26400 & 0.28000 & 0.27149 & 1.976 & 0.026 & 0.088 & 0.092 \\
0.28000 & 0.29800 & 0.28841 & 1.369 & 0.023 & 0.069 & 0.073 \\
0.29800 & 0.31500 & 0.30597 & 0.920 & 0.022 & 0.052 & 0.056 \\
0.31500 & 0.35000 & 0.33016 & 0.456 & 0.014 & 0.030 & 0.033 \\
0.35000 & 0.38500 & 0.36504 & 0.196 & 0.012 & 0.020 & 0.024 \\
0.38500 & 0.42000 & 0.40012 & 0.069 & 0.011 & 0.013 & 0.017 \\
0.42000 & 0.46000 & 0.43760 & 0.033 & 0.018 & 0.010 & 0.020 \\
\hline
\end{tabular}
\end{center}
\end{table}
 
\clearpage

% End of text imported from the .//Appendix.tex input file

\printbibliography
 
\clearpage
% ATLAS Collaboration author list
% Reference date of STDM-2018-08 is 2022-05-10
% Author list last updated on date 24-MAR-23
% Data extracted on 24-Mar-2023 for paper reference STDM-2018-08
% at 4:21pm
 
\begin{flushleft}
\hypersetup{urlcolor=black}
{\Large The ATLAS Collaboration}

\bigskip

\AtlasOrcid[0000-0002-6665-4934]{G.~Aad}$^\textrm{\scriptsize 101}$,
\AtlasOrcid[0000-0002-5888-2734]{B.~Abbott}$^\textrm{\scriptsize 119}$,
\AtlasOrcid[0000-0002-7248-3203]{D.C.~Abbott}$^\textrm{\scriptsize 102}$,
\AtlasOrcid[0000-0002-1002-1652]{K.~Abeling}$^\textrm{\scriptsize 55}$,
\AtlasOrcid[0000-0002-8496-9294]{S.H.~Abidi}$^\textrm{\scriptsize 29}$,
\AtlasOrcid[0000-0002-9987-2292]{A.~Aboulhorma}$^\textrm{\scriptsize 35e}$,
\AtlasOrcid[0000-0001-5329-6640]{H.~Abramowicz}$^\textrm{\scriptsize 150}$,
\AtlasOrcid[0000-0002-1599-2896]{H.~Abreu}$^\textrm{\scriptsize 149}$,
\AtlasOrcid[0000-0003-0403-3697]{Y.~Abulaiti}$^\textrm{\scriptsize 116}$,
\AtlasOrcid[0000-0003-0762-7204]{A.C.~Abusleme~Hoffman}$^\textrm{\scriptsize 136a}$
% ...
\bigskip
\\

% ...
$^{29}$Physics Department, Brookhaven National Laboratory, Upton NY; United States of America.\\
% ...
$^{35}$$^{(a)}$Facult\'e des Sciences Ain Chock, R\'eseau Universitaire de Physique des Hautes Energies - Universit\'e Hassan II, Casablanca;$^{(b)}$Facult\'{e} des Sciences, Universit\'{e} Ibn-Tofail, K\'{e}nitra;$^{(c)}$Facult\'e des Sciences Semlalia, Universit\'e Cadi Ayyad, LPHEA-Marrakech;$^{(d)}$LPMR, Facult\'e des Sciences, Universit\'e Mohamed Premier, Oujda;$^{(e)}$Facult\'e des sciences, Universit\'e Mohammed V, Rabat;$^{(f)}$Institute of Applied Physics, Mohammed VI Polytechnic University, Ben Guerir; Morocco.\\
% ...
$^{55}$II. Physikalisches Institut, Georg-August-Universit\"{a}t G\"ottingen, G\"ottingen; Germany.\\
% ...
$^{101}$CPPM, Aix-Marseille Universit\'e, CNRS/IN2P3, Marseille; France.\\
$^{102}$Department of Physics, University of Massachusetts, Amherst MA; United States of America.\\
% ...
$^{116}$Department of Physics, New York University, New York NY; United States of America.\\
% ...
$^{119}$Homer L. Dodge Department of Physics and Astronomy, University of Oklahoma, Norman OK; United States of America.\\
% ...
$^{136}$$^{(a)}$Departamento de F\'isica, Pontificia Universidad Cat\'olica de Chile, Santiago;$^{(b)}$Millennium Institute for Subatomic physics at high energy frontier (SAPHIR), Santiago;$^{(c)}$Instituto de Investigaci\'on Multidisciplinario en Ciencia y Tecnolog\'ia, y Departamento de F\'isica, Universidad de La Serena;$^{(d)}$Universidad Andres Bello, Department of Physics, Santiago;$^{(e)}$Instituto de Alta Investigaci\'on, Universidad de Tarapac\'a, Arica;$^{(f)}$Departamento de F\'isica, Universidad T\'ecnica Federico Santa Mar\'ia, Valpara\'iso; Chile.\\
% ...
$^{149}$Department of Physics, Technion, Israel Institute of Technology, Haifa; Israel.\\
$^{150}$Raymond and Beverly Sackler School of Physics and Astronomy, Tel Aviv University, Tel Aviv; Israel.\\
% ...
$^{a}$ Also Affiliated with an institute covered by a cooperation agreement with CERN.\\
$^{b}$ Also at Borough of Manhattan Community College, City University of New York, New York NY; United States of America.\\
$^{c}$ Also at Bruno Kessler Foundation, Trento; Italy.\\
$^{d}$ Also at Center for High Energy Physics, Peking University; China.\\
$^{e}$ Also at Centro Studi e Ricerche Enrico Fermi; Italy.\\
$^{f}$ Also at CERN, Geneva; Switzerland.\\
% ...

\end{flushleft}

% Created with Glance <Atlas.Glance@cern.ch>

\clearpage
% The next lines are included from the .//ANA-STDM-2018-08-PAPER-hepdata.tex input file
\section{HepData material}
 
The HEPData record \href{https://www.hepdata.net/record/128017}{128017} contains the following tables:
\begin{enumerate}
\item The total cross section
\item The $\rho$-parameter
\item The nuclear slope parameter $B$
\item The nuclear slope parameter $C$
\item The nuclear slope parameter $D$
\item The measured total elastic cross section inside the fiducial volume
\item The nuclear part of the total elastic cross section extrapolated to full phase space
\item The total inelastic cross section
\item The ratio of the elastic to total cross section
\item The differential elastic cross section with statistical uncertainties and 20 experimental systematic signed relative uncertainties associated to nuisance parameters used in the profile fit
\item The statistical covariance matrix for the differential elastic cross section
\end{enumerate}

% End of text imported from the .//ANA-STDM-2018-08-PAPER-hepdata.tex input file
\clearpage

\end{document}